\documentclass[12pt]{article}

\usepackage{enumitem}
\usepackage{graphicx}
\usepackage{float}
\usepackage{cite}
\usepackage{amsfonts}
\usepackage{amssymb}
\usepackage{amsmath}

\def\gtwid{\mathrel{\raise.3ex\hbox{$>$\kern-.75em\lower1ex\hbox{$\sim$}}}}
\def\ltwid{\mathrel{\raise.3ex\hbox{$<$\kern-.75em\lower1ex\hbox{$\sim$}}}}
\def\square{\kern1pt\vbox{\hrule height 1.2pt\hbox{\vrule width 1.2pt\hskip 3pt
   \vbox{\vskip 6pt}\hskip 3pt\vrule width 0.6pt}\hrule height 0.6pt}\kern1pt}

\begin{document}

\begin{titlepage}

\begin{flushright}
UFIFT-QG-24-01
\end{flushright}

\vskip 0.2cm

\begin{center}
{\bf Gauge Independent Logarithms from Inflationary Gravitons}
\end{center}

\vskip 0.2cm

\begin{center}
D. Glavan$^{1*}$, S. P. Miao$^{2\star}$, T. Prokopec$^{3\dagger}$ and
R. P. Woodard$^{4\ddagger}$
\end{center}
\vspace{0.cm}
\begin{center}
\it{$^{1}$ CEICO, Institute of Physics of the Czech Academy of Sciences (FZU), \\
Na Slovance 1999/2, 182 21 Prague 8, CZECH REPUBLIC}
\end{center}
\vspace{-0.3cm}
\begin{center}
\it{$^{2}$ Department of Physics, National Cheng Kung University, \\
No. 1 University Road, Tainan City 70101, TAIWAN}
\end{center}
\vspace{-0.3cm}
\begin{center}
\it{$^{3}$ Institute for Theoretical Physics, Spinoza Institute \& EMME$\Phi$, \\
Utrecht University, Postbus 80.195, 3508 TD Utrecht, THE NETHERLANDS}
\end{center}
\vspace{-0.3cm}
\begin{center}
\it{$^{4}$ Department of Physics, University of Florida,\\
Gainesville, FL 32611, UNITED STATES}
\end{center}

\vspace{0.2cm}

\begin{center}
ABSTRACT
\end{center}

\begin{flushleft}
\end{flushleft}
Dependence on the graviton gauge enters the conventional effective field 
equations because they fail to account for quantum gravitational correlations
with the source which excites the effective field and with the observer who
measures it. Including these correlations has been shown to eliminate gauge
dependence in flat space background. We generalize the technique to de Sitter
background for the case of the 1-loop graviton corrections to the exchange
potential of a massless, minimally coupled scalar. 

\vspace{0.2cm}

\begin{flushleft}
$^{*}$ e-mail: glavan@fzu.cz \\
$^{\star}$ e-mail: spmiao5@mail.ncku.edu.tw \\
$^{\dagger}$ e-mail: T.Prokopec@uu.nl \\
$^{\ddagger}$ e-mail: woodard@phys.ufl.edu
\end{flushleft}

\end{titlepage}

\section{Introduction}

The background geometry of cosmology with scale factor $a(t)$ is,
\begin{equation}
ds^2 = -dt^2 + a^2(t) d\vec{x} \!\cdot\! d\vec{x} \qquad \Longrightarrow
H(t) \equiv \frac{\dot{a}}{a} \quad , \quad \epsilon(t) \equiv 
-\frac{\dot{H}}{H^2} \; . \label{geometry}
\end{equation}
The Hubble parameter $H(t)$ measures the expansion rate while the first slow
roll parameter $\epsilon(t)$ measures the rate at which $H(t)$ changes.
Inflation is characterized by positive $H(t)$ and $0 \leq \epsilon(t) < 1$,
which means that both the first and second time derivatives of $a(t)$ are
positive. The most highly accelerated geometry is de Sitter, with 
$\epsilon(t) = 0$ and $H$ constant, which serves as a paradigm for primordial
inflation.

The accelerated expansion of inflation can rip massless, not conformally
invariant virtual quanta out of the vacuum. That process is responsible for 
the primordial power spectra of scalars \cite{Mukhanov:1981xt} and gravitons 
\cite{Starobinsky:1979ty}, but these quanta must, at some level, interact with 
themselves and with other particles to affect kinematics and long-range 
forces. One studies these changes on a field by computing its 1PI 
(one-particle-irreducible) 2-point function and then using that to 
quantum-correct the linearized field equation. For a scalar field the 1PI 
2-point function is known as the self-mass $-i M^2(x;x')$. If the scalar is 
massless and minimally coupled (MMC) then its quantum-corrected, linearized
field equation is,
\begin{equation}
\partial_{\mu} \Bigl[\sqrt{-g} \, g^{\mu\nu} \partial_{\nu} \varphi(x)\Bigr]
- \int \!\! d^4x' \, M^2(x;x') \varphi(x') = J(x) \; . \label{EFE}
\end{equation}
Here $g_{\mu\nu}(x)$ is the background metric and $J(x)$ is the source. Zero
source solutions tell one about modifications to the kinematics of single
particle solutions whereas setting $J$ to a static point source determines 
the exchange potential.

Effects mediated by inflationary scalars are generally stronger and simpler 
to compute, but they suffer from being model dependent. In contrast, 
graviton-mediated effects are weaker, and much more difficult to work out, 
but are completely generic because general relativity is the unique low 
energy effective field theory of gravity \cite{Donoghue:1994dn,Donoghue:1995cz,
Burgess:2003jk,Donoghue:2012zc,Donoghue:2017,Donoghue:2022eay}. Despite the
complexity of working with gravitons on an inflationary background, six cases
have been found of significant effects on de Sitter:
\begin{itemize}[topsep=0pt,itemsep=-1ex,partopsep=1ex]
\item{Growth of the fermion field strength \cite{Miao:2006gj};}
\item{Temporal and spatial growth of the electric force \cite{Glavan:2013jca};}
\item{Growth of the of the photon field strength \cite{Wang:2014tza};}
\item{Growth of the graviton field strength \cite{Tan:2021lza};}
\item{Spatial suppression of the force carried by an MMC scalar 
\cite{Glavan:2021adm}; and}
\item{Temporal suppression of the gravitational force \cite{Tan:2022xpn}.}
\end{itemize}

A standard concern when studying vector-mediated or tensor-mediated corrections 
to effective field equations is how to distinguish physical effects from gauge 
artifacts such as the acausality apparently implied by the instantaneous Coulomb 
potential. Exhibiting the gauge dependence of 1PI $n$-point functions is 
relatively simple in flat space \cite{Jackiw:1974cv}, but very difficult for 
inflationary backgrounds. The simplest gauge \cite{Tsamis:1992xa,Woodard:2004ut}
happens to break de Sitter invariance, which has been a point of contention for
decades \cite{Allen:1986tt,Hawking:2000ee,Higuchi:2001uv,Higuchi:2002sc,
Miao:2011fc,Higuchi:2011vw,Miao:2011ng,Morrison:2013rqa,Miao:2013isa}. However,
other gauges are so difficult to use that, of the ten graviton loops so far
evaluated \cite{Tsamis:1996qm,Tsamis:1996qk,Tsamis:2005je,Miao:2005am,
Kahya:2007bc,Miao:2012bj,Leonard:2013xsa,Glavan:2015ura,Miao:2017vly,
Glavan:2020gal,Glavan:2021adm},\footnote{An independent computation of graviton 
corrections to the self-mass of a massless, conformally coupled scalar 
\cite{Boran:2014xpa,Boran:2017fsx} also used the simplest gauge, but disagrees
with \cite{Glavan:2020gal}.} only one was performed using a different gauge. 
This one was of the graviton contributions to the vacuum polarization 
\cite{Glavan:2015ura} in a 1-parameter family of de Sitter invariant gauges 
\cite{Mora:2012zi}. When it was used to check the growth of the photon field 
strength, the same time dependence was found as in the simplest gauge 
\cite{Wang:2014tza}, but with a different multiplicative factor 
\cite{Glavan:2016bvp}. One can therefore conclude that there is nothing 
particularly misleading about the simplest gauge, but that 1PI functions on 
de Sitter harbor the same sort of gauge dependence as they do on flat space.

It was recently shown that the usual effective field equations are gauge 
dependent because they ignore quantum gravitational correlations with the
source which excites the effective field and with the observer who measures 
it \cite{Miao:2017feh}. When these correlations are restored on flat space
background, the effective field equations for both an MMC scalar 
\cite{Miao:2017feh} and for electromagnetism \cite{Katuwal:2020rkv} become 
gauge independent. One restores source and observer correlations by writing 
down the position-space diagrams which contribute to $t$-channel scattering 
between two massive source particles. These diagrams consist of 4-point, 
3-point and 2-point Green's functions. A series of identities due to Donoghue 
and collaborators \cite{Donoghue:1993eb,Donoghue:1994dn,Bjerrum-Bohr:2002aqa,
Donoghue:1996mt} permits one to extract the $t$-channel poles from the 3-point 
and 4-point diagrams, which reduces them to 2-point form. Then the propagator
equation is used to regard these 2-point forms as corrections to the 1PI 
2-point function, and hence to the effective field equation. Our goal here is 
to generalize this technique to de Sitter for graviton corrections to an MMC 
scalar.

This paper contains seven sections. In section 2 we review the Feynman rules,
including the massive source-observer field, along with gravity and the MMC 
scalar. We also generalize the Donoghue identities from flat space to de 
Sitter. It turns out that five classes of diagrams contribute to $t$-channel 
scattering, in addition to the self-mass. These diagrams are presented in 
section 3. Section 4 demonstrates an important cancellation which occurs for 
any background geometry and in any gauge. The remaining diagrams are reduced 
in section 5, and incorporated as corrections to the renormalized self-mass. 
Section 6 gives our conclusions, including the crucial issue of whether or 
not the 1-loop correction to the scalar exchange potential contains a large
logarithm.

\section{Feynman Rules and Reduction Strategy}

The purpose of this section is to present the basics of our computation.
We begin with the full action which describes the massive source-observer 
field $\Psi$, in addition to gravity and the massless, minimally coupled 
scalar whose exchange potential it modifies. The necessary Feynman rules 
are presented, then the general strategy is described. The section closes 
with some useful derivative identities.

\subsection{Propagators and Vertices}

We model the source and observer as a massive, minimally coupled scalar $\Psi$
which has a cubic coupling to the massless, minimally coupled scalar $\varphi$.
The bare Lagrangian is,
\begin{eqnarray}
\lefteqn{\mathcal{L} = \frac{[R \!-\! (D\!-\!2) \Lambda] \sqrt{-g}}{16 \pi G} 
-\frac12 \partial_{\mu} \varphi \partial_{\nu} \varphi g^{\mu\nu} \sqrt{-g} }
\nonumber \\
& & \hspace{4.3cm} - \frac12 \partial_{\mu} \Psi \partial_{\nu} \Psi g^{\mu\nu} 
\sqrt{-g} - \frac12 \Bigl(m^2 \!+\! \lambda \varphi\Bigr) \Psi^2 \sqrt{-g} \; .
\qquad \label{Lagrangian} 
\end{eqnarray}
Here $G$ is Newton's constant, $\Lambda$ is the positive cosmological constant 
and $D$ is the dimension of spacetime, left arbitrary to facilitate the use
of dimensional regularization. It is convenient to change the time variable 
from co-moving time $t$ to conformal time $\eta$, with $d\eta \equiv dt/a(t)$. 
This makes the de Sitter background conformal to flat space,
\begin{equation}
ds^2 = a^2 \Bigl[-d\eta^2 + d\vec{x} \!\cdot\! d\vec{x}\Bigr] \qquad , \qquad
a = -\frac1{ H \eta} \quad , \quad H \equiv \sqrt{\frac{\Lambda}{D\!-\!1}} \; .
\end{equation}
We adopt the convention that the indices of partial derivatives are raised and 
lowered with the Minkowski metric, $\partial^{\mu} \equiv \eta^{\mu\nu} 
\partial_{\nu}$.

Our analysis employs four scalar propagators. The source and observer ($\Psi$)
propagator $i\Delta_m(x;x')$ has a mass $m \gg H$ which is assumed much larger 
than the Hubble parameter and spatial momenta. The massless scalar ($\varphi$) 
propagator $i\Delta_A(x;x')$ has zero mass, while the graviton propagator 
requires three scalar propagators $i \Delta_I(x;x')$ whose masses are,
\begin{equation}
M^2_A = 0 \qquad , \qquad M^2_B = (D\!-\!2) H^2 \qquad , \qquad M^2_C = 2 
(D\!-\!3) H^2 \; . 
\end{equation}
All four propagators obey equations involving the differential operator 
$\mathcal{D} \equiv \partial_{\mu} [\sqrt{-g} \, g^{\mu\nu} \partial_{\nu} ] =
\partial^{\mu} a^{D-2} \partial_{\mu}$,
\begin{equation}
\Bigl(\mathcal{D} \!-\! a^D m^2\Bigr) i\Delta_{m}(x;x') = i\delta^D(x \!-\! x') =
\Bigl( \mathcal{D} \!-\! a^D M^2_{I}\Bigr) i\Delta_I(x;x') \; .
\end{equation}
The three propagators $i\Delta_I(x;x')$ can be usefully expanded as differences
of a series in $D$-dependent powers of $a a' \Delta x^2 \equiv a a' (x - x')^{\mu}
(x - x')^{\nu} \eta_{\mu\nu} + i \epsilon$ and a series of integer powers of the 
same quantity,
\begin{eqnarray}
\lefteqn{i\Delta_A = \frac{\Gamma(\frac{D}2 \!-\! 1)}{4 \pi^{\frac{D}2}} \Biggl\{
\frac1{[a a' \Delta x^2]^{\frac{D}2 - 1}} + \frac{D (D \!-\! 2)}{8 (D \!-\! 4)}
\frac{H^2}{[a a' \Delta x^2]^{\frac{D}2 - 2}} + \ldots \Biggr\} } \nonumber \\
& & \hspace{1.4cm} - \frac{(\frac{H}{2})^{D-4}}{4 \pi^{\frac{D}2}} 
\frac{\Gamma(D \!-\! 1)}{2 \Gamma(\frac{D}2)} \Biggl\{0 + \tfrac12 H^2 \Bigl[
\pi \cot\Bigl(\frac{D \pi}{2}\Bigr) \!-\! \ln(a a') \Bigr] + \ldots \Biggr\} ,
\qquad \label{DAexp} \\
\lefteqn{i\Delta_B = \frac{\Gamma(\frac{D}2 \!-\! 1)}{4 \pi^{\frac{D}2}} \Biggl\{
\frac1{[a a' \Delta x^2]^{\frac{D}2 - 1}} + \Bigl(\frac{D \!-\! 2}{8}\Bigr) 
\frac{H^2}{[a a' \Delta x^2]^{\frac{D}2 - 2}} + \ldots \Biggr\} } \nonumber \\
& & \hspace{4.7cm} - \frac{(\frac{H}{2})^{D-4}}{4 \pi^{\frac{D}2}} 
\frac{\Gamma(D \!-\! 1)}{2 \Gamma(\frac{D}2)} \Biggl\{0 + \frac{H^2}{2 (D \!-\! 2)}
+ \ldots \Biggr\} , \qquad \label{DBexp} \\
\lefteqn{i\Delta_C = \frac{\Gamma(\frac{D}2 \!-\! 1)}{4 \pi^{\frac{D}2}} \Biggl\{
\frac1{[a a' \Delta x^2]^{\frac{D}2 - 1}} + \Bigl(\frac{D \!-\! 6}{8}\Bigr) 
\frac{H^2}{[a a' \Delta x^2]^{\frac{D}2 - 2}} + \ldots \Biggr\} } \nonumber \\
& & \hspace{3.5cm} - \frac{(\frac{H}{2})^{D-4}}{4 \pi^{\frac{D}2}} 
\frac{\Gamma(D \!-\! 1)}{2 \Gamma(\frac{D}2)} \Biggl\{0 - \frac{H^2}{2 
(D \!-\! 2) (D \!-\! 3)} + \ldots \Biggr\} . \qquad \label{DCexp}
\end{eqnarray}
The neglected terms in expressions (\ref{DAexp}-\ref{DCexp}) are not only less
singular at coincidence (${x'}^{\mu} = x^{\mu} \Longrightarrow \Delta x^2 = 0$),
they also vanish in $D = 4$ dimensions. Indeed, even summing the order $[a a'
\Delta x^2]^{2-\frac{D}2}$ and $[a a' \Delta x^2]^0$ terms of $i\Delta_{B}$ and
$i\Delta_C$ vanishes in $D = 4$. This fact has great significance for our 
calculation.

The great thing about the simplest gauge is that the graviton propagator (in
conformal coordinates) consists of a sum of spacetime constant tensor factors 
multiplying the three scalar propagators $i \Delta_I(x;x')$ \cite{Tsamis:1992xa,
Woodard:2004ut},
\begin{eqnarray}
\lefteqn{ i[ \mbox{}_{\mu\nu} \Delta_{\rho\sigma}](x;x') = 
\Bigl[ 2 \overline{\eta}_{\mu (\rho} \overline{\eta}_{\sigma) \nu} \!-\! 
\frac{2 \overline{\eta}_{\mu\nu} \overline{\eta}_{\rho\sigma}}{D\!-\!3} \Bigr] 
i\Delta_A(x;x') - 4 \delta^{0}_{~(\mu} \overline{\eta}_{\nu) (\rho} 
\delta^{0}_{~\sigma)} i \Delta_B(x;x') } \nonumber \\
& & \hspace{-0.1cm} + \frac{2}{(D\!-\!3) (D\!-\!2)} \Bigl[ \overline{\eta}_{\mu\nu} 
\!+\! (D\!-\!3) \delta^{0}_{~\mu} \delta^{0}_{~\nu} \Bigr] \Bigl[ 
\overline{\eta}_{\rho\sigma} \!+\! (D\!-\!3) \delta^{0}_{~\rho} \delta^{0}_{~\sigma}
\Bigr] i \Delta_C(x;x') \; . \qquad \label{gravprop1}
\end{eqnarray}
(Parenthesized indices are symmetrized.) The three tensor factors involve the 
Minkowski metric $\eta_{\mu\nu}$, the Kronecker delta function $\delta^{0}_{~\mu}$, 
and a combination $\overline{\eta}_{\mu\nu} \equiv \eta_{\mu\nu} + \delta^{0}_{~\mu} 
\delta^{0}_{~\nu}$ which gives the purely spatial part of the Minkowski metric. The 
only contraction we require is,
\begin{equation}
\eta^{\rho\sigma} \!\times\! i[ \mbox{}_{\mu\nu} \Delta_{\rho\sigma}\Bigr](x;x') 
= -\frac{4 \eta_{\mu\nu}}{D\!-\!2} \, i\Delta_C - \frac{4 \overline{\eta}_{\mu\nu}
}{D\!-\!3} \Bigl(i\Delta_A \!-\! i\Delta_C\Bigr) \; .
\end{equation}

Note that summing the three tensor factors in expression (\ref{gravprop1}) 
produces the tensor factor of the de Donder gauge propagator in flat space,
\begin{eqnarray}
\lefteqn{\Bigl[ 2 \overline{\eta}_{\mu (\rho} \overline{\eta}_{\sigma) \nu} \!-\! 
\frac{2 \overline{\eta}_{\mu\nu} \overline{\eta}_{\rho\sigma}}{D\!-\!3} \Bigr] 
- 4 \delta^{0}_{~(\mu} \overline{\eta}_{\nu) (\rho} \delta^{0}_{~\sigma)} +
\frac{2}{(D\!-\!3) (D\!-\!2)} \Bigl[ \overline{\eta}_{\mu\nu} \!+\! (D\!-\!3) 
\delta^{0}_{~\mu} \delta^{0}_{~\nu} \Bigr] } \nonumber \\
& & \hspace{3.9cm} \times \Bigl[ \overline{\eta}_{\rho\sigma} \!+\! (D\!-\!3) 
\delta^{0}_{~\rho} \delta^{0}_{~\sigma} \Bigr] = 2 \eta_{\mu (\rho} \eta_{\sigma) \nu} 
- \frac{2 \eta_{\mu\nu} \eta_{\rho\sigma}}{D \!-\! 2}  \; . \qquad \label{deDonder}
\end{eqnarray}
This suggests the advantages of singling out a particular scalar propagator --- call 
it $i \Delta_J$ --- and then expanding the other two propagators around it,
\begin{equation}
i\Delta_I = i\Delta_J + (i\Delta_I \!-\! i\Delta_J) \; . \label{expIJ}
\end{equation}
Substituting (\ref{expIJ}) into (\ref{gravprop1}) will then produce $i\Delta_J$ 
times the flat space tensor (\ref{deDonder}), plus a series of differences of
propagators. These differences are less singular at coincidence because expressions
(\ref{DAexp}-\ref{DCexp}) show that all three propagators have the same leading
singularity $\Gamma(\frac{D}2 - 1)/4\pi^{\frac{D}2} [a a' \Delta x^2]^{\frac{D}2-1}$.
The best choice for our calculation is $J = C$, which gives,
\begin{eqnarray}
\lefteqn{ i [ \mbox{}_{\mu\nu} \Delta_{\rho\sigma}](x;x') = \Bigl[ 2 \eta_{\mu (\rho} 
\eta_{\sigma) \nu} \!-\! \frac{2 \eta_{\mu\nu} \eta_{\rho\sigma}}{D\!-\!2}  \Bigr] 
i\Delta_C + 4 \eta_{\mu) (\rho} \overline{\eta}_{\sigma) (\nu} \Bigl(i \Delta_B \!-\! 
i \Delta_C\Bigr) } \nonumber \\
& & \hspace{1.9cm} + 2 \overline{\eta}_{\mu (\rho} \overline{\eta}_{\sigma) \nu} 
\Bigl( i\Delta_A \!-\! 2 i \Delta_B + i \Delta_C\Bigr) - 
\frac{2 \overline{\eta}_{\mu\nu} \overline{\eta}_{\rho\sigma}}{D \!-\! 3} 
 \Bigl( i\Delta_A \!-\! i \Delta_C \Bigr) \; . \qquad \label{gravprop2}
\end{eqnarray}
This form is especially effective for us because the 4 indices of the graviton
propagator are typically contracted into derivatives from vertices and any 
spatial derivatives are assumed small compared to the source observer mass $m$.
In the 3-point and 4-point contributions these vertex derivatives must supply 
at least two factors of $m$ to cancel inverse factors from the Donoghue 
Identities. Hence we can neglect all the terms on the last line of (\ref{gravprop2})
which have only spatial indices. It turns out that even the last term on the
first line of (\ref{gravprop2}) drops out because it is proportional to the 
difference $(i\Delta_B - i\Delta_C)$, which lacks the leading singularity and
actually vanishes in $D = 4$ dimensions. 

\begin{figure}[H]
\centering
\includegraphics[width=2.5cm]{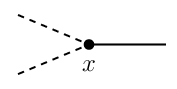}
\includegraphics[width=2.5cm]{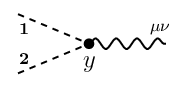}
\includegraphics[width=2.5cm]{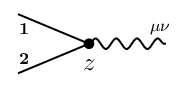}
\includegraphics[width=2.5cm]{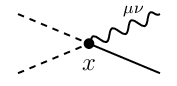}
\includegraphics[width=2.5cm]{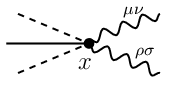}
\caption{\footnotesize Vertices A-E from left to right, given in expressions
(\ref{VertexA}-\ref{VertexE}). Solid lines represent the massless, minimally 
coupled scalar, dashed lines represent the massive source-observer field, and 
wavy lines represent the graviton.}
\label{FeynRules}
\end{figure}

Figure~\ref{FeynRules} depicts the vertices we require. From left to right 
they are,
\begin{eqnarray}
{\rm Vertex\ A} &\!\!\! = \!\!\!& -i\lambda a_x^D \; , \qquad 
\label{VertexA} \\
{\rm Vertex\ B} &\!\!\! = \!\!\!& -i\kappa a_y^{D-2} \Bigl[-\partial^{\mu}_1
\partial^{\nu}_2 + \tfrac12 \eta^{\mu\nu} (\partial_1 \!\cdot\! \partial_2 \!+\! 
a_{y}^2 m^2) \Bigr] \; , \qquad \label{VertexB} \\
{\rm Vertex\ C} &\!\!\! = \!\!\!& -i\kappa a_z^{D-2} \Bigl[-\partial^{\mu}_1
\partial^{\nu}_2 + \tfrac12 \eta^{\mu\nu} \partial_1 \!\cdot\! \partial_2 \Bigr]
\; , \qquad \label{VertexC} \\
{\rm Vertex\ D} &\!\!\! = \!\!\!& -\tfrac{i}{2} \kappa \lambda a_x^{D} 
\eta^{\mu\nu} \; , \qquad \label{VertexD} \\
{\rm Vertex\ E} &\!\!\! = \!\!\!& -i \kappa^2 \lambda a_x^{D} \Bigl[\tfrac18
\eta^{\mu\nu} \eta^{\rho\sigma} - \tfrac14 \eta^{\mu \rho} \eta^{\sigma \nu}
\Bigr] \; . \qquad \label{VertexE}
\end{eqnarray}
The subscripts on scale factors indicate the coordinate argument, for example,
$a_x \equiv -\frac1{H x^0}$ (with $x^0 < 0$). Subscripts on derivatives indicate 
which leg is differentiated, and we have not bothered to symmetrize indices 
because they are contracted into the symmetric graviton propagator.

\subsection{Reduction Strategy}

As discussed in section 1, the gauge dependence of a 1PI 2-point function can
be removed by first writing down (in position space) all the diagrams which
contribute to the $t$-channel scattering process $\Psi_1(x)$ + $\Psi_3(x') 
\longrightarrow \Psi_2(y) + \Psi_4(y')$. This consists of a collection of 
2-point, 3-point and 4-point Green's functions,
\begin{equation}
-i V(x;x') \;\; , \;\; -i V(x;x';y) \;\; , \;\; -i V(x;x';y') \;\; , \;\; -i
V(x;x';y;y') \; . \label{GreenF}
\end{equation}
The 2-point and 3-point Green's functions must be multiplied by appropriate
delta functions to fix $y^{\mu}$ and $y^{\prime \mu}$, for example,
\begin{equation}
-i V(x;x') \longrightarrow \delta^D(x \!-\! y) \delta^D(x' \!-\! y') \!\times\!
-i V(x;x') \; . 
\end{equation}

One then extracts the $t$-channel poles from 3-point and 4-point contributions 
using identities derived by Donoghue and collaborators \cite{Donoghue:1993eb,
Donoghue:1994dn,Bjerrum-Bohr:2002aqa,Donoghue:1996mt} and easy to check from 
general results \cite{Beenakker:1988jr,Ellis:2007qk}. In flat position space 
one can reduce 3-point Green's functions to 2-point form using the relation 
\cite{Miao:2017feh},
\begin{equation}
i\Delta_m(x;y) i\Delta(x;x') i\Delta(y;x') \longrightarrow 
\frac{i \delta^D(x \!-\! y)}{2 m^2} \Bigl[i \Delta(x;x')\Bigr]^2 \; ,
\label{Dono3flat}
\end{equation}
where $i\Delta(x;x')$ denotes the massless scalar propagator in flat space.
Reducing the flat 4-point Green's functions to 2-point form requires two 
such identities \cite{Donoghue:1996mt,Bjerrum-Bohr:2002aqa},
\begin{eqnarray}
& & \hspace{-0.5cm} i\Delta_m(x;y) i\Delta_m(x';y') i\Delta(x;x') i\Delta(y;y') 
\longrightarrow -\frac{i}{m^2} \Bigl[ 1 - \Bigl( \frac{\overline{\partial}_x 
\!\cdot\! \overline{\partial}_{x'} \!-\! m^2}{3 m^2}\Bigr) \Bigr] \nonumber \\
& & \hspace{2.8cm} \times \Biggl\{ \delta^D(x \!-\! y)
\delta^D(x' \!-\! y') \! \int \!\! d^Dz \Bigl[ i\Delta(x;z)\Bigr]^2 i\Delta(z;x')
\Biggr\} \; , \qquad \label{Dono4aflat} \\
& & \hspace{-0.5cm} i\Delta_m(x;y) i\Delta_m(x';y') i\Delta(x;y') i\Delta(x';y) 
\longrightarrow +\frac{i}{m^2} \Bigl[ 1 + \Bigl( \frac{\overline{\partial}_x 
\!\cdot\! \overline{\partial}_{y'} \!+\! m^2}{3 m^2}\Bigr) \Bigr] \nonumber \\
& & \hspace{2.8cm} \times \Biggl\{ \delta^D(x \!-\! y)
\delta^D(x' \!-\! y') \! \int \!\! d^Dz \Bigl[ i\Delta(x;z)\Bigr]^2 i\Delta(z;x')
\Biggr\} \; , \qquad \label{Dono4bflat}
\end{eqnarray}
where the over-lined derivatives are understood to act on the external wave
functions. Note that the integrals at the end of expressions (\ref{Dono4aflat}) 
and (\ref{Dono4bflat}) is symmetric under interchange of $x^{\mu}$ and 
$x^{\prime \mu}$.

\begin{figure}[H]
\centering
\includegraphics[width=3cm]{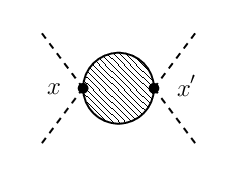}
\includegraphics[width=3cm]{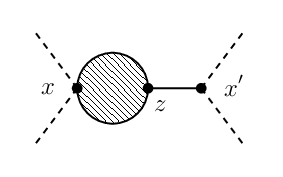}
\includegraphics[width=3cm]{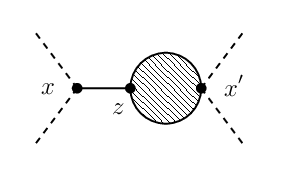}
\includegraphics[width=3cm]{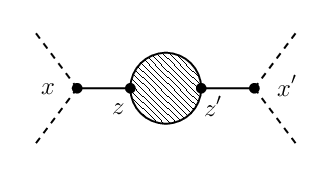}
\caption{\footnotesize The four classes of diagrams ($\alpha$ through $\delta$, 
from left to right) produced after reducing diagrams with source and observer 
correlations to 2-point form. The shaded circles are $-i \lambda^2 (a_{x} a_{x'})^D
f_{\alpha}(x;x')$, $-i \lambda a_x^D f_{\beta}(x;z')$, $-i \lambda a_{x'}^D 
f_{\gamma}(z;x')$ and $-i f_{\delta}(z;z')$, respectively.}
\label{Reduced}
\end{figure}
Figure~\ref{Reduced} shows the generic 2-point forms produced by these
reductions. We can regard each of these 2-point contributions as a correction
to the 1PI 2-point function by using the propagator equation $i \delta^D(x-z)
= \mathcal{D}_z i\Delta_A(x;z)$ and then partially integrating. We will give 
all the steps for the leftmost diagram,
\begin{eqnarray}
\lefteqn{{\rm Class} \, \alpha \equiv \lambda^2 (a_x a_{x'})^D \!\times\! 
-i f_{\alpha}(x;x') \; , } \\
& & \hspace{-0.5cm} = -i\lambda a_x^D \!\! \int \!\! d^Dz \, i\delta^D(x \!-\! z)
\!\times\! -i\lambda a_{x'}^D \!\! \int \!\! d^Dz' \, i\delta^D(x' \!-\! z')
\!\times\! -i f_{\alpha}(z;z') \; , \qquad \\
& & \hspace{-0.5cm} = -i\lambda a_x^D \!\! \int \!\! d^Dz \, \mathcal{D}_{z}
i \Delta_A(x;z) \!\times\! -i\lambda a_{x'}^D \!\! \int \!\! d^Dz' \, 
\mathcal{D}_{z'} i\Delta_A(x';z') \!\times\! -i f_{\alpha}(z;z') \; , \qquad \\
& & \hspace{-0.5cm} = -i\lambda a_x^D \!\! \int \!\! d^Dz \, i \Delta_A(x;z) 
\!\times\! -i\lambda a_{x'}^D \!\! \int \!\! d^Dz' \, i\Delta_A(x';z') 
\!\times\! -i \mathcal{D}_{z} \mathcal{D}_{z'} f_{\alpha}(z;z') \; . \qquad
\end{eqnarray}
We merely give the results for the remaining diagrams,
\begin{eqnarray}
\lefteqn{{\rm Class} \, \beta \equiv -i \lambda a_{x'}^D \!\! \int \!\!
d^Dz' \, i\Delta_A(x';z') \!\times -i \lambda a_{x}^D f_{\beta}(x;z') \; , } \\
& & \hspace{0cm} = -i\lambda a_x^D \!\! \int \!\! d^Dz \, i \Delta_A(x;z) 
\!\times\! -i\lambda a_{x'}^D \!\! \int \!\! d^Dz' \, i\Delta_A(x';z') 
\!\times\! -i \mathcal{D}_{z} f_{\beta}(z;z') \; \, \qquad \\
\lefteqn{{\rm Class} \, \gamma \equiv -i \lambda a_x^D \!\! \int \!\!
d^Dz \, i\Delta_A(x;z) \!\times -i \lambda a_{x'}^D f_{\gamma}(z;x') \; , } \\
& & \hspace{0cm} = -i\lambda a_x^D \!\! \int \!\! d^Dz \, i \Delta_A(x;z) 
\!\times\! -i\lambda a_{x'}^D \!\! \int \!\! d^Dz' \, i\Delta_A(x';z') 
\!\times\! -i \mathcal{D}_{z'} f_{\gamma}(z;z') \; , \qquad \\
\lefteqn{{\rm Class} \, \delta \equiv -i\lambda a_x^D \!\! \int \!\! d^Dz \, 
i \Delta_A(x;z) \!\times\! -i\lambda a_{x'}^D \!\! \int \!\! d^Dz' \, 
i\Delta_A(x';z') \!\times\! -i f_{\delta}(z;z') \; .}
\end{eqnarray}
The gauge invariant self-mass is the sum of all four classes,
\begin{equation}
M^2_{\rm inv}(x;x') = \mathcal{D}_{x} \mathcal{D}_{x'} f_{\alpha}(x;x') +
\mathcal{D}_{x} f_{\beta}(x;x') + \mathcal{D}_{x'} f_{\gamma}(x;x') + 
f_{\delta}(x;x') \; . \label{invM2}
\end{equation}

It remains to generalize the Donoghue Identities (\ref{Dono3flat}-\ref{Dono4bflat})
from flat space to de Sitter. This is a matter of preserving general coordinate 
invariance and taking account of the scalar propagator $i\Delta_A$ possibly differing
from the scalar propagators $i\Delta_I$ in the graviton propagator. For the 3-point 
identity (\ref{Dono3flat}) the required generalization is,
\begin{equation}
i\Delta_m(x;y) i\Delta_A(x;x') i\Delta_I(y;x') \longrightarrow 
\frac{i \delta^D(x \!-\! y)}{2 m^2 a_x^D} \, i\Delta_A(x;x') i\Delta_I(x;x') \; . 
\label{Dono3dS}
\end{equation}
We also need a 3-point identity with a single derivative,
\begin{equation}
i\Delta_m(x;y) i\Delta_A(x;z) \partial_y^{\mu} i\Delta_C(y;z) \longrightarrow
(\overline{\partial}_y^{\mu} \!+\! \overline{\partial}_z^{\mu}) \frac{i \delta^D(
x \!-\! y)}{2 m^2 a_x^D} \, i\Delta_A(x;z) i\Delta_C(x;z) \; . \label{der3ptDon}
\end{equation}
The 4-point identities (\ref{Dono4aflat}-\ref{Dono4bflat}) have the complication
of contracting derivatives at different points. This can be made invariant using 
the parallel transport matrix $[\mbox{}^{\mu} g^{\nu}](x;x')$ whose $3+1$ 
decomposition in de Sitter conformal coordinates is,
\begin{eqnarray}
[\mbox{}^{\mu} g^{\nu}](x;x') & \equiv & \left( \begin{matrix} [\mbox{}^0 g^0] & 
[\mbox{}^0 g^{n}] \\ [\mbox{}^m g^0] & [\mbox{}^m g^n] \\ \end{matrix} \right)
\; , \\
& = & \frac{\eta^{\mu\nu}}{a_{x} a_{x'}} + \frac{2}{4 \!-\! y} 
\left( \begin{matrix} - H^2 \Delta r^2 & -(\frac1{a_{x}} \!+\! \frac1{a_{x'}}) H 
\Delta r^n \\ (\frac1{a_{x}} \!+\! \frac1{a_{x'}}) H \Delta r^m & H^2 \Delta r^m 
\Delta r^n \\ \end{matrix} \right) \; , \qquad \label{parallel} 
\end{eqnarray}
where $\Delta r^i \equiv (x - x')^i$ and $y \equiv a_{x} a_{x'} H^2 \Delta x^2$.
Note that $[\mbox{}^{\mu} g^{\nu}](x;x') \simeq \eta^{\mu\nu}/a_{x} a_{x'}$ near
spatial coincidence. The appropriate generalizations of the 4-point Donoghue
Identities (\ref{Dono4aflat}-\ref{Dono4bflat}) are,
\begin{eqnarray}
& & \hspace{-0.7cm} i\Delta_m(x;y) i\Delta_m(x';y') i\Delta_A(x;x') 
i\Delta_C(y;y') \nonumber \\
& & \hspace{0cm} \longrightarrow -\frac{i}{m^2} \Bigl[ 1 - \Bigl( 
\frac{\overline{\partial}^{x}_{\mu} \overline{\partial}^{x'}_{\nu} 
[\mbox{}^{\mu} g^{\nu}](x;x') \!-\! m^2}{3 m^2}\Bigr) \Bigr] 
\frac{\delta^D(x \!-\! y) \delta^D(x' \!-\! y')}{(a_{x} a_{x'})^D} \nonumber \\
& & \hspace{1cm} \times \frac12 \Biggl\{ \! \int \!\! d^Dz \, a_z^D 
i \Delta_A(x;z) i \Delta_C(x;z) i\Delta_A(z;x') + \Bigl( x^{\mu} 
\longleftrightarrow {x'}^{\mu} \Bigr) \! \Biggr\} , \qquad \label{Dono4dS} \\
& & \hspace{-0.7cm} i\Delta_m(x;y) i\Delta_m(x';y') i\Delta_A(x;y') 
i\Delta_C(x';y) \nonumber \\
& & \hspace{0cm} \longrightarrow +\frac{i}{m^2} \Bigl[ 1 + \Bigl( 
\frac{\overline{\partial}^{x}_{\mu} \overline{\partial}^{y'}_{\nu}
[\mbox{}^{\mu} g^{\nu}](x;y') \!+\! m^2}{3 m^2}\Bigr) \Bigr]
\frac{\delta^D(x \!-\! y) \delta^D(x' \!-\! y')}{(a_{x} a_{x'})^D} \nonumber \\
& & \hspace{1cm} \times \frac12 \Biggl\{ \!\int \!\! d^Dz \, a_z^D 
i\Delta_A(x;z) i\Delta_C(x;z) i\Delta_A(z;y') + \Bigl( x^{\mu}
\longleftrightarrow {y'}^{\mu} \Bigr) \! \Biggr\} . \qquad \label{Dono4bdS}
\end{eqnarray}

\subsection{Derivative Identities}

Momentum conservation at a 3-point vertex (\ref{VertexA}-\ref{VertexC})
is complicated by the factors of $a^{D-2}$ (for derivative vertices) and
$a^D$ (for non-derivative vertices). If we label derivatives of the three
propagators (or external wave functions) by ``$\partial_{1}^{\mu}$'',
``$\partial_{2}^{\mu}$'' and ``$\partial_{3}^{\mu}$'' then we might use
``$\partial_{4}^{\mu}$ to stand for differentiation of the scale factors.
With this understanding, momentum conservation reads,
\begin{equation}
\Bigl(\partial_1 + \partial_2 + \partial_3 + \partial_4 \Bigr)^{\mu} = 0 \; .
\label{momcons}
\end{equation}
We often require contractions of two derivatives $\partial_1 \cdot 
\partial_2 \equiv \eta_{\mu\nu} \partial_{1}^{\mu} \partial_{2}^{\nu}$.
When the vertex contains a factor of $a^{D-2}$ such contractions can be 
expressed in terms of the scalar covariant d'Alembertian,
\begin{equation}
\mathcal{D}_i \equiv \eta_{\mu\nu} \partial_{i}^{\mu} a^{D-2} 
\partial_{i}^{\nu} = a^{D-2} \partial_{i} \!\cdot\! (\partial_{i} \!+\!
\partial_{4}) \; . \label{calD_i}
\end{equation}
The derivation is,
\begin{eqnarray}
\partial_{1} \!\cdot\! \partial_{2} &\!\!\! = \!\!\!& \tfrac12 (\partial_{1}
\!+\! \partial_{2})^2 - \tfrac12 \partial_{1}^2 - \tfrac12 \partial_{2}^2 
\; , \\
&\!\!\! = \!\!\!& \tfrac12 (\partial_{3} \!+\! \partial_{4})^2 - \tfrac12 
\partial_{1}^2 - \tfrac12 \partial_{2}^2 \; , \\
&\!\!\! = \!\!\!& \tfrac12 \partial_{3} \!\cdot\! (\partial_{3} \!+\! 
\partial_{4}) + \tfrac12 \partial_{4} \!\cdot\! (\partial_{3} \!+\! 
\partial_{4}) - \tfrac12 \partial_{1}^2 - \tfrac12 \partial_{2}^2 \; , 
\qquad \\
&\!\!\! = \!\!\!& \frac1{2 a^{D-2}} \Bigl[ \mathcal{D}_{3} - \mathcal{D}_{1}
- \mathcal{D}_{2}\Bigr] \; . \label{DID1}
\end{eqnarray}
A related identity of great utility is,
\begin{equation}
a^{D-2} \partial_1 \!\cdot\! \partial_2 + a^D m^2 = \frac12 \mathcal{D}_3 - 
\frac12 \Bigl( \mathcal{D}_1 \!-\! a^D m^2\Bigr) - \frac12 \Bigl( \mathcal{D}_2 
\!-\! a^D m^2 \Bigr) \; . \label{DID2}
\end{equation}

\section{The Five New Diagrams}

\begin{figure}[H]
\centering
\includegraphics[width=4cm]{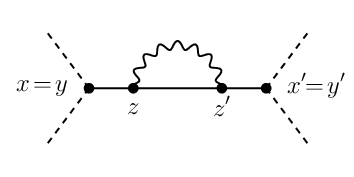}
\includegraphics[width=4cm]{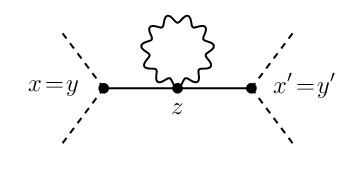}
\includegraphics[width=4cm]{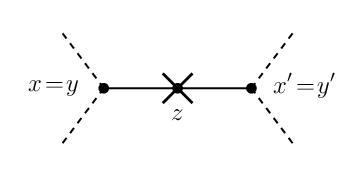}
\caption{\footnotesize Diagrams $0a$ (left) and $0b$ (right) included in the 
uncorrected, self-mass.}
\label{Diagram0}
\end{figure}

The scattering process $\Psi_1(x) + \Psi_3(x') \longrightarrow \Psi_2(y) + 
\Psi_4(y')$ receives a $t$-channel contribution from the 1-loop self-mass 
$-i M_{0}^2(x;x')$ as shown in Figure~\ref{Diagram0}. The analytic expression 
is,
\begin{equation}
-i V_0(x;x') = -\lambda^2 (a_{x} a_{x'})^D \!\! \int \!\! d^Dz \, i\Delta_A(x;z)
\!\! \int \!\! d^Dz' \, i \Delta_A(x';z') \!\times\! -i M_{0}^2(z;z') \; .
\label{V0def}
\end{equation}
The purpose of this section is to give the five other classes of diagrams which
make $t$-channel contributions to the same 1-loop scattering amplitude. We use 
subscripts to distinguish scale factors at different spacetime points, as in 
$a_x$ and $a_{x'}$ in expression (\ref{V0def}). The same convention applies to 
derivative operators, whose indices are raised and lowered using the Minkowski 
metric, as in $\partial^{\mu}_{y} \equiv \eta^{\mu\nu} \frac{\partial}{\partial 
y^{\nu}}$. Because there are sometimes too many propagators to distinguish how 
derivatives act by order, we employ a special convention for a $h_{\mu\nu} 
\Psi_{\rm int} \Psi_{\rm ext}$ vertex: a bar indicates that the derivative acts 
on the external $\Psi$ leg, a tilde indicates that it acts on the graviton leg, 
and a derivative without either distinction acts on the internal $\Psi$ leg.

\subsection{Vertex-Vertex}

\begin{figure}[H]
\centering
\includegraphics[width=4cm]{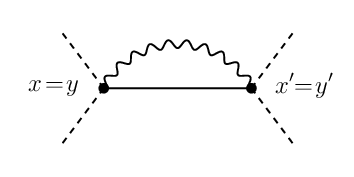}
\includegraphics[width=4cm]{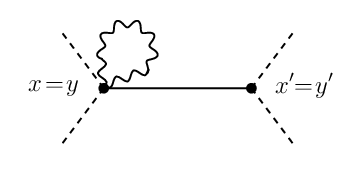}
\includegraphics[width=4cm]{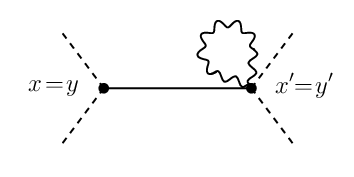}
\caption{\footnotesize Diagrams $1a$ (left), $1b$ (middle) and $1c$ (right) in 
which the vertex is corrected.}
\label{Diagram1}
\end{figure}

The first of the new diagrams derive from correlators of gravitons which emerge
from the vertex. Figure~\ref{Diagram1} shows the three possibilities. The first
case is where a single graviton emerges from each vertex,
\begin{equation}
-i V_{1a}(x;x') = \Bigl(-\tfrac{i}{2} \kappa \lambda a_{x}^{D} \eta^{\mu\nu} \Bigr) 
i [\mbox{}_{\mu\nu} \Delta_{\rho\sigma}](x;x') \Bigl(-\tfrac{i}{2} \kappa \lambda 
a_{x'}^D \eta^{\rho\sigma} \Bigr) i \Delta_A(x;x') \; . \label{V1adef}
\end{equation}
The 2nd and 3rd are where two gravitons emerge from the same vertex,
\begin{eqnarray}
\lefteqn{-i V_{1b}(x;x') } \nonumber \\
& & \hspace{-0.3cm} = \Bigl[-i \kappa^2 \lambda a_{x}^{D} \Big( \tfrac18 \eta^{\mu\nu}
\eta^{\rho\sigma} \!-\! \tfrac14 \eta^{\mu\rho} \eta^{\nu\sigma}\Bigr) \Bigr] 
i [\mbox{}_{\mu\nu} \Delta_{\rho\sigma}](x;x) \Bigl(-i \lambda a_{x'}^D\Bigr)
i \Delta_{A}(x;x') \; , \qquad \label{V1bdef} \\
\lefteqn{-i V_{1c}(x;x') } \nonumber \\
& & \hspace{-0.3cm} = \Bigl(-i \lambda a_{x}^D\Bigr) i [\mbox{}_{\mu\nu} 
\Delta_{\rho\sigma}](x';x') \Bigl[-i \kappa^2 \lambda a_{x'}^{D} \Big( \tfrac18 
\eta^{\mu\nu} \eta^{\rho\sigma} \!-\! \tfrac14 \eta^{\mu\rho} \eta^{\nu\sigma}\Bigr) 
\Bigr] i \Delta_{A}(x;x') \; . \qquad \label{V1cdef}
\end{eqnarray}

\subsection{Source(Observer)-Vertex}

\begin{figure}[H]
\centering
\includegraphics[width=3.3cm]{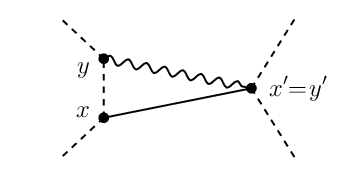}
\includegraphics[width=3.3cm]{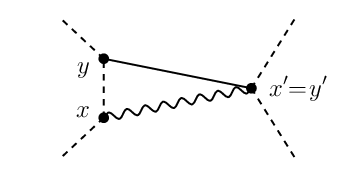}
\includegraphics[width=3.3cm]{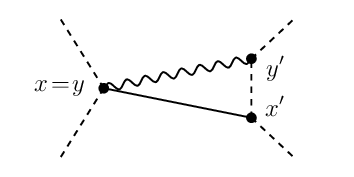}
\includegraphics[width=3.3cm]{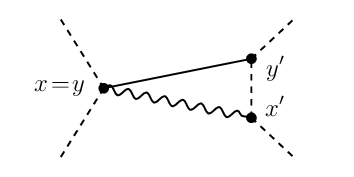}
\caption{\footnotesize Diagrams $2a$ (left), $2b$ (2nd), $2c$ (3rd) and $2d$ (right)
including a correlation from the propagation of the source or observer with the vertex.}
\label{Diagram2}
\end{figure}

Figure~\ref{Diagram2} shows the second class of new diagrams: those which involve a
graviton from the propagation of the source or observer correlated with a graviton 
from the vertex on the opposite side. (Having the vertex is on the same side
produces an irrelevant external leg correction.) The leftmost diagram is,
\begin{eqnarray}
\lefteqn{-i V_{2a}(x;x';y) = \Bigl[-i \kappa a_{y}^{D-2} \Bigl( -\overline{\partial}_{y}^{\mu}
\partial_{y}^{\nu} + \tfrac12 \eta^{\mu\nu} [\overline{\partial}_{y} \!\cdot\! \partial_{y} 
\!+\! a_{y}^2 m^2]\Bigr) \Bigr] i [\mbox{}_{\mu\nu} \Delta_{\rho\sigma}](y;x') } \nonumber \\
& & \hspace{3.5cm} \times \Bigl(-\tfrac{i}{2} \kappa \lambda a_{x'}^D \eta^{\rho\sigma}\Bigr)
i \Delta_m(y;x) \Bigl(-i \lambda a_{x}^D\Bigr) i \Delta_A(x;x') \; , \qquad \\
& & \hspace{0cm} = \frac{i}{2} \kappa^2 \lambda^2 (a_{x} a_{x'})^D a_{y}^{D-2} \Bigl[ -
\overline{\partial}_{y}^{\mu} \partial_{y}^{\nu} + \tfrac12 \eta^{\mu\nu} 
(\overline{\partial}_{y} \!\cdot\! \partial_{y} \!+\! a_{y}^2 m^2) \Bigr] \nonumber \\
& & \hspace{5.5cm} \times i [\mbox{}_{\mu\nu} \Delta^{\rho}_{~\rho}](y;x') i \Delta_m(x;y)
i \Delta_A(x;x') \; . \qquad \label{V2a}
\end{eqnarray}
We merely give final results for the other three diagrams,
\begin{eqnarray}
\lefteqn{-i V_{2b}(x;x';y) = \frac{i}{2} \kappa^2 \lambda^2 (a_{y} a_{x'})^D 
a_{x}^{D-2} \Bigl[ -\overline{\partial}_{x}^{\mu} \partial_{x}^{\nu} + \tfrac12 
\eta^{\mu\nu} (\overline{\partial}_{x} \!\cdot\! \partial_{x} \!+\! 
a_{x}^2 m^2) \Bigr] } \nonumber \\
& & \hspace{5.3cm} \times i [\mbox{}_{\mu\nu} \Delta^{\rho}_{~\rho}](x;x') i \Delta_m(x;y) 
i \Delta_A(y;x') \; , \qquad \label{V2b} \\
\lefteqn{-i V_{2c}(x;x';y') = \frac{i}{2} \kappa^2 \lambda^2 (a_{x} a_{x'})^D 
a_{y'}^{D-2} \Bigl[ -\overline{\partial}_{y'}^{\rho} \partial_{y'}^{\sigma} + \tfrac12 
\eta^{\rho\sigma} (\overline{\partial}_{y'} \!\cdot\! \partial_{y'} \!+\! 
a_{y'}^2 m^2) \Bigr] } \nonumber \\
& & \hspace{5.3cm} \times i [\mbox{}^{\mu}_{~\mu} \Delta_{\rho\sigma}](x;y') i 
\Delta_m(x';y') i \Delta_A(x;x') \; , \qquad \label{V2c} \\
\lefteqn{-i V_{2d}(x;x';y') = \frac{i}{2} \kappa^2 \lambda^2 (a_{x} a_{y'})^D 
a_{x'}^{D-2} \Bigl[ -\overline{\partial}_{x'}^{\rho} \partial_{x'}^{\sigma} + \tfrac12 
\eta^{\rho\sigma} (\overline{\partial}_{x'} \!\cdot\! \partial_{x'} \!+\! 
a_{x'}^2 m^2) \Bigr] } \nonumber \\
& & \hspace{5.3cm} \times i [\mbox{}^{\mu}_{~\mu} \Delta_{\rho\sigma}](x;x') i 
\Delta_m(x';y') i \Delta_A(x;y') \; , \qquad \label{V2d}
\end{eqnarray}

\subsection{Vertex-Exchange}

\begin{figure}[H]
\centering
\includegraphics[width=6cm]{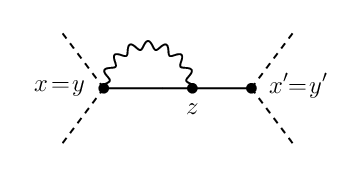}
\includegraphics[width=6cm]{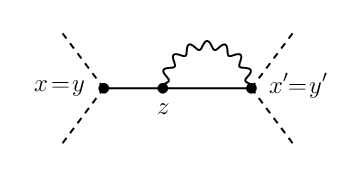}
\caption{\footnotesize Diagrams $3a$ (left) and $3b$ (right) included a 
correlation between a vertex and the propagation of the exchange scalar.}
\label{Diagram3}
\end{figure}

Another class of corrections is where a graviton from a vertex correlates with a
graviton from the propagation of the exchange scalar. Figure~\ref{Diagram3} shows
these diagrams. Their analytic expressions are,
\begin{eqnarray}
\lefteqn{-i V_{3a}(x;x') = \tfrac{i}{2} \kappa^2 \lambda^2 (a_{x} a_{x'})^D \!
\int \!\! d^Dz \, a_{z}^{D-2} \Bigl[-\overline{\partial}_{z}^{\rho} 
\partial_{z}^{\sigma} + \tfrac12 \eta^{\rho\sigma} \overline{\partial}_{z} 
\!\cdot\! \partial_{z} \Bigr] } \nonumber \\
& & \hspace{5.5cm} \times i [\mbox{}^{\mu}_{~\mu} \Delta_{\rho\sigma}](x;z)
i \Delta_A(x;z) i \Delta_A(z;x') \; , \qquad \label{V3a} \\
\lefteqn{-i V_{3b}(x;x') = \tfrac{i}{2} \kappa^2 \lambda^2 (a_{x} a_{x'})^D \!
\int \!\! d^Dz \, a_{z}^{D-2} \Bigl[-\overline{\partial}_{z}^{\mu} 
\partial_{z}^{\nu} + \tfrac12 \eta^{\mu\nu} \overline{\partial}_{z} 
\!\cdot\! \partial_{z} \Bigr] } \nonumber \\
& & \hspace{5.5cm} \times i [\mbox{}_{\mu\nu} \Delta^{\rho}_{~\rho}](z;x')
i \Delta_A(x;z) i \Delta_A(z;x') \; . \qquad \label{V3b}
\end{eqnarray}
Our notation in (\ref{V3a}-\ref{V3b}) is that $\overline{\partial}_{z}$
differentiates the scalar propagator external to the loop, while $\partial_{z}$
acts on the scalar propagator inside the loop.

\subsection{Source-Observer}

\begin{figure}[H]
\centering
\includegraphics[width=3.3cm]{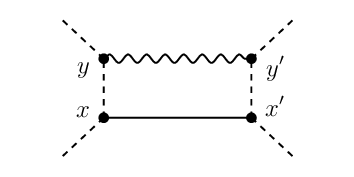}
\includegraphics[width=3.3cm]{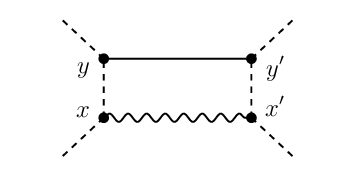}
\includegraphics[width=3.3cm]{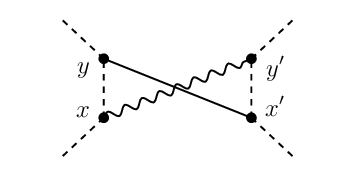}
\includegraphics[width=3.3cm]{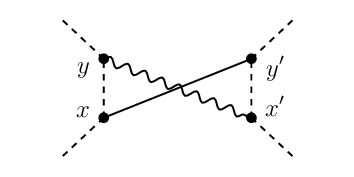}
\caption{\footnotesize Diagrams $4a$ (left), $4b$ (2nd), $4c$ (3rd)
and $4d$ (right) including a correlation between the propagation
of the source and the propagation of the observer.}
\label{Diagram4}
\end{figure}

Figure~\ref{Diagram4} shows diagrams which result from a graviton in the propagation
of the source correlating with a graviton from the observer. (Source-source and
Observer-observer correlations either give external state factors or else make vertex
corrections which are not enhanced.) The leftmost diagram involves two A vertices 
(\ref{VertexA}) and two B vertices (\ref{VertexB}),
\begin{eqnarray}
\lefteqn{-i V_{4a}(x;x';y;y') = \Bigl[-i \kappa a_{y}^{D-2} \Bigl( -
\overline{\partial}_{y}^{\mu} \partial_{y}^{\nu} + \tfrac12 \eta^{\mu\nu} 
[\overline{\partial}_{y} \!\cdot\! \partial_{y} \!+\! a_{y}^2 m^2]\Bigr) \Bigr]  
i [\mbox{}_{\mu\nu} \Delta_{\rho\sigma}](y;y') } \nonumber \\
& & \hspace{2.5cm} \times \Bigl[-i \kappa a_{y'}^{D-2} \Bigl( 
-\overline{\partial}_{y'}^{\rho} \partial_{y'}^{\sigma} + \tfrac12
\eta^{\rho\sigma} [\overline{\partial}_{y'} \!\cdot\! \partial_{y'} \!+\! 
a_{y'}^2 m^2]\Bigr)\Bigr] \nonumber \\
& & \hspace{2.5cm} \times i\Delta_{m}(x;y) \Bigl(- i\lambda a_{x}^{D}\Bigr) 
i\Delta_A(x;x') \Bigl(-i \lambda a_{x'}^D\Bigr) i \Delta_{m}(x';y') 
\; , \qquad \\
& & \hspace{-0.7cm} = \kappa^2 \lambda^2 (a_{x} a_{x'})^D (a_{y} a_{y'})^{D-2}
\Bigl[\overline{\partial}_{y}^{\mu} \partial_{y}^{\nu} \!-\! \tfrac12 \eta^{\mu\nu}
(\overline{\partial}_{y} \!\cdot\! \partial_{y} \!+\! a_{y}^2 m^2)\Bigr] 
\Bigl[\overline{\partial}_{y'}^{\rho} \partial_{y'}^{\sigma} \nonumber \\
& & \hspace{-0.3cm} - \tfrac12 \eta^{\rho\sigma} (\overline{\partial}_{y'} \!\cdot\! 
\partial_{y'} \!+\! a_{y'}^2 m^2)\Bigr] i\Delta_{m}(x;y) i \Delta_{m}(x';y') 
i \Delta_{A}(x;x') i [\mbox{}_{\mu\nu} \Delta_{\rho\sigma}](y;y') \; . \qquad 
\label{V4a}
\end{eqnarray}
For the remaining three diagrams we give only the final forms,
\begin{eqnarray}
\lefteqn{-i V_{4b}(x;x';y;y') = \kappa^2 \lambda^2 (a_{y} a_{y'})^D (a_{x} 
a_{x'})^{D-2} \Bigl[\overline{\partial}_{x}^{\mu} \partial_{x}^{\nu} \!-\! 
\tfrac{\eta^{\mu\nu}}{2} (\overline{\partial}_{x} \!\!\cdot\! \partial_{x} \!+\! 
a_{x}^2 m^2) \!\Bigr] \! \Bigl[\overline{\partial}_{x'}^{\rho} 
\partial_{x'}^{\sigma} } \nonumber \\
& & \hspace{0cm} - \tfrac{\eta^{\rho\sigma}}{2} (\overline{\partial}_{x'} 
\!\!\cdot\! \partial_{x'} \!+\! a_{x'}^2 m^2)\Bigr] i\Delta_{m}(x;y) i 
\Delta_{m}(x';y') i \Delta_{A}(y;y') i [\mbox{}_{\mu\nu} \Delta_{\rho\sigma}](x;x') 
\; , \qquad \label{V4b} \\
\lefteqn{-i V_{4c}(x;x';y;y') = \kappa^2 \lambda^2 (a_{y} a_{x'})^D (a_{x} 
a_{y'})^{D-2} \Bigl[\overline{\partial}_{x}^{\mu} \partial_{x}^{\nu} \!-\! 
\tfrac{\eta^{\mu\nu}}{2} (\overline{\partial}_{x} \!\!\cdot\! \partial_{x} \!+\! 
a_{x}^2 m^2) \!\Bigr] \! \Bigl[\overline{\partial}_{y'}^{\rho} 
\partial_{y'}^{\sigma} } \nonumber \\
& & \hspace{0cm} - \tfrac{\eta^{\rho\sigma}}{2} (\overline{\partial}_{y'} 
\!\!\cdot\! \partial_{y'} \!+\! a_{y'}^2 m^2)\Bigr] i\Delta_{m}(x;y) i 
\Delta_{m}(x';y') i \Delta_{A}(y;x') i [\mbox{}_{\mu\nu} \Delta_{\rho\sigma}](x;y') 
\; , \qquad \label{V4c} \\
\lefteqn{-i V_{4d}(x;x';y;y') = \kappa^2 \lambda^2 (a_{x} a_{y'})^D (a_{y} 
a_{x'})^{D-2} \Bigl[\overline{\partial}_{y}^{\mu} \partial_{y}^{\nu} \!-\! 
\tfrac{\eta^{\mu\nu}}{2} (\overline{\partial}_{y} \!\!\cdot\! \partial_{y} \!+\! 
a_{y}^2 m^2) \!\Bigr] \! \Bigl[\overline{\partial}_{x'}^{\rho} 
\partial_{x'}^{\sigma} } \nonumber \\
& & \hspace{0cm} - \tfrac{\eta^{\rho\sigma}}{2} (\overline{\partial}_{x'} 
\!\!\cdot\! \partial_{x'} \!+\! a_{x'}^2 m^2)\Bigr] i\Delta_{m}(x;y) i 
\Delta_{m}(x';y') i \Delta_{A}(x;y') i [\mbox{}_{\mu\nu} \Delta_{\rho\sigma}](y;x') 
\; . \qquad \label{V4d}
\end{eqnarray}

\subsection{Source(Observer)-Exchange}

\begin{figure}[H]
\centering
\includegraphics[width=3.3cm]{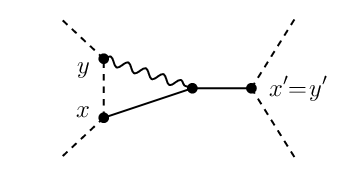}
\includegraphics[width=3.3cm]{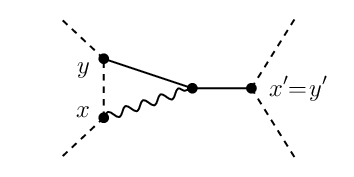}
\includegraphics[width=3.3cm]{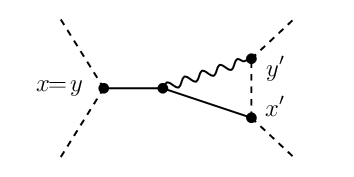}
\includegraphics[width=3.3cm]{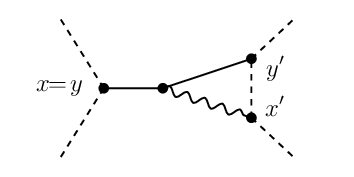}
\caption{\footnotesize Diagrams $5a$ (left), $5b$ (2nd), $5c$ (3rd) and
$5d$ (right) including a correlation between the propagation of the 
source or observer and the propagation of the exchange scalar.}
\label{Diagram5}
\end{figure}

The final class of contributions consists of gravitons from the propagation of 
the source or observer correlation with gravitons from the exchange scalar. 
Figure~\ref{Diagram5} shows these diagrams, which each involves two $\varphi \Psi^2$
vertices (\ref{VertexA}), a single $h_{\mu\nu} \Psi^2$ vertex (\ref{VertexB})
and a single $h_{\mu\nu} \varphi^2$ vertex (\ref{VertexC}). The leftmost diagram 
is,
\begin{eqnarray}
\lefteqn{-i V_{5a}(x;x';y) = \Bigl[-i \kappa a_{y}^{D-2} \Bigl(-
\overline{\partial}_{y}^{\mu} \partial_{y}^{\nu} + \tfrac12 \eta^{\mu\nu} 
[\overline{\partial}_{y} \!\!\cdot\! \partial_{y} \!+\! a_{y}^2 m^2]\Bigr)
\Bigr] i \Delta_{m}(y;x) } \nonumber \\
& & \hspace{3cm} \times \! \int \!\! d^Dz \, i [\mbox{}_{\mu\nu} 
\Delta_{\rho\sigma}](y;z) \Bigl[-i \kappa a_{z}^{D-2} \Bigl(-
\overline{\partial}_{z}^{\rho} \partial_{z}^{\sigma} + \tfrac12 \eta^{\rho\sigma} 
\overline{\partial}_{z} \!\!\cdot\! \partial_{z} \Bigr)\Bigr] \nonumber \\
& & \hspace{4.5cm} \times i\Delta_A(z;x) 
\Bigl(-i \lambda a_{x}^D\Bigr) i\Delta_A(z;x') \Bigl(-i \lambda a_{x'}^D\Bigr) 
\; , \qquad \\
& & = \kappa^2 \lambda^2 (a_{x} a_{x'})^D \!\! \int \!\! d^Dz (a_{y} a_{z})^{D-2} 
\Bigl[\overline{\partial}_{y}^{\mu} \partial_{y}^{\nu} - \tfrac{\eta^{\mu\nu}}{2} 
(\overline{\partial}_{y} \!\!\cdot\! \partial_{y} \!+\! a_{y}^2 m^2) \Bigr]
\Bigl[ \overline{\partial}_{z}^{\rho} \partial_{z}^{\sigma} \nonumber \\
& & \hspace{2.4cm} - \tfrac{\eta^{\rho\sigma}}{2} \overline{\partial}_{z} 
\!\!\cdot\! \partial_{z} \Bigr] i \Delta_{m}(x;y) i [\mbox{}_{\mu\nu} 
\Delta_{\rho\sigma}](y;z) i \Delta_{A}(x;z) i\Delta_{A}(z;x') \; . \qquad 
\label{V5a}
\end{eqnarray}
Recall that $\overline{\partial}_{z}$ indicates differentiation of the scalar
propagator external to the loop, where $\partial_{z}$ acts on the scalar propagator 
inside the loop. For the remaining three diagrams we give only the final forms,
\begin{eqnarray}
\lefteqn{-i V_{5b}(x;x';y) = \kappa^2 \lambda^2 (a_{y} a_{x'})^D \!\! \int \!\! 
d^Dz (a_{x} a_{z})^{D-2} \Bigl[\overline{\partial}_{x}^{\mu} \partial_{x}^{\nu} 
- \tfrac{\eta^{\mu\nu}}{2} (\overline{\partial}_{x} \!\!\cdot\! \partial_{x} 
\!+\! a_{x}^2 m^2) \Bigr]  } \nonumber \\
& & \hspace{0.5cm} \times \Bigl[ \overline{\partial}_{z}^{\rho} 
\partial_{z}^{\sigma} \!-\! \tfrac{\eta^{\rho\sigma}}{2} \overline{\partial}_{z} 
\!\!\cdot\! \partial_{z} \Bigr] i \Delta_{m}(x;y) i [\mbox{}_{\mu\nu} 
\Delta_{\rho\sigma}](x;z) i \Delta_{A}(y;z) i\Delta_{A}(z;x') \; , \qquad 
\label{V5b} \\
\lefteqn{-i V_{5c}(x;x';y') = \kappa^2 \lambda^2 (a_{x} a_{x'})^D \!\! \int \!\! 
d^Dz (a_{y'} a_{z})^{D-2} \Bigl[\overline{\partial}_{y'}^{\rho} 
\partial_{y'}^{\sigma} - \tfrac{\eta^{\rho\sigma}}{2} (\overline{\partial}_{y'} 
\!\!\cdot\! \partial_{y'} \!+\! a_{y'}^2 m^2) \Bigr]  } \nonumber \\
& & \hspace{0.5cm} \times \Bigl[ \overline{\partial}_{z}^{\mu} \partial_{z}^{\nu} 
\!-\! \tfrac{\eta^{\mu\nu}}{2} \overline{\partial}_{z} \!\!\cdot\! \partial_{z} 
\Bigr] i \Delta_{m}(x';y') i [\mbox{}_{\mu\nu} \Delta_{\rho\sigma}](z;y') 
i \Delta_{A}(x;z) i\Delta_{A}(z;x') \; , \qquad \label{V5c} \\
\lefteqn{-i V_{5d}(x;x';y') = \kappa^2 \lambda^2 (a_{x} a_{y'})^D \!\! \int \!\! 
d^Dz (a_{x'} a_{z})^{D-2} \Bigl[\overline{\partial}_{x'}^{\rho} 
\partial_{x'}^{\sigma} - \tfrac{\eta^{\rho\sigma}}{2} (\overline{\partial}_{x'} 
\!\!\cdot\! \partial_{x'} \!+\! a_{x'}^2 m^2) \Bigr]  } \nonumber \\
& & \hspace{0.5cm} \times \Bigl[ \overline{\partial}_{z}^{\mu} \partial_{z}^{\nu} 
\!-\! \tfrac{\eta^{\mu\nu}}{2} \overline{\partial}_{z} \!\!\cdot\! \partial_{z} 
\Bigr] i \Delta_{m}(x';y') i [\mbox{}_{\mu\nu} \Delta_{\rho\sigma}](z;x') 
i \Delta_{A}(x;z) i\Delta_{A}(z;y') \; . \qquad \label{V5d}
\end{eqnarray}

\section{Consolidation}

When the derivative identity (\ref{DID2}) is used on the $h_{\mu\nu} \Psi^2$ 
vertex (\ref{VertexB}) at an external point, and the external state condition 
$\overline{\mathcal{D}} = a^D m^2$ is employed, the vertex simplies to,
\begin{eqnarray}
\lefteqn{-i \kappa a^{D-2} \Bigl( -\overline{\partial}^{\mu} \partial^{\nu} + 
\tfrac12 \eta^{\mu\nu} \Bigl[\overline{\partial} \!\cdot\! \partial + a^2 m^2
\Bigr]\Bigr) } \nonumber \\
& & \hspace{3.5cm} = -i \kappa \Bigl( -a^{D-2} \overline{\partial}^{\mu} 
\partial^{\nu} + \tfrac14 \eta^{\mu\nu} \Bigl[\widetilde{\mathcal{D}} - 
(\mathcal{D} \!-\! a^D m^2) \Bigr] \Bigr) \; , \qquad \label{exthPsi2}
\end{eqnarray}
Recall that $\overline{\partial}^{\mu}$ acts on the external $\Psi$ leg, 
$\partial^{\mu}$ acts on the internal $\Psi$ line and $\widetilde{\partial}^{\mu}$ 
differentiates the graviton propagator. Of course the factor of $(\mathcal{D} - 
a^D m^2)$ acts on the internal $\Psi$ propagator to give a delta function, which 
effectively reduces the number of distinct points of that portion of the diagram. 
This part of the Source(Observer)-Exchange ($V_{5}$) diagrams turns out to cancel 
the Vertex-Exchange ($V_{3}$) corrections. Similarly, these parts of the 
Source-Observer ($V_{4}$) diagrams cancel the main Vertex-Vertex ($V_{1a}$) 
correction and all of the Source(Observer)-Vertex ($V_{2}$) corrections. The 
purpose of this section is to demonstrate these cancellations, which are especially
powerful because they pertain for any gauge. We close by listing the residual 
diagrams which remain.

\subsection{The 3-5 Complex}

It is convenient to label the three terms inside the large parentheses of 
expression (\ref{exthPsi2}) as ``$i$'', ``$ii$'' and ``$iii$'',
\begin{eqnarray}
-a^{D-2} \overline{\partial}^{\mu} \partial^{\nu} & \Longleftrightarrow &
i \; , \label{parti} \\
\tfrac14 \eta^{\mu\nu} \widetilde{\mathcal{D}} & \Longleftrightarrow & ii \; , 
\label{partii} \\
-\frac14 \eta^{\mu\nu} (\mathcal{D} \!-\! a^D m^2) & \Longleftrightarrow &
iii \; . \label{partiii}
\end{eqnarray}
We can decompose $-iV_{5}(x;x';y)$ accordingly,
\begin{eqnarray}
-i V_{5a_{i}}(x;x';y) &\!\!\! \equiv \!\!\!& \kappa^2 \lambda^2 (a_x a_{x'})^D 
\!\times\! a_{y}^{D-2} \overline{\partial}^{\mu} \partial_{y}^{\nu} \!\times\!
\!\! \int \!\! d^Dz \, a_{z}^{D-2} \Bigl[ \overline{\partial}_{z}^{\rho} 
\partial_{z}^{\sigma} \!-\! \tfrac12 \eta^{\mu\nu} \overline{\partial}_{z}
\!\cdot\! \partial_{z} \Bigr] \nonumber \\
& & \hspace{1.2cm} \times i\Delta_{m}(x;y) i [\mbox{}_{\mu\nu} 
\Delta_{\rho\sigma}](y;z) i\Delta_A(x;z) i\Delta_A(z;x') \; , \qquad 
\label{V5ai} \\
-i V_{5a_{ii}}(x;x';y) &\!\!\! \equiv \!\!\!& \kappa^2 \lambda^2 (a_x a_{x'})^D 
\!\times\! -\tfrac14 \eta^{\mu\nu} \widetilde{\mathcal{D}_{y}} \!\times\! \!\! 
\int \!\! d^Dz \, a_{z}^{D-2} \Bigl[ \overline{\partial}_{z}^{\rho} 
\partial_{z}^{\sigma} \!-\! \tfrac12 \eta^{\mu\nu} \overline{\partial}_{z} 
\!\cdot\! \partial_{z} \Bigr] \nonumber \\
& & \hspace{1.2cm} \times i\Delta_{m}(x;y) i [\mbox{}_{\mu\nu} 
\Delta_{\rho\sigma}](y;z) i\Delta_A(x;z) i\Delta_A(z;x') \; , \qquad 
\label{V5aii} \\
-i V_{5a_{iii}}(x;x';y) &\!\!\! \equiv \!\!\!& \kappa^2 \lambda^2 (a_x a_{x'})^D 
\!\times\! \tfrac14 \eta^{\mu\nu} (\mathcal{D}_y \!-\! a^D m^2) \!\times\! \!\!
\int \!\! d^Dz \, a_{z}^{D-2} \Bigl[ \overline{\partial}_{z}^{\rho} 
\partial_{z}^{\sigma} \nonumber \\
& & \hspace{-1.2cm} - \tfrac12 \eta^{\mu\nu} \overline{\partial}_{z} \!\cdot\! 
\partial_{z} \Bigr] \!\times\! i\Delta_{m}(x;y) i [\mbox{}_{\mu\nu} 
\Delta_{\rho\sigma}](y;z) i\Delta_A(x;z) i\Delta_A(z;x') \; . \qquad 
\label{V5aiii}
\end{eqnarray}
The same conventions pertain for diagrams $-i V_{5b}(x;x';y)$, $-i 
V_{5c}(x;x';y')$ and $-i V_{5d}(x;x';y')$. We use a bar to indicate the sum of 
the ``$i$'' and ``$ii$'' terms,
\begin{equation}
-i V_{\overline{5}}(x;x';y) \equiv -i V_{5_{i}}(x;x';y) -i V_{5_{ii}}(x;x';y) 
\; . \label{5abardef}
\end{equation}

Note that $-i V_{5a_{iii}}(x;x';y)$ can be simplified by acting the $\Psi$ 
kinetic operator $(\mathcal{D}_{y} - a^D m^2)$ on the $\Psi$ propagator
$i \Delta_{m}(x;y)$,
\begin{eqnarray}
-i V_{5a_{iii}}(x;x';y) &\!\!\! = \!\!\!& \delta^D(x \!-\! y) \!\times\! 
\tfrac{i}{4} \kappa^2 \lambda^2 (a_x a_{x'})^D \!\! \int \!\! d^Dz \, 
a_{z}^{D-2} \Bigl[ \overline{\partial}_{z}^{\rho} \partial_{z}^{\sigma} 
- \tfrac12 \eta^{\mu\nu} \overline{\partial}_{z} \!\cdot\! \partial_{z} 
\Bigr] \nonumber \\
& & \hspace{3cm} \times i [\mbox{}^{\mu}_{~\mu} \Delta_{\rho\sigma}](x;z) 
i\Delta_A(x;z) i\Delta_A(z;x') \; . \qquad \label{V5aiiisimp}
\end{eqnarray}
One gets the same result for $-i V_{5b_{iii}}(x;x';y)$. Taken together, 
they cancel $\delta^D(x - y) \times -i V_{3a}(x;x')$,
\begin{eqnarray}
\lefteqn{\delta^D(x \!-\! y) \!\times\! -i V_{3a}(x;x') = \delta^D(x \!-\! y) 
\!\times\! -\tfrac{i}{2} \kappa^2 \lambda^2 (a_x a_{x'})^D \!\! \int \!\! d^Dz 
\, a_{z}^{D-2} \Bigl[ \overline{\partial}_{z}^{\rho} \partial_{z}^{\sigma} }
\nonumber \\
& & \hspace{3.4cm} - \tfrac12 \eta^{\rho\sigma} \overline{\partial}_{z} \!\cdot\! 
\partial_{z} \Bigr] \!\times\! i [\mbox{}^{\mu}_{~\mu} \Delta_{\rho\sigma}](x;z) 
i\Delta_A(x;z) i\Delta_A(z;x') \; . \qquad \label{V3asimp}
\end{eqnarray}
Hence we can write,
\begin{equation}
-i V_{5a_{iii}}(x;x';y) - i V_{5b_{iii}}(x;x';y) + \delta^D(x \!-\! y) 
\!\times\! -i V_{3a}(x;x') = 0 \; . \label{5ab3a}
\end{equation}
Similarly, one can show,
\begin{equation}
-i V_{5c_{iii}}(x;x';y') - i V_{5d_{iii}}(x;x';y') + \delta^D(x' \!-\! y') 
\!\times\! -i V_{3b}(x;x') = 0 \; . \label{5cd3b}
\end{equation}
It follows that the Vertex-Exchange diagrams are completely canceled by the
``$iii$'' parts of the Source(Observer)-Exchange diagrams.

\subsection{The 1a-2-4 Complex}

The Source-Observer diagrams ($V_{4}$) each have two $h_{\mu\nu} \Psi^2$ 
vertices (\ref{VertexB}), each one of which can be decomposed into terms 
``$i$'', ``$ii$'' and ``$iii$'' as in (\ref{parti}-\ref{partiii}). For 
$V_{4a}$ we first expand on the left hand vertex, keeping the right hand 
vertex together,
\begin{equation}
V_{4a} = V_{\overline{4a}} + V_{4a_{iii}} \; . \label{V4aexpleft}
\end{equation}
We then expand on the right hand vertex,
\begin{equation}
V_{\overline{4a}} = V_{\overline{\overline{4a}}} + V_{\overline{4a}_{iii}}
\qquad , \qquad V_{4a_{iii}} = V_{\overline{4a_{iii}}} + V_{(4a_{iii})_{iii}}
\; , \label{V4aexpright}
\end{equation}
where the resulting four terms are,
\begin{eqnarray}
-i V_{\overline{\overline{4a}}} &\!\!\! \equiv \!\!\!& \kappa^2 \lambda^2 
(a_x a_{x'})^D \Bigl[a_{y}^{D-2} \overline{\partial}_{y}^{\mu} 
\partial_{y}^{\nu} \!-\! \tfrac14 \eta^{\mu\nu} \widetilde{\mathcal{D}}_{y}
\Bigr] \Bigl[ a_{y'}^{D-2} \overline{\partial}_{y'}^{\rho} 
\partial_{y'}^{\sigma} \!-\! \tfrac14 \eta^{\rho\sigma} \widetilde{D}_{y'} 
\Bigr] \nonumber \\
& & \hspace{1.7cm} \times i \Delta_{m}(x;y) i \Delta_{m}(x';y') i \Delta_{A}(x;x')
i [\mbox{}_{\mu\nu} \Delta_{\rho\sigma}](y;y') \; , \qquad \label{V4abarbar} \\
-i V_{\overline{4a}_{iii}} &\!\!\! \equiv \!\!\!& \tfrac14 \kappa^2 \lambda^2 
(a_x a_{x'})^D \Bigl[a_{y}^{D-2} \overline{\partial}_{y}^{\mu} 
\partial_{y}^{\nu} \!-\! \tfrac14 \eta^{\mu\nu} \widetilde{\mathcal{D}}_{y}
\Bigr] (\mathcal{D}_{y'} \!-\! a_{y'}^D m^2) \nonumber \\
& & \hspace{1.7cm} \times i \Delta_{m}(x;y) i \Delta_{m}(x';y') i \Delta_{A}(x;x')
i [\mbox{}_{\mu\nu} \Delta^{\rho}_{~\rho}](y;y') \; , \qquad 
\label{V4abariii} \\
-i V_{\overline{4a_{iii}}} &\!\!\! \equiv \!\!\!& \tfrac14 \kappa^2 \lambda^2 
(a_x a_{x'})^D (\mathcal{D}_{y} \!-\! a_{y}^D m^2) \Bigl[ a_{y'}^{D-2} 
\overline{\partial}_{y'}^{\rho} \partial_{y'}^{\sigma} \!-\! \tfrac14 
\eta^{\rho\sigma} \widetilde{D}_{y'} \Bigr] \nonumber \\
& & \hspace{1.7cm} \times i \Delta_{m}(x;y) i \Delta_{m}(x';y') i \Delta_{A}(x;x')
i [\mbox{}^{\mu}_{~\mu} \Delta_{\rho\sigma}](y;y') \; , \qquad 
\label{V4aiiibar} \\
-i V_{(4a_{iii})_{iii}} &\!\!\! \equiv \!\!\!& \tfrac1{16} \kappa^2 \lambda^2 
(a_x a_{x'})^D (\mathcal{D}_{y} \!-\! a_{y}^D m^2) (\mathcal{D}_{y'} \!-\! 
a_{y'}^D m^2) \nonumber \\
& & \hspace{1.7cm} \times i \Delta_{m}(x;y) i \Delta_{m}(x';y') i \Delta_{A}(x;x')
i [\mbox{}^{\mu}_{~\mu} \Delta^{\rho}_{~\rho}](y;y') \; . \qquad 
\label{V4aiiiiii}
\end{eqnarray}
For $V_{4b}$ we also expand first on the left hand vertex and then on the 
right hand vertex. However, it is best to expand $V_{4c}$ and $V_{4d}$ first on
the right hand vertex and then on the left hand one. This convention makes for
the simplest correspondence with the Source(Observer)-Vertex diagrams ($V_{2}$).

The Source(Observer)-Vertex diagrams (\ref{V2a}-\ref{V2d}) contain only one 
$h_{\mu\nu} \Psi^2$ vertex. We expand this as usual, for example,
\begin{equation}
V_{2a} = V_{\overline{2a}} + V_{2a_{iii}} \; .
\end{equation}
To be explicit, the two terms are,
\begin{eqnarray}
-i V_{\overline{2a}} &\!\!\! \equiv \!\!\!& -\tfrac{i}2 \kappa^2 \lambda^2 
(a_{x} a_{x'})^D \Bigl[a_{y}^{D-2} \overline{\partial}_{y}^{\mu} \partial_{y}^{\nu}
\!-\! \tfrac14 \eta^{\mu\nu} \widetilde{\mathcal{D}}_{y}\Bigr] \nonumber \\
& & \hspace{4cm} \times i \Delta_{m}(x;y) i \Delta_{A}(x;x') i [\mbox{}_{\mu\nu} 
\Delta^{\rho}_{~\rho}](y;x') \; , \qquad \label{V2abar} \\
-i V_{2a_{iii}} &\!\!\! \equiv \!\!\!& -\tfrac{i}{8} \kappa^2 \lambda^2 
(a_{x} a_{x'})^D (\mathcal{D}_{y} \!-\! a_{y}^D m^2) \nonumber \\
& & \hspace{4cm} \times i \Delta_{m}(x;y) i \Delta_{A}(x;x') i [\mbox{}^{\mu}_{\mu} 
\Delta^{\rho}_{~\rho}](y;x') \; . \qquad \label{V2aiii}
\end{eqnarray}

Let us first act the $\Psi$ kinetic operators on the $\Psi$ propagators in 
expressions (\ref{V4aiiiiii}) and (\ref{V2aiii}) to conclude,
\begin{eqnarray}
-i V_{(4a_{iii})_{iii}} &\!\!\! = \!\!\!& \delta^D(x \!-\! y) 
\delta^D(x' \!-\! y') \nonumber \\
& & \hspace{2.3cm} \!\times\! -\tfrac1{16} \kappa^2 \lambda^2 (a_{x} a_{x'})^D i 
\Delta_{A}(x;x') i [\mbox{}^{\mu}_{~\mu} \Delta^{\rho}_{~\rho}](x;x') \; , 
\qquad \\
-i V_{2a_{iii}} &\!\!\! = \!\!\!& \delta^D(x \!-\! y) \!\times\! \tfrac18 
\kappa^2 \lambda^2 (a_{x} a_{x'})^D i \Delta_{A}(x;x') i [\mbox{}^{\mu}_{~\mu} 
\Delta^{\rho}_{~\rho}](x;x') \; . \qquad 
\end{eqnarray}
Adding the two contributions (with appropriate delta functions) gives $-\tfrac14$ 
times the $-i V_{1a}$ contribution (\ref{V1adef}),
\begin{equation}
\delta^D(x' \!-\! y') \!\times\! -i V_{2a_{iii}} - i V_{(4a_{iii})_{iii}} =
-\frac14 \delta^D(x \!-\! y) \delta^D(x' \!-\! y') \!\times\! -i V_{1a} \; .
\end{equation}
The same relation pertains as well for the ``$b$'', ``$c$'' and ``$d$'' 
contributions, with the result that $-i V_{1a}$ is completely canceled.

Now act the $\Psi$ kinetic operator of expression (\ref{V4abariii}),
\begin{eqnarray}
-i V_{\overline{4a}_{iii}} &\!\!\! = \!\!\!& \delta^D(x' \!-\! y') \!\times\!
\tfrac{i}{4} \kappa^2 \lambda^2 (a_{x} a_{x'})^D \Bigl[a_{y}^{D-2} 
\overline{\partial}_{y}^{\mu} \partial_{y}^{\nu} \!-\! \tfrac14 \eta^{\mu\nu} 
\widetilde{\mathcal{D}}_{y} \Bigr] \nonumber \\
& & \hspace{4.3cm} \times i \Delta_{m}(x;y) i \Delta_{A}(x;x') i [\mbox{}_{\mu\nu} 
\Delta^{\rho}_{~\rho}](y;y') \; . \qquad 
\end{eqnarray}
One gets exactly the same thing from $-i V_{\overline{4c_{iii}}}$, and their sum
cancels $-i V_{\overline{2a}}$,
\begin{equation}
-i V_{\overline{4a}_{iii}} - i V_{\overline{4d_{iii}}} = -\delta^D(x' \!-\! y')
\!\times\! -i V_{\overline{2a}} \; .
\end{equation}
A number of similar relations pertain,
\begin{eqnarray}
-i V_{\overline{4b}_{iii}} - i V_{\overline{4c_{iii}}} &\!\!\! = \!\!\!& 
-\delta^D(x' \!-\! y') \!\times\! -i V_{\overline{2b}} \; , \qquad \\
-i V_{\overline{4c}_{iii}} - i V_{\overline{4a_{iii}}} &\!\!\! = \!\!\!& 
-\delta^D(x \!-\! y) \!\times\! -i V_{\overline{2c}} \; , \qquad \\
-i V_{\overline{4d}_{iii}} - i V_{\overline{4b_{iii}}} &\!\!\! = \!\!\!& 
-\delta^D(x \!-\! y) \!\times\! -i V_{\overline{2d}} \; . \qquad
\end{eqnarray}
Hence the $-i V_{2}$ diagrams are completely canceled.

\subsection{Surviving Residual Diagrams}

After all these consolidations there are just three classes of ``new'' diagrams 
which remain. The simplest are the two surviving $i=1$ diagrams 
(\ref{V1bdef}-\ref{V1cdef}),
\begin{eqnarray}
\lefteqn{-i V_{1b}(x;x') = -\kappa^2 \lambda^2 (a_x a_{x'})^D } \nonumber \\
& & \hspace{3cm} \times \Bigl\{ \tfrac18 i [\mbox{}^{\alpha}_{~\alpha} 
\Delta^{\beta}_{~\beta}](x;x) - \tfrac14 i [\mbox{}^{\alpha\beta} 
\Delta_{\alpha\beta}](x;x)\Bigr\} i\Delta_A(x;x') \; , \qquad \label{V1b} \\
\lefteqn{-i V_{1c}(x;x') = -\kappa^2 \lambda^2 (a_x a_{x'})^D } \nonumber \\
& & \hspace{3cm} \times i\Delta_A(x;x') \Bigl\{ \tfrac18 i 
[\mbox{}^{\alpha}_{~\alpha} \Delta^{\beta}_{~\beta}](x';x') - \tfrac14 
i [\mbox{}^{\alpha\beta} \Delta_{\alpha\beta}](x';x') \Bigr\} \; . \qquad
\label{V1c}
\end{eqnarray}
Next come the four residual $i=4$ diagrams,
\begin{eqnarray}
\lefteqn{-i V_{\overline{\overline{4a}}}(x;x';y;y') = \kappa^2 \lambda^2 (a_{x} a_{x'})^D
\Bigl[a_{y}^{D-2} \overline{\partial}^{\mu}_{y} \partial^{\nu}_{y} \!-\! \tfrac14 
\eta^{\mu\nu} \widetilde{\mathcal{D}}_{y}\Bigr] \Bigl[a_{y'}^{D-2} 
\overline{\partial}_{y'}^{\rho} \partial_{y'}^{\sigma} \!-\! \tfrac14 \eta^{\rho\sigma} 
\widetilde{\mathcal{D}}_{y'}\Bigr]} \nonumber \\
& & \hspace{3.6cm} \times 
i\Delta_{m}(x;y) i\Delta_{m}(x';y') i \Delta_A(x;x') i [\mbox{}_{\mu\nu}
\Delta_{\rho\sigma}](y;y') \; , \qquad \label{4abarbar} \\
\lefteqn{-i V_{\overline{\overline{4b}}}(x;x';y;y') = \kappa^2 \lambda^2 (a_{y} a_{y'})^D
\Bigl[a_{x}^{D-2} \overline{\partial}^{\mu}_{x} \partial^{\nu}_{x} \!-\! \tfrac14 
\eta^{\mu\nu} \widetilde{\mathcal{D}}_{x}\Bigr] \Bigl[a_{x'}^{D-2} 
\overline{\partial}_{x'}^{\rho} \partial_{x'}^{\sigma} \!-\! \tfrac14 \eta^{\rho\sigma} 
\widetilde{\mathcal{D}}_{x'}\Bigr]} \nonumber \\
& & \hspace{3.6cm} \times 
i\Delta_{m}(x;y) i\Delta_{m}(x';y') i \Delta_A(y;y') i [\mbox{}_{\mu\nu}
\Delta_{\rho\sigma}](x;x') \; , \qquad \label{4bbarbar}
\end{eqnarray}
\begin{eqnarray}
\lefteqn{-i V_{\overline{\overline{4c}}}(x;x';y;y') = \kappa^2 \lambda^2 (a_{y} a_{x'})^D
\Bigl[a_{x}^{D-2} \overline{\partial}^{\mu}_{x} \partial^{\nu}_{x} \!-\! \tfrac14 
\eta^{\mu\nu} \widetilde{\mathcal{D}}_{x}\Bigr] \Bigl[a_{y'}^{D-2} 
\overline{\partial}_{y'}^{\rho} \partial_{y'}^{\sigma} \!-\! \tfrac14 \eta^{\rho\sigma} 
\widetilde{\mathcal{D}}_{y'}\Bigr]} \nonumber \\
& & \hspace{3.6cm} \times 
i\Delta_{m}(x;y) i\Delta_{m}(x';y') i \Delta_A(y;x') i [\mbox{}_{\mu\nu}
\Delta_{\rho\sigma}](x;y') \; , \qquad \label{4cbarbar}  \\
\lefteqn{-i V_{\overline{\overline{4d}}}(x;x';y;y') = \kappa^2 \lambda^2 (a_{x} a_{y'})^D
\Bigl[a_{y}^{D-2} \overline{\partial}^{\mu}_{y} \partial^{\nu}_{y} \!-\! \tfrac14 
\eta^{\mu\nu} \widetilde{\mathcal{D}}_{y}\Bigr] \Bigl[a_{x'}^{D-2} 
\overline{\partial}_{x'}^{\rho} \partial_{x'}^{\sigma} \!-\! \tfrac14 \eta^{\rho\sigma} 
\widetilde{\mathcal{D}}_{x'}\Bigr]} \nonumber \\
& & \hspace{3.6cm} \times 
i\Delta_{m}(x;y) i\Delta_{m}(x';y') i \Delta_A(x;y') i [\mbox{}_{\mu\nu}
\Delta_{\rho\sigma}](y;x') \; . \qquad \label{4dbarbar}
\end{eqnarray}
Finally, there are the four $i=5$ residual diagrams,
\begin{eqnarray}
\lefteqn{-i V_{\overline{5a}}(x;x';y) = \kappa^2 \lambda^2 (a_{x} a_{x'})^D
\Bigl[a_{y}^{D-2} \overline{\partial}^{\mu}_{y} \partial^{\nu}_{y} \!-\! \tfrac14 
\eta^{\mu\nu} \widetilde{\mathcal{D}}_{y}\Bigr] \! \int \!\! d^Dz \, a_{z}^{D-2} 
\Bigl[ \overline{\partial}_{z}^{\rho} \partial_{z}^{\sigma} } \nonumber \\
& & \hspace{1.3cm} - \tfrac12 \eta^{\rho\sigma} \overline{\partial}_{z} \!\cdot\! 
\partial_{z} \Bigr] \!\times\! i\Delta_{m}(x;y) i [\mbox{}_{\mu\nu} 
\Delta_{\rho\sigma}](y;z) i \Delta_A(x;z) i \Delta_A(z;x') \; , \qquad 
\label{5abar} \\
\lefteqn{-i V_{\overline{5b}}(x;x';y) = \kappa^2 \lambda^2 (a_{y} a_{x'})^D
\Bigl[a_{x}^{D-2} \overline{\partial}^{\mu}_{x} \partial^{\nu}_{x} \!-\! \tfrac14 
\eta^{\mu\nu} \widetilde{\mathcal{D}}_{x}\Bigr] \! \int \!\! d^Dz \, a_{z}^{D-2} 
\Bigl[ \overline{\partial}_{z}^{\rho} \partial_{z}^{\sigma} } \nonumber \\
& & \hspace{1.3cm} - \tfrac12 \eta^{\rho\sigma} \overline{\partial}_{z} \!\cdot\! 
\partial_{z} \Bigr] \!\times\! i\Delta_{m}(x;y) i [\mbox{}_{\mu\nu} 
\Delta_{\rho\sigma}](x;z) i \Delta_A(y;z) i \Delta_A(z;x') \; , \qquad 
\label{5bbar} \\
\lefteqn{-i V_{\overline{5c}}(x;x';y) = \kappa^2 \lambda^2 (a_{x} a_{x'})^D
\Bigl[a_{y'}^{D-2} \overline{\partial}^{\mu}_{y'} \partial^{\nu}_{y'} \!-\! \tfrac14 
\eta^{\mu\nu} \widetilde{\mathcal{D}}_{y'}\Bigr] \! \int \!\! d^Dz \, a_{z}^{D-2} 
\Bigl[ \overline{\partial}_{z}^{\rho} \partial_{z}^{\sigma} } \nonumber \\
& & \hspace{1.3cm} - \tfrac12 \eta^{\rho\sigma} \overline{\partial}_{z} \!\cdot\! 
\partial_{z} \Bigr] \!\times\! i\Delta_{m}(x';y') i [\mbox{}_{\mu\nu} 
\Delta_{\rho\sigma}](y';z) i \Delta_A(x;z) i \Delta_A(z;x') \; , \qquad 
\label{5cbar} \\
\lefteqn{-i V_{\overline{5d}}(x;x';y) = \kappa^2 \lambda^2 (a_{x} a_{y'})^D
\Bigl[a_{x'}^{D-2} \overline{\partial}^{\mu}_{x'} \partial^{\nu}_{x'} \!-\! \tfrac14 
\eta^{\mu\nu} \widetilde{\mathcal{D}}_{x'}\Bigr] \! \int \!\! d^Dz \, a_{z}^{D-2} 
\Bigl[ \overline{\partial}_{z}^{\rho} \partial_{z}^{\sigma} } \nonumber \\
& & \hspace{1.3cm} - \tfrac12 \eta^{\rho\sigma} \overline{\partial}_{z} \!\cdot\! 
\partial_{z} \Bigr] \!\times\! i\Delta_{m}(x';y') i [\mbox{}_{\mu\nu} 
\Delta_{\rho\sigma}](x';z) i \Delta_A(x;z) i \Delta_A(z;y') \; . \qquad 
\label{5dbar}
\end{eqnarray}
It is important to note that the reductions to these residual forms are exact,
and independent of the graviton gauge.

\section{Gauge Independent Self-Mass}

The purpose of this section is to combine the $t$-channel contributions from 
the five new diagrams with $-i M_{0}(x;x')$ to produce a gauge independent
self-mass. We begin by discussing the general reduction strategy for the 
$\overline{5}$ and $\overline{\overline{4}}$ contributions. Then each of
these is worked out. The section closes by adding the $1b$ and $1c$ contributions
and renormalizing the final result. 

\subsection{General Reduction Strategy}

A number of simplifications concern the graviton propagator. First, because the
propagator's indices are typically contracted into derivatives, only the temporal 
components of which can be significant for the infrared, $t$-channel exchange 
contributions of interest to us, we need retain just the first term in expression 
(\ref{gravprop2}),
\begin{equation}
i[\mbox{}_{\mu\nu} \Delta_{\rho\sigma}] \longrightarrow \Bigl[\eta_{\mu\rho} 
\eta_{\nu\sigma} + \eta_{\mu\sigma} \eta_{\nu\rho} - \frac{2}{D \!-\! 2} 
\eta_{\mu\nu} \eta_{\rho\sigma}\Bigr] \!\times\! i \Delta_{C} \; . \label{gravprop3}
\end{equation}
Second, in acting $\widetilde{D}$ on $i\Delta_C$ in 4-point contributions we can 
ignore the delta function because that term will have one fewer massless propagator,
\begin{equation}
\widetilde{\mathcal{D}} i\Delta_C(x;x') = i \delta^D(x \!-\! x') + 2 (D\!-\!3) H^2 
i\Delta_C(x;x') \longrightarrow 2 (D\!-\!3) H^2 i\Delta_C(x;x') \; . \label{Csimp}
\end{equation}
Finally, single derivatives of $i\Delta_{C}$ can be approximately reflected from 
one coordinate to the other using the relation,\footnote{Recall that 
$\partial^{\mu}_{a_x} = -(D-2) H a_x \delta^{\mu}_{~0}$ indicates the derivative 
acting on the factor of $a^{D-2}_{x}$ which accompanies vertices from the kinetic 
energy.}
\begin{equation}
\Bigl( \widetilde{\partial}^{\mu}_{x} + \frac12 \partial^{\mu}_{a_x}\Bigr) 
i\Delta_{C}(x;x') \simeq -\Bigl( \widetilde{\partial}^{\mu}_{x'} + \frac12 
\partial^{\mu}_{a_{x'}} \Bigr) i\Delta_{C}(x;x') \; . \label{reflection}
\end{equation}
This approximation is exact on the first term of the $i\Delta_C(x;x')$ expansion;
and the higher terms should not matter because they vanish for $D=4$.

Our basic strategy is to follow the same reductions as in flat space
\cite{Miao:2017feh}, collecting de Sitter obstructions as they occur.
However, we need not retain terms with more than two factors of $H$ because
any additional factors of $H$ will necessarily come in the dimensionless 
form of $H/m$ which vanishes when the source and observer masses become 
large.

One major change from the flat space reduction concerns the external state 
factors for the source and observer. In flat space these are simple phase 
factors; in de Sitter they are spatial plane waves times Hankel functions
with imaginary arguments. Their series expansions in powers of $k^2/a^2$ 
are more useful for our purposes,
\begin{equation}
u_m(x;\vec{k}) = \frac{e^{-i\mu t + i \vec{k} \cdot \vec{x}}}{\sqrt{2 \mu 
a^{D-1}}} \sum_{n=0}^{\infty} \frac{\Gamma(1 \!+\! \frac{i\mu}{H})}{n! \,
\Gamma(1 \!+\! n \!+\! \frac{i \mu}{H})} \Bigl( \frac{-k^2}{4 H^2 a^2}
\Bigr)^{n} \; , \; \mu^2 \equiv m^2 \!-\! (\tfrac{D-1}{2})^2 H^2 \; . 
\label{external}
\end{equation}
Note that this expression employs the co-moving time $t = \ln(a)/H$, rather 
than the conformal time. We can expand external first derivatives as,
\begin{equation}
\partial_0 u_m(x;\vec{k}) = -i a \Bigl[\mu - i (\tfrac{D-1}{2}) H + 
\tfrac{k^2}{2 m a^2} + \ldots \Bigr] u_m \quad , \quad \partial_i 
u_m(x;\vec{k}) = i k_i u_m \; . \label{extderiv} 
\end{equation}
We sometimes also need second time derivatives,
\begin{equation}
\partial_0^2 u_{m}(x;\vec{k}) = -a^2 \Bigl[ m^2 - i (D\!-\!2) \mu H 
- \tfrac12 (D\!-\!1) (D \!-\!2) H^2 + \tfrac{k^2}{a^2} + \ldots \Bigr]
u_{m}(x;\vec{k}) \; . \label{extdsq}
\end{equation}
We regard the two in-coming wave functions as,
\begin{equation}
u_{m}(x;\vec{k}_1) \qquad , \qquad u_{m}(x';\vec{k}_3) \; . 
\label{incoming}
\end{equation}
The corresponding out-going wave functions are,
\begin{equation}
u_{m}^*(y;\vec{k}_2) \qquad , \qquad u_{m}^*(y';\vec{k}_4) \; . 
\label{outgoing}
\end{equation}
And spatial momentum conservation implies,
\begin{equation}
\vec{k}_1 + \vec{k}_3 = \vec{k}_2 + \vec{k}_4 \qquad \Longrightarrow \qquad
\vec{k}_1 - \vec{k}_2 = \vec{k}_4 - \vec{k}_3 \equiv \vec{q} \; .
\label{qdef}
\end{equation}

\subsection{$V_{\overline{5}}(x;x';y;y')$ Diagrams}

Here we give the reduction for the residual diagrams of type $\overline{5}$
listed in (\ref{5abar}-\ref{5dbar}). The details are given
for part $\overline{5a}$, and the remaining three parts are inferred from symmetry.

The initial step of the reduction is contracting the simplified graviton
propagator (\ref{gravprop3}) into the tensor structures and derivatives of the
two 3-vertices in (\ref{5abar}),
\begin{eqnarray}
&&
\hspace{-1cm}
\Bigl[ a_{y}^{D-2} \overline{\partial}^{\mu}_{y} \partial^{\nu}_{y} \!-\! \tfrac{1}{4} 
	\eta^{\mu\nu} \widetilde{\mathcal{D}}_{y}\Bigr] 
	a_{z}^{D-2} 
	\Bigl[ \overline{\partial}_{z}^{\rho} \partial_{z}^{\sigma}  
	- \tfrac{1}{2} \eta^{\rho\sigma} 
		\overline{\partial}_{z} \!\cdot\! \partial_{z} \Bigr] 
	\Bigl[ \eta_{\mu\rho} \eta_{\nu\sigma} 
		+ \eta_{\mu\sigma} \eta_{\nu\rho} 
		- \frac{2 \eta_{\mu\nu} \eta_{\rho\sigma} }{D \!-\! 2}  \Bigr]
\nonumber \\
&&	\hspace{1cm}
	=
	a_z^{D-2}
	\Bigl[
	a_y^{D-2}
	\overline{\partial}_y \!\cdot\! \overline{\partial}_z
	\partial_y \!\cdot\! \partial_z
	+
	a_y^{D-2}
	\overline{\partial}_y \!\cdot\! \partial_z
	\partial_y \!\cdot\! \overline{\partial}_z
	-
	\tfrac{1}{2} \overline{\partial}_z \!\cdot\! \partial_z \widetilde{\mathcal{D}}_y
	\Bigr]
	\, .
\label{initial5bar}
\end{eqnarray}
The derivatives in the resulting three terms above are then partially integrated 
using (\ref{momcons}), and reflected over the graviton propagator using (\ref{reflection})
so as to extract as many of them from the triangular loop.
Not all derivatives can be extracted this way. The ones remaining inside 
we arrange when possible to be contracted with another derivative at the same 
vertex. This procedure gives the following for the first term in  (\ref{initial5bar}),
\begin{eqnarray}
&&
\hspace{-0.7cm}
(a_y a_z)^{D-2}
	\overline{\partial}_y \!\cdot \! \overline{\partial}_z
	\partial_y \!\cdot\! \partial_z
	\longrightarrow
	-
	\tfrac{1}{2}
	\overline{\partial}_y \!\cdot \! \overline{\partial}_z
	\bigr(
	a_z^{D-2} \mathcal{D}_y
	+
	a_y^{D-2} \mathcal{D}_z
	\bigr)
\label{flat11}
\\
&&
\hspace{0.5cm}
	+
	\tfrac{1}{2} (a_y a_z)^{D-2}
	\overline{\partial}_y \!\cdot \! \overline{\partial}_z
	\Bigl[
	\overline{\partial}_y \!\cdot\! ( \overline{\partial}_y \!+\! \partial_{a_y} ) 
	+
	\overline{\partial}_z \!\cdot\! ( \overline{\partial}_z \!+\! \partial_{a_z} ) 
	+
	2\overline{\partial}_y \!\cdot \! \overline{\partial}_z
	\Bigr]
	\qquad
\label{flat12}
\\
&&
\hspace{0.5cm}
	+
	\tfrac{1}{4}
	(D\!-\!2)
	(a_y a_z)^{D-2}
	\overline{\partial}_y \!\cdot \! \overline{\partial}_z
	\Bigl\{
	H ( a_y \!-\! a_z) 
	\Bigl[
	2 \bigl( {\overline{\partial}_{y^0} }
		-
		\overline{\partial}_{z^0} \bigr)
\nonumber 
\\
&&
\hspace{2.cm}
	+
	\widetilde{\partial}_{y^0} 
	-
	\widetilde{\partial}_{z^0}
	+
	(D\!-\!2) H ( a_y \!-\! a_z )
	\Bigr]
	+
	H^2 ( a_y^2 \!+\! a_z^2 )
	\Bigr\}
	\, .
\label{puredS1}
\qquad
\end{eqnarray}
Note that we recognized scalar d'Alembertians (\ref{calD_i})
where that was not impeded by other derivatives or powers of the scale factor.
They are defined to act directly onto the corresponding propagator or external leg.
The first two lines in the expression above
contain terms that are de Sitter generalizations of flat 
space contributions, that we refer to as flat space descendants,
while the remaining lines contain only pure de Sitter terms.

The second term in (\ref{initial5bar}), upon reorganizing the derivatives, doubles
the contribution of the first term, and provides additional contributions,
\begin{eqnarray}
&&	\hspace{-0.8cm}
	(a_y a_z)^{D-2}
	\overline{\partial}_y \!\cdot\! \partial_z
	\partial_y \!\cdot\! \overline{\partial}_z
	\longrightarrow
	(a_y a_z)^{D-2}
	\overline{\partial}_y \!\cdot \! \overline{\partial}_z
	\partial_y \!\cdot\! \partial_z
\\
&&	\hspace{-0.5cm}
	+
	\tfrac{ 1 }{2}
	\overline{\partial}_y \!\cdot\! \overline{\partial}_z
	\bigl(
	a_z^{D-2}
	\widetilde{\mathcal{D}}_y
	+
	a_y^{D-2}
	\widetilde{\mathcal{D}}_z
	\bigr)
	+
	\tfrac{1}{4}
	\bigl(
	\mathcal{D}_y
	-
	\overline{\mathcal{D}}_y
	-
	\widetilde{\mathcal{D}}_y
	\bigr)
	\bigl(
	\mathcal{D}_z
	-
	\overline{\mathcal{D}}_z
	-
	\widetilde{\mathcal{D}}_z
	\bigr)
\qquad
\label{flat21}
\\
&&	\hspace{-0.5cm}
	+
	\tfrac{1}{4} (D\!-\!2)
		(a_y a_z)^{D-2}
	\Bigl\{
	H ( a_y \!-\! a_z) 
	\Bigl[
	2
	\overline{\partial}_y \!\cdot\! \widetilde{\partial}_y
	\overline{\partial}_{z^0}
	- 
	2 
	\overline{\partial}_z \!\cdot\! \widetilde{\partial}_z
	\overline{\partial}_{y^0} 
\label{puredS2}
\\
&&	\hspace{-0.2cm}
	-
	\overline{\partial}_y \!\cdot\! \overline{\partial}_z
	\bigl(
	\widetilde{\partial}_{y^0} \!-\! \widetilde{\partial}_{z^0} 
	\bigr)
	-
	(D\!-\!2)
	H (a_y \!-\! a_z)
	\overline{\nabla}_{y} \!\cdot\! \overline{\nabla}_{z}
	\Bigr]
	- H^2( a_y^2 \!+\! a_z^2 )
	\overline{\partial}_y \!\cdot\! \overline{\partial}_z
	\Bigr\}
\label{puredS3}
	 .
	\qquad
\end{eqnarray}
The latter can again be recognized as flat space descendants 
in line (\ref{flat21}), and as pure de Sitter contributions 
in lines (\ref{puredS2}) and (\ref{puredS3}).
Finally, the third term in (\ref{initial5bar}) produces only flat
space descendants,
\begin{eqnarray}
	-
	\tfrac{1}{2} 
	a_z^{D-2}
	\overline{\partial}_z \!\cdot\! \partial_z \widetilde{\mathcal{D}}_y
	\longrightarrow
	-
	\tfrac{1}{4} 
	\widetilde{\mathcal{D}}_y
	\bigl(
	\widetilde{\mathcal{D}}_z
	-
	\overline{\mathcal{D}}_z
	-
	\mathcal{D}_z
	\bigr)
	\, .
\label{flat31}
\end{eqnarray}

Further reduction of terms in (\ref{flat11}-\ref{flat31})
is accomplished by acting the scalar kinetic operators where 
possible, applying 3-point Donoghue identities (\ref{Dono3dS}) 
and (\ref{der3ptDon}), and finally acting the remaining 
derivatives onto external legs. We consider these reductions separately for 
flat space descendants and for pure de Sitter contributions.

\subsubsection{Reduction of flat space descendants}
\label{subsubsec: ReductionFlat}

Flat space descendant contributions comprises twice the terms in 
lines (\ref{flat11}) and (\ref{flat12}),
and terms in lines (\ref{flat21}) and (\ref{flat31}),
that added together read,
\begin{eqnarray}
\hspace{-0.5cm}
&&
	(a_y a_z)^{D-2}
	\overline{\partial}_y \!\cdot \! \overline{\partial}_z
	\Bigl[
	2\overline{\partial}_y \!\cdot \! \overline{\partial}_z
	+
	\overline{\partial}_y \!\cdot\! ( \overline{\partial}_y \!+\! \partial_{a_y} ) 
	+
	\overline{\partial}_z \!\cdot\! ( \overline{\partial}_z \!+\! \partial_{a_z} ) 
	\Bigr]
\label{descendant1}
\\
\hspace{-0.5cm}
&&
	-
	\tfrac{1}{2}
	\overline{\partial}_y \!\cdot \! \overline{\partial}_z
	\Bigr[
	2a_z^{D-2} \mathcal{D}_y
	+
	2a_y^{D-2} \mathcal{D}_z
	-
	a_z^{D-2}
	\widetilde{\mathcal{D}}_y
	-
	a_y^{D-2}
	\widetilde{\mathcal{D}}_z
	\Bigr]
\label{descendant2}
\\
\hspace{-0.5cm}
&&
	+
	\tfrac{1}{4}
	\bigl(
	\mathcal{D}_y
	-
	\overline{\mathcal{D}}_y
	\bigr)
	\bigl(
	\mathcal{D}_z
	-
	\overline{\mathcal{D}}_z
	-
	\widetilde{\mathcal{D}}_z
	\bigr)
	+
	\tfrac{1}{2}
	\widetilde{\mathcal{D}}_y
	\overline{\mathcal{D}}_z
	\, .
\label{descendant3}
\end{eqnarray}
Simplest to reduce are terms in the last line~(\ref{descendant3}), 
whose contribution we label by ``$(A)$'',
that consist of products of scalar d'Alembertians.
Only terms containing
combination~$(\mathcal{D}_y \!-\! \overline{\mathcal{D}}_y)$ that contracts the
massive propagator contribute in the end,
\begin{equation}
(A): \quad
	\tfrac{1}{4}
	\bigl(
	\mathcal{D}_y
	\!-\!
	\overline{\mathcal{D}}_y
	\bigr)
	\bigl(
	\mathcal{D}_z
	\!-\!
	\overline{\mathcal{D}}_z
	\!-\!
	\widetilde{\mathcal{D}}_z
	\bigr)
	i \Delta_m(x;y)
	\longrightarrow
	\tfrac{1}{4} 
	i \delta^D(x\!-\!y)
	\bigl(
	\mathcal{D}_z
	\!-\!
	\overline{\mathcal{D}}_z
	\!-\!
	\widetilde{\mathcal{D}}_z
	\bigr)
	\, ,
	\ \
\label{(A)contribution}
\end{equation}
while the remaining ones vanish in the large mass limit.

The terms in line (\ref{descendant2}) that have two external derivatives
are best considered together with the latter two terms in first line (\ref{descendant1})
that contain two external would-be scalar d'Alembertians. 
We label this contribution by ``$(B)$''.
It is only the middle
term in (\ref{descendant1}) and the first term in (\ref{descendant2}) that contain
sufficient number of derivatives acting on either external mode functions or
on the massive propagator to generate a nonvanishing contribution. This contribution
is obtained by first acting the scalar d'Alembertian, then applying the
3-point Donoghue identity (\ref{Dono3dS}), followed by acting derivatives
on external mode functions using (\ref{extderiv}),
\begin{eqnarray}
&& \hspace{-1.cm}
(B): \qquad
	a_z^{D-2}
	\overline{\partial}_y \!\cdot \! \overline{\partial}_z
	\Bigl[
	a_y^{D-2}
	\overline{\partial}_y \!\cdot\! ( \overline{\partial}_y \!+\! \partial_{a_y} ) 
	-
	\mathcal{D}_y
	\Bigr]
	i \Delta_m(x;y)
\\
&& \hspace{-0.7cm}
	\longrightarrow
	a_z^{D-2}
	\overline{\partial}_y \!\cdot \! \overline{\partial}_z
	\Bigl\{
	a_y^{D-2}
	\Bigl[
	\overline{\partial}_y \!\cdot\! ( \overline{\partial}_y \!+\! \partial_{a_y} ) 
	-
	m^2a_y^2 
	\Bigr]
	i \Delta_m(x;y)
	-
	i \delta^D(x\!-\!y)
	\Bigr\}
	\qquad
\\
&& \hspace{-0.7cm}
	\longrightarrow
	a_z^{D-2}
	\overline{\partial}_y \!\cdot \! \overline{\partial}_z
	\Bigl[
	\tfrac{1}{2m^2a_y^2}
	\overline{\partial}_y \!\cdot\! ( \overline{\partial}_y \!+\! \partial_{a_y} ) 
	-
	\tfrac{3}{2}
	\Bigr]
	i\delta^D(x\!-\!y)
\\
&& \hspace{-0.7cm}
	\longrightarrow
	a_z^{D-2}
	\Bigl[
	- \overline{\partial}_y \!\cdot \! \overline{\partial}_z
	-
	H a_y \overline{\partial}_{z^0}
	\Bigr]
	i\delta^D(x\!-\!y)
	\, .
\label{(B)contribution}
\end{eqnarray}
In the last step above one needs to account for all the terms generated by
multiple derivatives acting on external mode functions,
\begin{equation}
\overline{\partial}_z \!\cdot\! \overline{\partial}_y \,
	\overline{\partial}_y \!\cdot\! ( \overline{\partial}_y \!+\! \partial_{a_y} )
	\longrightarrow
	m^2 a_y^2 \Bigl[
	\overline{\partial}_y \!\cdot\! \overline{\partial}_z
	- 
	2 H a_y \overline{\partial}_{z^0}
	-
	 \tfrac{(D-2) H^2}{m^2} 
		\overline{\partial}_{y^0} \overline{\partial}_{z^0}
		\Bigr]
	\, ,
\end{equation}
and keep only ones contributing in the large mass limit.

The last flat space descendant contribution,
that we label by ``$(C)$'', is the first term in 
line (\ref{descendant1}) with all four derivatives extracted.
Reducing it is straightforward and again follows from applying the 
3-point Donoghue identity, followed by
acting the derivatives on the external mode function,
\begin{eqnarray}
&&	\hspace{-2cm}
(C): \quad
2(a_ya_z)^{D-2} 
	( \overline{\partial}_y \!\cdot\! \overline{\partial}_z )^2
	i \Delta_m(x;y)
\nonumber \\
&&
	\longrightarrow
a_z^{D-2} 
	( \overline{\partial}_y \!\cdot\! \overline{\partial}_z )^2
	\frac{ i \delta^D(x\!-\!y)}{m^2 a_y^2}
	\longrightarrow
	-
	a_z^{D-2} 
	( \overline{\partial}_{z^0} )^2
	i \delta^D(x\!-\!y)
	\, .
	\quad
\label{(C)contribution}
\end{eqnarray}

\subsubsection{Reduction of pure de Sitter terms}
\label{subsubsec: ReductionDS}

Pure de Sitter contributions to $\overline{5a}$ residual diagram 
consist of twice the contributions in (\ref{puredS1}), and lines
(\ref{puredS2}) and (\ref{puredS3}), which added together read,
\begin{eqnarray}
&&	\hspace{-1cm}
	\tfrac{1}{4}
	(D\!-\!2)
	(a_y a_z)^{D-2}
	\times
	\Bigl\{
	H ( a_y \!-\! a_z) 
	\Bigl[
	4 \overline{\partial}_y \!\cdot \! \overline{\partial}_z
	\bigl(
		{\overline{\partial}_{y^0} }
		-
		\overline{\partial}_{z^0}
		\bigr)
\label{pure1}
\\
&&
	+
	\overline{\partial}_y \!\cdot \! \overline{\partial}_z
	\bigl(
	\widetilde{\partial}_{y^0} 
	-
	\widetilde{\partial}_{z^0}
	\bigr)
	+
	2
	\overline{\partial}_y \!\cdot\! \widetilde{\partial}_y \,
	\overline{\partial}_{z^0}
	- 
	2 
	\overline{\partial}_z \!\cdot\! \widetilde{\partial}_z \,
	\overline{\partial}_{y^0} 
\label{pure2}
\\
&&
	+
	(D\!-\!2) 
	H ( a_y \!-\! a_z )
	\bigl(
	2
	\overline{\partial}_y \!\cdot \! \overline{\partial}_z
	-
	\overline{\nabla}_{y} \!\cdot\! \overline{\nabla}_{z}
	\bigr)
	\Bigr]
	+ H^2 (a_y^2 \!+\! a_z^2) \overline{\partial}_y \!\cdot\! \overline{\partial}_z
	\Bigr\}
	\, .
	\qquad
\label{pure3}
\end{eqnarray}
Terms in the first line (\ref{pure1}) with all derivatives extracted,
whose contribution we label by ``$(D)$'',
are immediately reduced by the Donoghue 3-point identity followed
by acting derivatives on external legs. 
Only the first term in (\ref{pure1}) manages to
contribute,
\begin{eqnarray}
&&	\hspace{-2.cm}
(D): \qquad
	(D\!-\!2)
	H ( a_y \!-\! a_z) 
	(a_y a_z)^{D-2}
	\overline{\partial}_y \!\cdot \! \overline{\partial}_z
	{\overline{\partial}_{y^0} }
	i \Delta_m(x;y)
\\
&&
	\hspace{0.5cm}
	\longrightarrow
	(D\!-\!2)
	H ( a_y \!-\! a_z) 
	a_z^{D-2}
	\overline{\partial}_y \!\cdot \! \overline{\partial}_z
	{\overline{\partial}_{y^0} }
	\frac{ i \delta^D(x\!-\!y) }{ 2m^2 a_y^2 }
	\qquad
\\
&&
	\hspace{0.5cm}
	\longrightarrow
	\tfrac{1}{2}
	(D\!-\!2)
	H ( a_y \!-\! a_z) 
	a_z^{D-2}
	\overline{\partial}_{z^0}
	i \delta^D(x\!-\!y)
	\, .
\label{(D)contribution}
\end{eqnarray}

The second line~(\ref{pure2}) contains terms with single internal derivatives that
cannot be extracted from the triangular loop. Reduction of such terms requires
the derivative 3-point Donoghue identity (\ref{der3ptDon}).
For the first two terms, their contribution labeled by $(E)$, we can apply this identity
immediately after reflecting the internal derivative to act on the~$y$ coordinate
of the graviton propagator.
Acting derivatives on external legs then concludes the reduction,
\begin{eqnarray}
&&	\hspace{-0.7cm}
	(E):
	\quad
	\tfrac{1}{2}
	(D\!-\!2)
	(a_y a_z)^{D-2}
	H ( a_y \!-\! a_z) 
	\Bigl[
	\overline{\partial}_y \!\cdot \! \overline{\partial}_z
	\widetilde{\partial}_{y^0}
	\!+\!
	\overline{\partial}_y \!\cdot\! \widetilde{\partial}_y
	\,
	\overline{\partial}_{z^0}
	\Bigr]
	i \Delta_m(x;y)
\qquad
\\
&&	\hspace{0.9cm}
	\longrightarrow
	\tfrac{1}{2}
	(D\!-\!2)
	a_z^{D-2}
	H ( a_y \!-\! a_z) 
	\!
	\Bigl[
	\overline{\partial}_y \!\cdot \! \overline{\partial}_z
	\,
	\overline{\partial}_{y^0}
	+
	\overline{\partial}_y \!\cdot\! \overline{\partial}_y
	\,
	\overline{\partial}_{z^0}
	\Bigr]
	\!
	\frac{ i \delta^D(x\!-\!y) }{ 2 m^2 a_y^2 }
	\qquad
	\
\\
&&	\hspace{0.9cm}
	\longrightarrow
	\tfrac{1}{2}
	(D\!-\!2)
	a_z^{D-2}
	H ( a_y \!-\! a_z) 
	\overline{\partial}_{z^0}
	i \delta^D(x\!-\!y)
	\, .
\label{(E)contribution}
\end{eqnarray}

For the remaining term in line~(\ref{pure2}), labeled by ``$(F)$'',
the derivative Donoghue identity should not be applied directly in order not 
to lose the subleading terms. 
We first write out the derivatives contracted
at the same vertex in terms of d'Alembertian operators keeping track of
correction terms up to order $H^2$,
\begin{eqnarray}
&&	\hspace{-1.3cm}
(F):	\quad
	- \tfrac{1}{2}
	(D\!-\!2)
	(a_y a_z)^{D-2}
	H ( a_y \!-\! a_z) 
	\overline{\partial}_z \!\cdot\! \widetilde{\partial}_z \,
	\overline{\partial}_{y^0} 
	i \Delta_m(x;y)
\qquad
\\
&&	\hspace{0.5cm}
\longrightarrow
	- \tfrac{1}{4}
	(D\!-\!2)
	a_y^{D-2}
	H ( a_y \!-\! a_z) 
	\bigl(
	\mathcal{D}_z 
	-
	\overline{\mathcal{D}}_z
	-
	\widetilde{\mathcal{D}}_z
	\bigr)
	\overline{\partial}_{y^0} 
	i \Delta_m(x;y)
	\qquad
\label{inter1}
\\
&&	\hspace{1.7cm}
	- \tfrac{1}{4}
	(D\!-\!2)
	(a_y a_z)^{D-2}
	H^2 a_z^2
	\widetilde{\partial}_{y^0} 
	\overline{\partial}_{y^0} 
	i \Delta_m(x;y)
	\, .
\label{inter2}
\end{eqnarray}
None of the terms in line (\ref{inter1}) can contribute, but the term in 
line (\ref{inter2}) produces a nonvanishing contribution after applying the
derivative 3-point Donoghue identity and acting the external leg derivatives,
\begin{equation}
(F):	\quad \
	\longrightarrow
	- \tfrac{1}{4}
	(D\!-\!2)
	H^2 a_z^D
	( \overline{\partial}_{y^0} )^2
	\frac{ i \delta^D(x\!-\!y)}{ 2 m^2 a_y^2}
	\longrightarrow
	\tfrac{1}{4}
	(D\!-\!2)
	H^2 a_z^D  i \delta^D(x\!-\!y)
	\, .
	\
	\
\label{(F)contribution}
\end{equation}
Lastly, the remaining terms, given in line (\ref{pure3}), do not have enough external 
derivatives to contribute.

\subsubsection{Self-mass contributions}

The reduced diagram~$\overline{5a}$ collects $(A)$--$(F)$ contributions
from (\ref{(A)contribution}), (\ref{(B)contribution}), (\ref{(C)contribution}),
(\ref{(D)contribution}), (\ref{(E)contribution}), and (\ref{(F)contribution}).
It is natural to consider it together with the reduced diagram~$\overline{5b}$,
that is obtained from $\overline{5a}$ by reflecting
$x \! \leftrightarrow \! y$ in the final results of reductions 
in sections \ref{subsubsec: ReductionFlat} and \ref{subsubsec: ReductionDS}. 
Together they make up diagrams of the same topology,
\begin{eqnarray}
&&	\hspace{-1.5cm}
\bigl[ - i V_{\overline{5a+b}} \bigr]^{\rm 3pt}_{(I)}
	=
	(-i\lambda)^2 \delta^{D}(x\!-\!y) \delta^{D}(x'\!-\!y') (a_xa_{x'})^D
\nonumber \\
&&	\hspace{0.5cm}
	\times
	i \kappa^2 \!\!  \int\! d^{D\!}z \, 
	\Theta_{I} \Bigl[ a_z^{D-2} i \Delta_C(x;z) i \Delta_A(x;z) \Bigr]
	\times i \Delta_A(z;x')
	\, ,
\label{V53pt}
\end{eqnarray}
where contributions $(A)$--$(F)$ are accounted for by derivative operators
that act to their right,
\begin{eqnarray}
&&
\hspace{-0.7cm}
\Theta_{(A)}
	=
	\tfrac{ 1 }{2} \mathcal{D}_z a_z^{2-D}
	\!+\!
	(D\!-\!3) H^2 a_z^2
	\, ,
\ \ \
\Theta_{(B)}
	=
	\partial_x \!\cdot\! \partial_z 
	\!-\! 
	(D\!+\!2) H a_x \partial_0^z
	\, ,
\label{Thetas}
\\
&&
\hspace{-0.7cm}
\Theta_{(C)}
	=
	2 (\partial_0^z )^2
	\, ,
\ \ \
\Theta_{(D+E)} \!=\! 2 (D\!-\!2) \partial_0^z H (a_x\!-\!a_z) 
	\, ,
\ \ \
\Theta_{(F)} = - \tfrac{1}{2} (D\!-\!2) H^2 a_z^2
\, .
\nonumber
\end{eqnarray}
In addition $(A)$ contributions generate diagrams of different topology when scalar
d'Alembertians in (\ref{(A)contribution}) contract either the graviton or the massless 
scalar propagator,
\begin{eqnarray}
&&	\hspace{-2cm}
\bigl[ - i V_{\overline{5a+b}} \bigr]^{\rm 4pt}_{(A)}
	=
	(-i\lambda)^2 \delta^{D}(x\!-\!y) \delta^{D}(x'\!-\!y') (a_xa_{x'})^D
\nonumber \\
&&	\hspace{1.5cm}
	\times
	\tfrac{ 1 }{2} \kappa^2
	\bigl[ i \Delta_C(x;x) - i \Delta_A(x;x) \bigr]
	\times i \Delta_A(x;x')
	\, .
\end{eqnarray}

Some of the differential operators in (\ref{Thetas}) can be further rewritten
owing to the topology of diagram (\ref{V53pt}) and the identity,
\begin{eqnarray}
&&	\hspace{-0.8cm}
a_x^{D-n_x} \partial_0^z \Bigl[ a_z^{D-n_z} i \Delta_C(x;z) i \Delta_{A}(x;z) \Bigr]
\nonumber \\
&&
=
a_x^{D-n_x} \bigl( - \partial_0^x + \partial_0^x + \partial_0^z \bigr)
	\Bigl[ a_z^{D-n_z} i \Delta_C(x;z) i \Delta_{A}(x;z) \Bigr]
\nonumber \\
&&
\longrightarrow
a_x^{D-n_x} \Bigl[ (1\!-\!n_x) H a_x + \partial_0^x + \partial_0^z \Bigr]
	\Bigl[ a_z^{D-n_z} i \Delta_C(x;z) i \Delta_{A}(x;z) \Bigr]
	\, ,
	\qquad
\label{zeroID}
\end{eqnarray}
that is derived by partial integration and accounting for action of time
derivatives on external legs,
\begin{equation}
\partial_0 \bigl[ u_m(x,\vec{k}) u_m^*(x,\vec{k}') \bigr]
	\xrightarrow{m^2 \to \infty} - (D\!-\!1) H a_x \times u_m(x,\vec{k}) u_m^*(x,\vec{k}')
	\, .
\end{equation}
It is $(B)$ and $(C)$ contributions that benefit from reorganizing derivatives,
\begin{eqnarray}
&&
\hspace{-0.55cm}
\Theta_{(B)} \!\longrightarrow\!
	-
	\mathcal{D}_z a_z^{2-D}
	\!\!-\!
	\bigl( \partial_0^z \!+\! 2 H a_x \bigr)
		\bigl( \partial_0^x \!+\! \partial_0^z \!+\! 2 H a_x \bigr)
	\!-\!
	(D\!-\!2) \partial_0^z H (a_x \!-\! a_z)
	\, ,
	\quad
\nonumber \\
&&
\hspace{-0.55cm}
\Theta_{(C)} \!\longrightarrow\!
	2 \bigl( \partial_0^z \!-\! H a_x \bigr)
	\bigl( \partial_0^x \!+\! \partial_0^z \!+\! 2 H a_x \bigr)
	\, .
	\qquad
\end{eqnarray}
The utility of this form is that only~$\mathcal{D}_z a_z^{2-D}$ term will contain
divergences, while all the other terms will be manifestly finite. 
Collecting all the contributions $(A)$--$(F)$ to
 diagrams $\overline{5a}\!+\!\overline{5b}$ finally produces a contribution to
the $\beta$-type diagram from Fig.~\ref{Reduced},
\begin{eqnarray}
\bigl[ - i f_\beta(x;x') \bigr]_{\overline{5}}
	&\!\!\!=\!\!\!&
	\tfrac{\kappa^2}{2}
	\Bigl[
	- \mathcal{D}_{x'} a_{x'}^{2-D}
	+
	2\bigl( \partial_0^{x'} \!-\! 4 H a_x \bigr)
	\bigl( \partial_0^x \!+\! \partial_0^{x'} \!+\! 2 H a_x \bigr)
\nonumber \\
&&	\hspace{-2cm}
	+
	2(D\!-\!2) \partial_0^{x'} ( a_x \!-\! a_{x'} )
	+
	(D\!-\!4) H^2 a_{x'}^2
	\Bigr]
	\Bigl[ a_{x'}^{D-2} i \Delta_C(x;x') i \Delta_A(x;x') \Bigr]
\nonumber \\
&&
	-
	\tfrac{ \kappa^2 }{2}
	i \delta^D(x\!-\!x')
	\Bigl[ i \Delta_C(x';x') - i \Delta_A(x';x') \Bigr] 
	\, .
 \label{fbeta5}
\end{eqnarray}
The two remaining diagrams $\overline{5c} \!+\! \overline{5d}$ are then inferred
by reflecting~$x \!\leftrightarrow\! x'$ in the~$\beta$-type contribution above, 
yielding a~$\gamma$-type contribution from Fig.~\ref{Reduced},
\begin{eqnarray}
\bigl[ - i f_\gamma (x;x') \bigr]_{\overline{5}}
	&\!\!\!=\!\!\!&
	\tfrac{\kappa^2}{2}
	\Bigl[
	- \mathcal{D}_{x} a_{x}^{2-D}
	+
	2\bigl(  \partial_0^{x} \!-\! 4 H a_{x'} \bigr)
	\bigl( \partial_0^x \!+\! \partial_0^{x'} \!+\! 2 H a_{x'} \bigr)
\nonumber \\
&&	\hspace{-2cm}
	-
	2(D\!-\!2) \partial_0^{x} ( a_x \!-\! a_{x'} )
	+
	(D\!-\!4) H^2 a_{x}^2
	\Bigr]
	\Bigl[ a_{x}^{D-2} i \Delta_C(x;x') i \Delta_A(x;x') \Bigr]
\nonumber \\
&&
	-
	\tfrac{ \kappa^2 }{2}
	i \delta^D(x\!-\!x')
	\Bigl[ i \Delta_C(x;x) - i \Delta_A(x;x) \Bigr] 
	\, .
\label{fgamma5}
\end{eqnarray}

\subsection{$V_{\overline{\overline{4}}}(x;x';y;y')$ Diagrams}

Each of the residual diagrams (\ref{4abarbar}-\ref{4dbarbar}) contains
a product of truncated vertices from a single graviton coupled to the 
massive scalar kinetic term. For the case of $-i V_{\overline{\overline{4a}}}$
these are,
\begin{eqnarray}
\lefteqn{ \Bigl[a_{y}^{D-2} \overline{\partial}^{\mu}_{y} \partial^{\nu}_{y} 
\!-\! \tfrac14 \eta^{\mu\nu} \widetilde{\mathcal{D}}_{y}\Bigr] \Bigl[a_{y'}^{D-2} 
\overline{\partial}_{y'}^{\rho} \partial_{y'}^{\sigma} \!-\! \tfrac14 
\eta^{\rho\sigma} \widetilde{\mathcal{D}}_{y'}\Bigr] = (a_y a_{y'})^{D-2} 
\overline{\partial}^{\mu}_{y} \partial^{\nu}_{y} \overline{\partial}_{y'}^{\rho} 
\partial_{y'}^{\sigma} } \nonumber \\
& & \hspace{1.5cm} - \tfrac14 a_y^{D-2} \overline{\partial}^{\mu}_{y} 
\partial^{\nu}_{y} \widetilde{\mathcal{D}}_{y'} \eta^{\rho\sigma} - \tfrac14 
a_{y'}^{D-2} \widetilde{\mathcal{D}}_{y} \eta^{\mu\nu} 
\overline{\partial}_{y'}^{\rho} \partial_{y'}^{\sigma} + \tfrac1{16} 
\widetilde{\mathcal{D}}_{y} \widetilde{\mathcal{D}}_{y'} \eta^{\mu\nu} 
\eta^{\rho\sigma} \; . \qquad \label{product}
\end{eqnarray}
With the simplification (\ref{gravprop3}) these factors will be contracted into
the flat space tensor,
\begin{eqnarray}
\lefteqn{ \Bigl[a_{y}^{D-2} \overline{\partial}^{\mu}_{y} \partial^{\nu}_{y} 
\!-\! \tfrac14 \eta^{\mu\nu} \widetilde{\mathcal{D}}_{y}\Bigr] \Bigl[a_{y'}^{D-2} 
\overline{\partial}_{y'}^{\rho} \partial_{y'}^{\sigma} \!-\! \tfrac14 
\eta^{\rho\sigma} \widetilde{\mathcal{D}}_{y'}\Bigr] \Bigl[ \eta_{\mu\rho}
\eta_{\nu\sigma} + \eta_{\mu\sigma} \eta_{\nu\rho} - \frac{2}{D\!-\!2} 
\eta_{\mu\nu} \eta_{\rho\sigma} \Bigr] } \nonumber \\
& & \hspace{-0.5cm} = (a_y a_{y'})^{D-2} \Bigl[ \overline{\partial}_{y} 
\!\cdot\! \overline{\partial}_{y'} \partial_{y} \!\cdot\! \partial_{y'} + 
\overline{\partial}_{y} \!\cdot\! \partial_{y'} \overline{\partial}_{y'} \!\cdot\! 
\partial_{y} \Bigr] - \frac{2}{D\!-\!2} \Bigl[ m^2 a_{y}^D + \tfrac12 (
\mathcal{D}_{y} \!-\! m^2 a_{y}^{D}) \Bigr] \nonumber \\ 
& & \hspace{5cm} \times \Bigl[ m^2 a_{y'}^D + \tfrac12 (\mathcal{D}_{y'} \!-\! 
m^2 a_{y'}^{D}) \Bigr] - \tfrac14 \widetilde{\mathcal{D}}_{y} 
\widetilde{\mathcal{D}}_{y'} \; . \qquad \label{prodreduc}
\end{eqnarray}
Note that the last term can be dropped. We label the remaining terms ``$A$'', 
``$B$'' and ``$C$'', respectively.

The ``$C$'' terms proportional to $-2/(D-2)$ in expression (\ref{prodreduc}) can
already be recognized as making 2-point, 3-point and 4-point contributions,
\begin{eqnarray}
{\rm 2\!-\!Point} &\!\!\! \Longrightarrow \!\!\!& -\tfrac{1}{2 (D-2)} 
\!\times\! (\mathcal{D}_{y} \!-\! m^2 a_y^D) \!\times\! (\mathcal{D}_{y'} 
\!-\! m^2 a_{y'}^D) \; , \qquad \label{C2point} \\
{\rm 3\!-\!Point} &\!\!\! \Longrightarrow \!\!\!& -\tfrac{1}{D-2} \!\times\! 
(\mathcal{D}_{y} \!-\! m^2 a_y^D) \!\!\times\! m^2 a_{y'}^{D} \!-\!
\tfrac{1}{D-2} \!\times\! m^2 a_{y}^D \!\!\times\! (\mathcal{D}_{y'} \!-\! 
m^2 a_{y'}^D) \; , \qquad \label{C3point} \\
{\rm 4\!-\!Point} &\!\!\! \Longrightarrow \!\!\!& -\tfrac{2}{D-2} \!\times\!
m^2 a_{y}^{D} \!\times\! m^2 a_{y'}^{D} \; . \qquad \label{C4point}
\end{eqnarray}
To reach the four classes of 2-point forms described in Figure~\ref{Reduced}
we first recall that each term in $-i V_{\overline{\overline{4a}}}$ shares a
common factor of,
\begin{equation}
\kappa^2 \lambda^2 (a_x a_{x'})^D i\Delta_m(x;y) i\Delta_m(x';y') 
i\Delta_A(x;x') i\Delta_C(y;y') \; . \label{4afactor}
\end{equation}
One now acts the massive scalar kinetic operators and invokes the appropriate
Donoghue Identity. For example, the 2-point contribution (\ref{C2point}) gives
\begin{eqnarray}
\lefteqn{-\tfrac{\kappa^2 \lambda^2}{2 (D - 2)} (a_x a_{x'})^D \!\times\!
i\delta^D(x \!-\! y) \!\times\! i\delta^D(x' \!-\! y') \!\times\! 
i\Delta_A(x;x') i\Delta_C(y;y') } \nonumber \\
& & \hspace{0cm} = \delta^D(x \!-\! y) \delta^D(x' \!-\! y') \!\times\!
-i \lambda^2 (a_x a_{x'})^D \!\times\! \tfrac{i \kappa^2}{2 (D-2)} 
i\Delta_A(x;x') i\Delta_C(x;x') \; . \qquad \label{C2alpha}
\end{eqnarray}
The final factor of (\ref{C2alpha}) contributes to $f_{\alpha}(x;x')$.
The 3-point contribution (\ref{C3point}) requires the 3-point Donoghue
Identity (\ref{Dono3dS}) after acting the scalar kinetic operators,
\begin{eqnarray}
\lefteqn{-\tfrac{\kappa^2 \lambda^2}{D - 2} (a_x a_{x'})^D \!\times\!
i\delta^D(x \!-\! y) \!\times\! m^2 a_{y'}^D \!\times\! i\Delta_m(x';y') 
i\Delta_A(x;x') i\Delta_C(y;y') } \nonumber \\
& & \hspace{0.5cm} -\tfrac{\kappa^2 \lambda^2}{D - 2} (a_x a_{x'})^D 
\!\times\! i\delta^D(x' \!-\! y') \!\times\! m^2 a_{y}^D \!\times\!
i\Delta_m(x;y) i\Delta_A(x;x') i\Delta_C(y;y') \nonumber \\
& & \hspace{0cm} \longrightarrow -\tfrac{\kappa^2 \lambda^2}{D - 2} 
(a_x a_{x'})^D \!\times\! i\delta^D(x \!-\! y) \!\times\! i\delta^D(x'
\!-\! y') \!\times\! i\Delta_A(x;x') i\Delta_C(x;x') \; , \qquad \\
& & \hspace{0cm} = \delta^D(x \!-\! y) \delta^D(x' \!-\! y') 
\!\times\! -i \lambda^2 (a_x a_{x'})^D \!\times\! \tfrac{i \kappa^2}{D-2} 
\, i\Delta_A(x;x') i\Delta_C(x;x') \; . \qquad \label{C3alpha}
\end{eqnarray}
This is a contribution to $f_{\alpha}(x;x')$ of twice the size as
(\ref{C2alpha}).

The 4-point contribution (\ref{C4point}) has no scalar kinetic operators
but it requires the 4-point Donoghue Identity (\ref{Dono4dS}). The first
line contains the factor,
\begin{equation}
\Bigl[-\frac{i}{m^2} + \frac{i}{3 m^4} \Bigl( \frac{\overline{\partial}_{x}
\!\cdot\! \overline{\partial}_{x'}}{a_x a_{x'}} - m^2\Bigr) \Bigr] 
\frac{\delta^D(x \!-\! y) \delta^D(x' \!-\! y')}{(a_x a_{x'})^D} \; . 
\label{ID4factor}
\end{equation}
The external state derivatives can be expanded using (\ref{extderiv}),
\begin{equation}
\frac{\overline{\partial}_{x} \!\cdot\! \overline{\partial}_{x'}}{
a_x a_{x'}} = m^2 - i(D\!-\!1) m H - \tfrac12 (D\!-\!1)^2 H^2 + \tfrac12
\Bigl\Vert \frac{\vec{k}_1}{a_x} - \frac{\vec{k}_3}{a_{x'}} \Bigr\Vert^2 
+ O\Bigl(\frac1{m}\Bigr) \; . \label{factorexp}
\end{equation}
It follows that the 4-point contribution (\ref{C4point}) reduces to,
\begin{eqnarray}
\lefteqn{-\tfrac{2 m^4 (a_y a_{y'})^D}{D - 2} \!\times\! \kappa^2 \lambda^2
(a_x a_{x'})^D i\Delta_m(x;y) i\Delta_m(x';y') i\Delta_A(x;x') 
i\Delta_C(y;y') } \nonumber \\
& & \hspace{-0.5cm} \longrightarrow \delta^D(x \!-\! y) \delta^D(x' \!-\! y')
\!\times\! -i \lambda^2 (a_x a_{x'})^D \nonumber \\
& & \hspace{0cm} \times -\tfrac{\kappa^2}{3 (D-2)} \Bigl[ 3 m^2 + i (D\!-\!1) 
m H + \tfrac12 (D\!-\!1)^2 H^2 - \tfrac12 \Vert \tfrac{\vec{k}_1}{a_x} \!-\! 
\tfrac{\vec{k}_3}{a_{x'}} \Vert^2 \Bigr] \nonumber \\
& & \hspace{1cm} \times \Biggl\{ \int \!\! d^Dz \, a_{z}^D i\Delta_A(x;z)
i\Delta_C(x;z) i\Delta_A(z;x') + \Bigl( x^{\mu} \longleftrightarrow 
{x'}^{\mu} \Bigr) \Biggr\} . \qquad \label{C4betagamma}
\end{eqnarray}

The 4-point ``$C$'' contribution from $-i V_{\overline{\overline{4b}}}$
is almost the same as (\ref{C4betagamma}) except the square-bracketed term
in the 3rd line is,
\begin{equation}
-i V_{\overline{\overline{4b}}} \Longrightarrow 3 m^2 - i (D\!-\!1) 
m H + \tfrac12 (D\!-\!1)^2 H^2 - \tfrac12 \Vert \tfrac{\vec{k}_2}{a_x} \!-\! 
\tfrac{\vec{k}_4}{a_{x'}} \Vert^2 \; . \label{line3b} 
\end{equation}
The other contributions are simpler,
\begin{equation}
-i V_{\overline{\overline{4c}}} \Longrightarrow -3 m^2 + \tfrac12 \Vert 
\tfrac{\vec{k}_1}{a_x} \!-\! \tfrac{\vec{k}_4}{a_{x'}} \Vert^2 \qquad ,
\qquad -i V_{\overline{\overline{4d}}} \Longrightarrow -3 m^2  + \tfrac12 
\Vert \tfrac{\vec{k}_2}{a_x} \!-\! \tfrac{\vec{k}_3}{a_{x'}} \Vert^2 \; .
\label{line3cd}
\end{equation}
Adding (\ref{line3b}) and (\ref{line3cd}) to the 3rd line of 
(\ref{C4betagamma}), and using spatial momentum conservation (\ref{qdef})
gives,
\begin{equation}
-i V_{\overline{\overline{4a,b,c,d}}} \Longrightarrow (D\!-\!1)^2 H^2
- \frac{\Vert \vec{q} \, \Vert^2}{a_x a_{x'}} \; . \label{line3}
\end{equation}
This corresponds to Group $C$ contributions to the Class $\beta$ and 
$\gamma$ diagrams of Figure~\ref{Reduced},
\begin{eqnarray}
-i f_{\overline{\overline{4C}}\beta}(x;x') &\!\!\!\! = \!\!\!\!& 
\tfrac{\kappa^2 a_{x'}^D}{3 (D-2)} \Bigl[ \tfrac{q^2}{a_x a_{x'}} - 
(D\!-\!1)^2 H^2 \Bigr] i\Delta_A(x;x') i\Delta_C(x;x') \; , \qquad 
\label{C4beta} \\
-i f_{\overline{\overline{4C}}\gamma}(x;x') &\!\!\!\! = \!\!\!\!& 
\tfrac{\kappa^2 a_{x}^D}{3 (D-2)} \Bigl[ \tfrac{q^2}{a_x a_{x'}} -
(D\!-\!1)^2 H^2 \Bigr] i\Delta_A(x;x') i\Delta_C(x;x') \; . \qquad 
\label{C4gamma}
\end{eqnarray}
In contrast, the $C$ contributions to the Class $\alpha$ diagram of
Figure~\ref{Reduced} is just four times the sum of the final factors
of expressions (\ref{C2alpha}) and (\ref{C3alpha}),
\begin{equation}
-i f_{\overline{\overline{4C}}\alpha}(x;x') = \tfrac{6 \kappa^2}{D-2}
\, i\Delta_A(x;x') i\Delta_C(x;x') \; . \label{C4alpha}
\end{equation}
There is no contribution to the Class $\delta$ diagram.

The ``$A$'' and ``$B$'' terms in expression (\ref{prodreduc}) require 
extensive reduction using momentum conservation (\ref{momcons}), the 
graviton reflection identity (\ref{reflection}) and action of the 
external derivatives (\ref{extderiv}). This is simplest for the two 
factors of the ``$B$'' term $\overline{\partial}_{y} \cdot \partial_{y'} 
\times \overline{\partial}_{y'} \cdot \partial_{y}$. The first factor 
gives,
\begin{eqnarray}
\overline{\partial}_{y} \!\cdot\! \partial_{y'} &\!\!\! = \!\!\!&
-\overline{\partial}_{y} \!\cdot\! \Bigl( \overline{\partial}_{y'} +
\widetilde{\partial}_{y'} + \partial_{a_{y'}}\Bigr) \; , \qquad \\
&\!\!\! \simeq \!\!\!& -\overline{\partial}_{y} \!\cdot\! 
\overline{\partial}_{y'} + \overline{\partial}_{y} \!\cdot\! \Bigl(
\widetilde{\partial}_{y} + \tfrac12 \partial_{a_{y}} - \tfrac12
\partial_{a_{y'}} \Bigr) \; , \qquad \\
&\!\!\! \simeq \!\!\!& -\overline{\partial}_{y} \!\cdot\! 
\overline{\partial}_{y'} + \tfrac12 \Bigl[ \partial_{y} \!\cdot\! 
(\partial_{y} \!+\! \partial_{a_y}) - m^2 a_{y}^2\Bigr] - \tfrac12
\widetilde{\partial}_{y} \!\cdot\! (\widetilde{\partial}_{y} \!+\!
\partial_{a_{y}}) \nonumber \\
& & \hspace{6.5cm} - (\tfrac{D-2}{2}) a_y a_{y'} H^2 \Delta \eta_{y y'}
\overline{\partial}_{y^0} , \qquad \label{factor2a}
\end{eqnarray}
where $\Delta \eta_{y y'} \equiv y^{0} - y^{\prime 0}$. The second 
factor produces a similar result,
\begin{eqnarray}
\lefteqn{\overline{\partial}_{y'} \!\cdot\! \partial_{y} \simeq 
-\overline{\partial}_{y} \!\cdot\! \overline{\partial}_{y'} + \tfrac12 
\Bigl[ \partial_{y'} \!\cdot\! (\partial_{y'} \!+\! \partial_{a_{y'}}) 
- m^2 a_{y'}^2\Bigr] - \tfrac12 \widetilde{\partial}_{y'} \!\cdot\! 
(\widetilde{\partial}_{y'} \!+\! \partial_{a_{y'}}) } \nonumber \\
& & \hspace{7.5cm} + (\tfrac{D-2}{2}) a_y a_{y'} H^2 \Delta \eta_{y y'}
\overline{\partial}_{{y'}^0} . \qquad \label{factor2b}
\end{eqnarray}
If we use $\widetilde{\mathcal{D}}_{y} i \Delta_C(y;y') = 
\widetilde{\mathcal{D}}_{y'} i \Delta_C(y;y') \longrightarrow 2 (D-3)
i \Delta_C(y;y')$ the ``$B$'' term in (\ref{prodreduc}) gives,
\begin{eqnarray}
\lefteqn{ (a_y a_{y'})^{D-2} \overline{\partial}_{y} \!\cdot\! 
\partial_{y'} \overline{\partial}_{y'} \!\cdot\! \partial_{y} 
\longrightarrow \Biggl\{a_{y}^{D-2} \overline{\partial}_{y} \!\cdot\! 
\overline{\partial}_{y'} + (\tfrac{D-2}{2}) a^{D-1}_y a_{y'} H^2 \Delta 
\eta_{y y'} \overline{\partial}_{y^0} } \nonumber \\
& & \hspace{-0.5cm} + (D\!-\!3) a_{y}^D H^2 \!-\! \tfrac12 \Bigl[ 
\mathcal{D}_{y} \!-\! m^2 a_{y}^D\Bigr] \Biggr\} \Biggl\{ a_{y'}^{D-2} 
\overline{\partial}_{y} \!\cdot\! \overline{\partial}_{y'} \!-\! 
(\tfrac{D-2}{2}) a_y a^{D-1}_{y'} H^2 \Delta \eta_{y y'} 
\overline{\partial}_{{y'}^0} \nonumber \\
& & \hspace{6cm} + (D\!-\!3) a_{y'}^D H^2 \!-\! \tfrac12 \Bigl[ 
\mathcal{D}_{y'} \!-\! m^2 a_{y'}^D\Bigr] \Biggr\} . \qquad 
\label{term2}
\end{eqnarray}

The 2-point contribution from (\ref{term2}) is easy to read off,
\begin{equation}
{\rm 2\!-\!Point} \Longrightarrow \tfrac14 \!\times\! (\mathcal{D}_{y} 
\!-\! m^2 a_y^D) \!\times\! (\mathcal{D}_{y'} \!-\! m^2 a_{y'}^D) \; . 
\label{B2point}
\end{equation}
This differs from the ``$C$'' contribution (\ref{C2point}) only by the
replacement of $-\frac1{2(D-2)}$ by $\frac14$, so we can immediately
conclude,
\begin{equation}
-i f_{\overline{\overline{4B}}\alpha}(x;x') = - \kappa^2
i\Delta_A(x;x') i\Delta_C(x;x') \; . \label{B4alpha}
\end{equation}
Because the 3-point Donoghue Identity (\ref{Dono3dS}) goes like $1/m^2$,
which can only be compensated by two external derivatives. For $-i
V_{\overline{\overline{4a}}}$ the only 3-point contribution which 
survives for large $m$ is,
\begin{equation} 
\overline{\overline{4aB}} \Longrightarrow -\tfrac12 a_{y'}^{D-2} 
\overline{\partial}_{y} \!\cdot\! \overline{\partial}_{y'}
(\mathcal{D}_{y} \!-\! m^2 a_y^D) - \tfrac12 a_{y}^{D-2}
\overline{\partial}_{y} \!\cdot\! \overline{\partial}_{y'} 
(\mathcal{D}_{y'} \!-\! m^2 a_{y'}^D) \; . \label{Ba3point}
\end{equation}
The other three variations give,
\begin{eqnarray} 
\overline{\overline{4bB}} &\!\!\! \Longrightarrow \!\!\!& 
-\tfrac12 a_{x'}^{D-2} \overline{\partial}_{x} \!\cdot\! 
\overline{\partial}_{x'} (\mathcal{D}_{x} \!-\! m^2 a_x^D) - \tfrac12 
a_{x}^{D-2} \overline{\partial}_{x} \!\cdot\! \overline{\partial}_{x'} 
(\mathcal{D}_{x'} \!-\! m^2 a_{x'}^D) \; , \qquad \label{Bb3point} \\
\overline{\overline{4cB}} &\!\!\! \Longrightarrow \!\!\!& 
-\tfrac12 a_{y'}^{D-2} \overline{\partial}_{x} \!\cdot\! 
\overline{\partial}_{y'} (\mathcal{D}_{x} \!-\! m^2 a_x^D) - \tfrac12 
a_{x}^{D-2} \overline{\partial}_{x} \!\cdot\! \overline{\partial}_{y'} 
(\mathcal{D}_{y'} \!-\! m^2 a_{y'}^D) \; , \qquad \label{Bc3point} \\
\overline{\overline{4dB}} &\!\!\! \Longrightarrow \!\!\!& 
-\tfrac12 a_{x'}^{D-2} \overline{\partial}_{x} \!\cdot\! 
\overline{\partial}_{x'} (\mathcal{D}_{y} \!-\! m^2 a_y^D) - \tfrac12 
a_{y}^{D-2} \overline{\partial}_{y} \!\cdot\! \overline{\partial}_{x'} 
(\mathcal{D}_{x'} \!-\! m^2 a_{x'}^D) \; , \qquad \label{Bd3point}
\end{eqnarray}
Because the external derivatives in (\ref{Ba3point}-\ref{Bb3point})
are either both incoming or outgoing they cancel against the mixed
incoming-outgoing derivatives of (\ref{Bc3point}-\ref{Bd3point}) when
account is taken of the factors of $\delta^D(x - y) \delta^D(x' - y')$,
\begin{eqnarray}
\overline{\partial}_{y} \!\cdot\! \overline{\partial}_{y'} 
\longrightarrow +m^2 a_y a_{y'} + O(m) & , & 
\overline{\partial}_{x} \!\cdot\! \overline{\partial}_{x'} 
\longrightarrow +m^2 a_x a_{x'} + O(m) \; , \qquad \\
\overline{\partial}_{x} \!\cdot\! \overline{\partial}_{y'} 
\longrightarrow -m^2 a_x a_{y'} + O(m) & , & 
\overline{\partial}_{y} \!\cdot\! \overline{\partial}_{x'} 
\longrightarrow -m^2 a_y a_{x'} + O(m) \; . \qquad
\end{eqnarray}

The 4-point ``$B$'' contributions are more complicated because they can
involve as many as four external derivatives, and because the 4-point 
Donoghue Identities (\ref{Dono4dS}-\ref{Dono4bdS}) contain factors of 
both $1/m^2$ and $1/m^4$, as well as up to two external derivatives. It 
is therefore best to make a preliminary reduction based on each external
derivative potentially contributing a factor of $m$,
\begin{eqnarray}
\lefteqn{{\rm 4\!-\!Point} \Longrightarrow (a_y a_{y'})^{D} \Biggl\{ \!
\Bigl(\frac{\overline{\partial}_{y} \!\cdot\! \overline{\partial}_{y'}}{
a_y a_{y'}} \Bigr)^2 \!\!+ H^2 \Bigl(\frac{\overline{\partial}_{y} \!\cdot\! 
\overline{\partial}_{y'}}{a_y a_{y'}} \Bigr) \Bigl[ (\tfrac{D-2}{2}) 
\Delta \eta_{y y'} (\overline{\partial}_{y^0} \!-\! 
\overline{\partial}_{{y'}^0}) } \nonumber \\
& & \hspace{3.9cm} + (D\!-\!3) (\tfrac{a_y}{a_{y'}} \!+\! \tfrac{a_{y'}}{a_y}) 
\Bigr] \!-\! (\tfrac{D-2}{2})^2 H^4 \Delta \eta^2_{y y'} \overline{\partial
}_{y^0} \overline{\partial}_{{y'}^0} \! \Biggr\} . \qquad \label{B4point}
\end{eqnarray}
The appropriate factors for the three other permutations are given by
making the following coordinate replacements in (\ref{B4point}),
\begin{eqnarray}
-i V_{\overline{\overline{4b}}} & \Longrightarrow & x^{\mu}
\longleftrightarrow y^{\mu} \qquad {\rm and} \qquad {x'}^{\mu} 
\longleftrightarrow {y'}^{\mu} \; , \\
-i V_{\overline{\overline{4c}}} & \Longrightarrow & {x'}^{\mu}
\longleftrightarrow {y'}^{\mu} \; , \\
-i V_{\overline{\overline{4d}}} & \Longrightarrow & x^{\mu}
\longleftrightarrow y^{\mu} \; .
\end{eqnarray}
Each of these factors must be multiplied into a factor from the 
appropriate 4-point Donoghue Identity,
\begin{eqnarray}
-i V_{\overline{\overline{4a}}} & \Longrightarrow & \Bigl[-\frac{i}{m^2} 
+ \frac{i}{3 m^4} \Bigl( \frac{\overline{\partial}_{x} \!\cdot\! 
\overline{\partial}_{x'}}{a_x a_{x'}} - m^2\Bigr) \Bigr] 
\frac{\delta^D(x \!-\! y) \delta^D(x' \!-\! y')}{(a_x a_{x'})^D} \; , 
\label{Dono4a} \\
-i V_{\overline{\overline{4b}}} & \Longrightarrow & \Bigl[-\frac{i}{m^2} 
+ \frac{i}{3 m^4} \Bigl( \frac{\overline{\partial}_{y} \!\cdot\! 
\overline{\partial}_{y'}}{a_y a_{y'}} - m^2\Bigr) \Bigr] 
\frac{\delta^D(x \!-\! y) \delta^D(x' \!-\! y')}{(a_y a_{y'})^D} \; , 
\label{Dono4b} \\
-i V_{\overline{\overline{4c}}} & \Longrightarrow & \Bigl[\frac{i}{m^2} 
+ \frac{i}{3 m^4} \Bigl( \frac{\overline{\partial}_{y} \!\cdot\! 
\overline{\partial}_{x'}}{a_y a_{x'}} + m^2\Bigr) \Bigr] 
\frac{\delta^D(x \!-\! y) \delta^D(x' \!-\! y')}{(a_y a_{x'})^D} \; , 
\label{Dono4c} \\
-i V_{\overline{\overline{4d}}} & \Longrightarrow & \Bigl[\frac{i}{m^2} 
+ \frac{i}{3 m^4} \Bigl( \frac{\overline{\partial}_{x} \!\cdot\! 
\overline{\partial}_{y'}}{a_x a_{y'}} + m^2\Bigr) \Bigr] 
\frac{\delta^D(x \!-\! y) \delta^D(x' \!-\! y')}{(a_x a_{y'})^D} \; .
\label{Dono4d}
\end{eqnarray}
Then all four permutations are summed and one discards terms which 
vanish in the large $m$ limit.

It is best to proceed sequentially with each of the four terms in
(\ref{B4point}), and its permutations, times the two terms from 
expressions (\ref{Dono4a}-\ref{Dono4d}). The product of the first 
factors gives,
\begin{eqnarray}
\lefteqn{ \frac{i}{m^2} \Biggl\{ -\Bigl(\frac{\overline{\partial}_{y} 
\!\cdot\! \overline{\partial}_{y'}}{a_y a_{y'}} \Bigr)^2 -\Bigl(
\frac{\overline{\partial}_{x} \!\cdot\! \overline{\partial}_{x'}}{
a_x a_{x'}} \Bigr)^2 + \Bigl( \frac{\overline{\partial}_{x} \!\cdot\! 
\overline{\partial}_{y'}}{a_x a_{y'}} \Bigr)^2  + \Bigl(
\frac{\overline{\partial}_{y} \!\cdot\! \overline{\partial}_{x'}}{
a_y a_{x'}} \Bigr)^2 \Biggr\} } \nonumber \\
& & \hspace{7cm} \longrightarrow i \Bigl[ 4 (D\!-\!2)^2 H^2 - 
\tfrac{2 q^2}{a_x a_{x'}} \Bigr] . \qquad \label{factor11}
\end{eqnarray}
The product of the first factor from (\ref{B4point}) and its 
permutations with the final factor of (\ref{Dono4a}-\ref{Dono4d}) is,
\begin{eqnarray}
\lefteqn{ \frac{i}{3 m^4} \Biggl\{ \Bigl(\frac{\overline{\partial}_{y} 
\!\cdot\! \overline{\partial}_{y'}}{a_y a_{y'}} \Bigr)^2 \Bigl[
\frac{\overline{\partial}_{x} \!\cdot\! \overline{\partial}_{x'}}{a_x 
a_{x'}} \!-\! m^2\Bigr] + \Bigl(\frac{\overline{\partial}_{x} \!\cdot\! 
\overline{\partial}_{x'}}{a_x a_{x'}} \Bigr)^2 \Bigl[
\frac{\overline{\partial}_{y} \!\cdot\! \overline{\partial}_{y'}}{a_y 
a_{y'}} \!-\! m^2\Bigr] } \nonumber \\
& & \hspace{3cm} + \Bigl( \frac{\overline{\partial}_{x} \!\cdot\! 
\overline{\partial}_{y'}}{a_x a_{y'}} \Bigr)^2 \Bigl[
\frac{\overline{\partial}_{y} \!\cdot\! \overline{\partial}_{x'}}{a_y 
a_{x'}} \!+\! m^2\Bigr] + \Bigl(\frac{\overline{\partial}_{y} \!\cdot\! 
\overline{\partial}_{x'}}{a_y a_{x'}} \Bigr)^2 \Bigl[
\frac{\overline{\partial}_{x} \!\cdot\! \overline{\partial}_{y'}}{a_x 
a_{y'}} \!+\! m^2\Bigr] \Biggr\} \nonumber \\
& & \hspace{0cm} \longrightarrow \frac{i}{3} \Bigl[ 4 (D\!-\!1) 
(D \!-\! 2) H^2 - (D\!-\!1)^2 H^2 + \tfrac{q^2}{a_x a_{x'}} \Bigr] 
\; . \qquad \label{factor12}
\end{eqnarray}
The product of the second factor of (\ref{B4point}) and its permutations
with the first factor of (\ref{Dono4a}-\ref{Dono4d}) is,
\begin{eqnarray}
\lefteqn{ i (\tfrac{D-2}{2}) \frac{H^2}{m^2} \Biggl\{ -\Bigl( \frac{
\overline{\partial}_{y} \!\cdot\! \overline{\partial}_{y'}}{a_y a_{y'}}\Bigr) 
\Delta \eta_{y y'} (\overline{\partial}_{y^0} \!-\! \overline{\partial}_{{y'}^0})
-\Bigl( \frac{\overline{\partial}_{x} \!\cdot\! \overline{\partial}_{x'}}{
a_x a_{x'}}\Bigr) \Delta \eta_{x x'} (\overline{\partial}_{x^0} \!-\! 
\overline{\partial}_{{x'}^0}) } \nonumber \\
& & \hspace{1.5cm} +\Bigl( \frac{\overline{\partial}_{y} \!\cdot\! 
\overline{\partial}_{x'}}{a_y a_{x'}}\Bigr) \Delta \eta_{y x'} 
(\overline{\partial}_{y^0} \!-\! \overline{\partial}_{{x'}^0})
+\Bigl( \frac{\overline{\partial}_{x} \!\cdot\! \overline{\partial}_{y'}}{
a_x a_{y'}}\Bigr) \Delta \eta_{x y'} (\overline{\partial}_{x^0} \!-\! 
\overline{\partial}_{{y'}^0}) \Biggr\} \nonumber \\
& & \hspace{5.5cm} \longrightarrow i \tfrac32 (D\!-\!1) (D\!-\!2) a_x a_{x'}
H^4 \Delta \eta^2_{x x'} \; . \qquad \label{factor21}
\end{eqnarray}
For the second factor of (\ref{B4point}) times the second factor of
(\ref{Dono4a}-\ref{Dono4d}) is is only the $a$ and $b$ permutations 
which survive the large $m$ limit,
\begin{eqnarray}
\lefteqn{ i (\tfrac{D-2}{2}) \frac{H^2}{3 m^4} \Biggl\{ \Bigl( \frac{
\overline{\partial}_{y} \!\cdot\! \overline{\partial}_{y'}}{a_y a_{y'}}
\Bigr) \Delta \eta_{y y'} (\overline{\partial}_{y^0} \!-\! 
\overline{\partial}_{{y'}^0}) \Bigl[ \frac{\overline{\partial}_{x} 
\!\cdot\! \overline{\partial}_{x'}}{a_x a_{x'}} \!-\! m^2 \Bigr] } 
\nonumber \\
& & \hspace{3cm} + \Bigl( \frac{\overline{\partial}_{x} \!\cdot\! 
\overline{\partial}_{x'}}{a_x a_{x'}} \Bigr) \Delta \eta_{x x'} 
(\overline{\partial}_{x^0} \!-\! \overline{\partial}_{{x'}^0}) \Bigl[ 
\frac{\overline{\partial}_{y} \!\cdot\! \overline{\partial}_{y'}}{
a_y a_{y'}} \!-\! m^2 \Bigr] \Biggr\} \nonumber \\
& & \hspace{5.5cm} \longrightarrow -\tfrac23 i (D\!-\!1)
(D\!-\!2) a_x a_{x'} H^4 \Delta \eta_{x x'}^2 . \qquad 
\label{factor22}
\end{eqnarray}
The 3rd and 4th factors of (\ref{B4point}), and their permutations, 
contain only two external derivatives so only the first factor of
(\ref{Dono4a}-\ref{Dono4d}) survives the large $m$ limit,
\begin{eqnarray}
\lefteqn{\frac{i H^2}{m^2} \Biggl\{ -\Bigl( \frac{\overline{\partial}_{y} 
\!\cdot\! \overline{\partial}_{y'}}{a_y a_{y'}} \Bigr) (D\!-\!3) 
(\tfrac{a_y}{a_{y'}} \!+\! \tfrac{a_{y'}}{a_y}) - (\tfrac{D-2}{2})^2
\overline{\partial}_{y^0} \overline{\partial}_{{y'}^0} H^2 \Delta 
\eta_{y y'}^2 + \ldots \Biggr\} } \nonumber \\
& & \hspace{2.4cm} \longrightarrow i 4 H^2 \Bigl[ -(D\!-\!3) 
(\tfrac{a_x}{a_{x'}} \!+\! \tfrac{a_{x'}}{a_x}) + (\tfrac{D-2}{2})^2
a_x a_{x'} H^2 \Delta \eta_{x x'}^2 \Bigr] , \qquad \\
& & \hspace{0cm} = i 4 H^2 \Bigl[ -2 (D\!-\!3) + \tfrac14 (D^2 \!-\! 8 D
\!+\! 28) a_x a_{x'} H^2 \Delta \eta_{x x'}^2 \Bigr] . \qquad 
\label{factor341}
\end{eqnarray}

The factors of $a_x a_{x'} H^2 \Delta \eta_{x x'}^2$ in expressions
(\ref{factor21}), (\ref{factor22}) and (\ref{factor341}) take us beyond 
the 4-point Donoghue Identities (\ref{Dono4dS}-\ref{Dono4bdS}). These 
factors must actually be internal to the $d^Dz$ integrations on the 
last lines of expressions (\ref{Dono4dS}-\ref{Dono4bdS}),
\begin{eqnarray}
\lefteqn{ \int \!\! d^Dz \, a_z^D \!\times\! a_x a_z H^2 \Delta \eta_{xz}^2
\!\times\! i\Delta_A(x;z) i\Delta_C(x;z) i\Delta_A(z;x') } \nonumber \\
& & \hspace{1.4cm} + \int \!\! d^Dz \, a_z^D i\Delta_A(x;z) \!\times\! 
a_z a_{x'} H^2 \Delta \eta_{zx'}^2 \!\times\! i\Delta_A(z;x') 
i\Delta_C(z;x') . \qquad \label{Dono4newB}
\end{eqnarray}
The same comments apply to the factors of $q^2/a_x a_{x'}$ which appear
in expressions (\ref{factor11}) and (\ref{factor12}). These factors must
give internal derivatives on the last lines of (\ref{Dono4dS}-\ref{Dono4bdS}),
\begin{eqnarray}
\lefteqn{ \int \!\! d^Dz \, \mathcal{D}_z \Bigl[i\Delta_A(x;z) i\Delta_C(x;z) 
\Bigr] \!\times\! i\Delta_A(z;x') } \nonumber \\
& & \hspace{4cm} + \int \!\! d^Dz \, i\Delta_A(x;z) \!\times\! \mathcal{D}_z
\Bigl[ i\Delta_A(z;x') i\Delta_C(z;x') \Bigr] . \qquad \label{Dono4qsq}
\end{eqnarray} 
With these understandings we can sum the ``$B$'' contributions from 
expressions (\ref{factor11}), (\ref{factor12}), (\ref{factor21}), 
(\ref{factor22}) and (\ref{factor341}) to give,
\begin{eqnarray}
-i f_{\overline{\overline{4B}}\beta}(x;x') &\!\!\! = \!\!\!& \kappa^2
\Bigl[ -\tfrac56 \mathcal{D}_{x'} + F_1(D) a_{x'}^D H^2 \nonumber \\
& & \hspace{1.5cm} + F_2(D) a_x a^{D+1}_{x'} H^4 \Delta \eta_{x x'}^2\Bigr] 
i\Delta_A(x;x') i\Delta_C(x;x') , \qquad \label{B4beta} \\
-i f_{\overline{\overline{4B}}\gamma}(x;x') &\!\!\! = \!\!\!& \kappa^2
\Bigl[ -\tfrac5{6} \mathcal{D}_{x} + F_1(D) a_x^D H^2 \nonumber \\
& & \hspace{1.5cm} + F_2(D) a^{D+1}_x a_{x'} H^4 \Delta \eta_{x x'}^2\Bigr] 
i\Delta_A(x;x') i\Delta_C(x;x') . \qquad \label{B4gamma}
\end{eqnarray}
The functions $F_1(D)$ and $F_2(D)$ are,
\begin{eqnarray}
F_1(D) &\!\!\! \equiv \!\!\!& 2 (D\!-\!2)^2 + \tfrac23 (D\!-\!1) 
(D\!-\!2) - \tfrac16 (D\!-\!1)^2 - 4 (D\!-\!3) \; , \qquad 
\label{F1def} \\
F_2(D) &\!\!\! \equiv \!\!\!& \tfrac18 (D\!-\!2)^2 + \tfrac{5}{12} 
(D\!-\!1) (D\!-\!2) + D \!-\! 3 \; . \label{F2def}
\end{eqnarray}

The ``$A$'' term of (\ref{prodreduc}) seems simpler than ``$B$'' 
because its first factor of $\overline{\partial}_{y} \cdot 
\overline{\partial}_{y'}$ is already reduced. However, the reduction 
of the second factor is complex,
\begin{eqnarray}
\partial_{y} \!\cdot\! \partial_{y'} &\!\!\! = \!\!\!& \tfrac12 
(\partial_{y} \!+\! \partial_{y'})^2 - \tfrac12 \partial_{y}^2 - \tfrac12
\partial_{y'}^2 \; , \\
&\!\!\! \simeq \!\!\!& \tfrac12 \Bigl[ \overline{\partial}_{y} \!+\! 
\tfrac12 \partial_{a_{y}} \!+\! \overline{\partial}_{y'} \!+\! \tfrac12
\partial_{a_{y'}} \Bigr]^2 - \tfrac12 \partial_{y}^2 - \tfrac12
\partial_{y'}^2 \; , \\
&\!\!\! = \!\!\!& \overline{\partial}_{y} \!\cdot\! \overline{\partial}_{y'}
- \tfrac12 \Bigl[\partial_{y} \!\cdot\! (\partial_{y} \!+\! \partial_{a_{y}})
- m^2 a_{y}^2\Bigr] - \tfrac12 \Bigl[\partial_{y'} \!\cdot\! (\partial_{y'} 
\!+\! \partial_{a_{y'}}) - m^2 a_{y'}^2\Bigr] \nonumber \\
& & \hspace{2cm} + \tfrac12 (\overline{\partial}_{y'} \!+\! \partial_{y})
\!\cdot\! \partial_{a_y} + \tfrac12 (\overline{\partial}_{y} \!+\! 
\partial_{y'}) \!\cdot\! \partial_{a_{y'}} + \tfrac18 (\partial_{a_y} \!+\!
\partial_{a_{y'}})^2 . \qquad \label{Afac2}
\end{eqnarray}
The terms on the last line can be usefully rearranged using momentum 
conservation,
\begin{eqnarray}
\tfrac12 (\overline{\partial}_{y'} \!+\! \partial_{y}) \!\cdot\! 
\partial_{a_y} &\!\!\! = \!\!\!& \tfrac12 (\overline{\partial}_{y'} \!-\! 
\overline{\partial}_{y}) \!\cdot\! \partial_{a_y} - \tfrac12 
(\widetilde{\partial}_{y} \!+\! \tfrac12 \partial_{a_y}) \!\cdot\! 
\partial_{a_y} - \tfrac14 \partial^2_{a_y} \; , \qquad \label{ID1} \\
\tfrac12 (\overline{\partial}_{y} \!+\! \partial_{y'}) \!\cdot\! 
\partial_{a_{y'}} &\!\!\! = \!\!\!& \tfrac12 (\overline{\partial}_{y} \!-\! 
\overline{\partial}_{y'}) \!\cdot\! \partial_{a_{y'}} - \tfrac12 
(\widetilde{\partial}_{y'} \!+\! \tfrac12 \partial_{a_{y'}}) \!\cdot\! 
\partial_{a_{y'}} - \tfrac14 \partial^2_{a_{y'}} \; . \qquad \label{ID2}
\end{eqnarray}
Substituting (\ref{ID1}-\ref{ID2}) in (\ref{Afac2}) and using the
reflection identity gives,
\begin{eqnarray}
\lefteqn{\partial_{y} \!\cdot\! \partial_{y'} \simeq \overline{\partial}_{y} 
\!\cdot\! \overline{\partial}_{y'} - \tfrac12 \Bigl[\partial_{y} \!\cdot\! 
(\partial_{y} \!+\! \partial_{a_{y}}) - m^2 a_{y}^2\Bigr] - \tfrac12 \Bigl[
\partial_{y'} \!\cdot\! (\partial_{y'} \!+\! \partial_{a_{y'}}) - m^2 
a_{y'}^2\Bigr] } \nonumber \\
& & \hspace{0cm} - \tfrac12 (\overline{\partial}_{y} \!-\! 
\overline{\partial}_{y'}) \!\cdot\! (\partial_{a_y} \!-\! \partial_{a_{y'}})
- \tfrac12 (\widetilde{\partial}_{y} \!+\! \tfrac12 \partial_{a_y}) 
\!\cdot\! (\partial_{a_y} \!-\! \partial_{a_{y'}}) - \tfrac18 (\partial_{a_y}
\!-\! \partial_{a_{y'}})^2 \; . \qquad \label{Afac2p}
\end{eqnarray}

Expression (\ref{Afac2p}) can be further simplified using the relation,
\begin{equation}
\partial_{a_y}^{\mu} \!-\! \partial_{a_{y'}}^{\mu} = -(D \!-\! 2) a_y a_{y'}
H^2 \Delta \eta_{y y'}^2 \; .
\end{equation}
Multiplying by $(a_y a_{y'})^{D-2} \overline{\partial}_{y} \cdot 
\overline{\partial}_{y'}$ gives the total ``$A$'' factor,
\begin{eqnarray}
\lefteqn{ (a_y a_{y'})^{D-2} \overline{\partial}_{y} \!\cdot\! 
\overline{\partial}_{y'} \, \partial_{y} \!\cdot\! \partial_{y'} \simeq (a_y
a_{y'})^{D} \Bigl( \frac{\overline{\partial}_{y} \!\cdot\! 
\overline{\partial}_{y'}}{a_y a_{y'}} \Bigr)^2 - \tfrac12 a_{y'}^{D-2} 
\overline{\partial}_{y} \!\cdot\! \overline{\partial}_{y'} \Bigl[ 
\widetilde{D}_{y} \!-\! m^2 a_{y}^D\Bigr] } \nonumber \\
& & \hspace{-0.5cm} - \tfrac12 a_{y}^{D-2} \overline{\partial}_{y} \!\cdot\! 
\overline{\partial}_{y'} \Bigl[ \widetilde{D}_{y'} \!-\! m^2 a_{y'}^D\Bigr]
+ (\tfrac{D-2}{2}) H^2 (a_y a_{y'})^D \Bigl( \frac{\overline{\partial}_{y} 
\!\cdot\! \overline{\partial}_{y'}}{a_y a_{y'}} \Bigr) \Biggl\{ \Delta 
\eta_{y y'} ( \overline{\partial}_{y^0} \!-\! \overline{\partial}_{{y'}^0}) 
\nonumber \\
& & \hspace{2.5cm} + \Delta \eta_{y y'} \Bigl[ \widetilde{\partial}_{y^0} \!+\! 
(\tfrac{D-2}{2}) a_y H\Bigr] \!+\! \tfrac12 \!+\! (\tfrac{D-1}{4}) a_{y} a_{y'} 
H^2 \Delta \eta_{y y'}^2 \Biggr\} . \qquad 
\end{eqnarray}
There are no 2-point contributions and, like the ``$B$'' case, the 3-point
contributions cancel. Most of the 4-point contributions are similar to those
of (\ref{B4point}) and can be read off from expressions (\ref{factor11}),
(\ref{factor12}), (\ref{factor21}), (\ref{factor22}) and (\ref{factor341}).
The only exception is penultimate term, which involves a derivative 
$\widetilde{\partial}_{y^0} + (\tfrac{D-2}{2}) a_y H \equiv \widetilde{D}_{y}$ 
of the $C$-type propagator. We understand this as acting on the internal 
propagator in the same sense as the scale factors. The final result is,
\begin{eqnarray}
\lefteqn{-i f_{\overline{\overline{4A}}\beta}(x;x') = \kappa^2 \Bigl[ 
-\tfrac5{6} \mathcal{D}_{x} + F_3(D) a_{x'}^D H^2 + F_4(D) a_x a^{D+1}_{x'} 
H^4 \Delta \eta_{x x'}^2\Bigr] } \nonumber \\
& & \hspace{-0.5cm} \times i\Delta_A(x;x') i\Delta_C(x;x') + (D\!-\!2) 
\kappa^2 H^2 a_{x'}^D i\Delta_A(x;x') \Delta \eta_{x x'} \widetilde{D}_{x} 
i\Delta_C(x;x') , \qquad \label{A4beta} \\
\lefteqn{-i f_{\overline{\overline{4A}}\gamma}(x;x') = \kappa^2 \Bigl[ 
-\tfrac5{6} \mathcal{D}_x + F_4(D) a_x^D H^2 + F_4(D) a^{D+1}_x a_{x'} 
H^4 \Delta \eta_{x x'}^2\Bigr] } \nonumber \\
& & \hspace{-0.5cm} \times i\Delta_A(x;x') i\Delta_C(x;x') + (D\!-\!2) 
\kappa^2 H^2 a_{x}^D i\Delta_A(x;x') \Delta \eta_{x x'} \widetilde{D}_{x} 
i\Delta_C(x;x') . \qquad \label{A4gamma}
\end{eqnarray}
The functions $F_3(D)$ and $F_4(D)$ are,
\begin{eqnarray}
F_3(D) &\!\!\! \equiv \!\!\!& 2 (D\!-\!2)^2 + \tfrac23 (D\!-\!1) 
(D\!-\!2) - \tfrac16 (D\!-\!1)^2 - \tfrac12 (D\!-\!2) \; , \qquad 
\label{F3def} \\
F_4(D) &\!\!\! \equiv \!\!\!& \tfrac{2}{3} (D\!-\!1) (D\!-\!2) \; . 
\label{F4def}
\end{eqnarray}

It is time to sum up all the contributions from $-i V_{\overline{\overline{4}}}$
to the four classes of 2-point forms in Figure~\ref{Reduced}. The ones of Class 
$\alpha$ come from expressions (\ref{C4alpha}) and (\ref{B4alpha}),
\begin{equation}
-i f_{\overline{\overline{4}}\alpha}(x;x') = -(\tfrac{D-8}{D-2}) \kappa^2
i\Delta_A(x;x') i\Delta_C(x;x') \; . \label{4alpha}
\end{equation}
Class $\beta$ diagrams derive from expressions (\ref{C4beta}), (\ref{B4beta})
and (\ref{A4beta}),
\begin{eqnarray}
\lefteqn{-i f_{\overline{\overline{4}}\beta}(x;x') = \kappa^2 \Bigl[ 
-\tfrac{(5 D - 11)}{3 (D-2)} \mathcal{D}_{x'} + G_1(D) a_{x'}^D H^2 + G_2(D) 
a_x a^{D+1}_{x'} H^4 \Delta \eta_{x x'}^2\Bigr] } \nonumber \\
& & \hspace{-0.5cm} \times i\Delta_A(x;x') i\Delta_C(x;x') + (D\!-\!2) 
\kappa^2 H^2 a_{x'}^D i\Delta_A(x;x') \Delta \eta_{x x'} \widetilde{D}_{x} 
i\Delta_C(x;x') , \qquad \label{4beta}
\end{eqnarray}
where the functions $G_1(D)$ and $G_2(D)$ are,
\begin{eqnarray}
G_1(D) &\!\!\! \equiv \!\!\!& -\tfrac{(D-1)^2}{3 (D-2)} + 4 (D-2)^2 + \tfrac43 
(D\!-\!1) (D\!-\!2) \nonumber \\
& & \hspace{4.5cm} - \tfrac13 (D\!-\!1)^2 - \tfrac12 (D\!-\!2) - 4 (D\!-\!3)
\; , \label{G1def} \qquad \\
G_2(D) &\!\!\! \equiv \!\!\!& \tfrac18 (D\!-\!2)^2 + \tfrac{13}{12} (D\!-\!1)
(D\!-\!2) + D \!-\! 3 \; . \label{G2def} \qquad
\end{eqnarray}
Expressions (\ref{C4gamma}), (\ref{B4gamma}) and (\ref{A4gamma}) give the
Class $\gamma$ diagrams,
\begin{eqnarray}
\lefteqn{-i f_{\overline{\overline{4}}\gamma}(x;x') = \kappa^2 \Bigl[ 
-\tfrac{(5 D - 11)}{3 (D-2)} \mathcal{D}_{x} + G_1(D) a_x^D H^2 + G_2(D) 
a^{D+1}_x a_{x'} H^4 \Delta \eta_{x x'}^2\Bigr] } \nonumber \\
& & \hspace{-0.5cm} \times i\Delta_A(x;x') i\Delta_C(x;x') + (D\!-\!2) 
\kappa^2 H^2 a_{x}^D i\Delta_A(x;x') \Delta \eta_{x x'} \widetilde{D}_{x} 
i\Delta_C(x;x') , \qquad \label{4gamma}
\end{eqnarray}
There are no Class $\delta$ diagrams from $-i V_{\overline{\overline{4}}}$.

\subsection{Renormalization}

Recall the four classes of diagrams described in Figure~\ref{Reduced}
which contribute to the gauge invariant self-mass,
\begin{equation}
M^2_{\rm inv}(x;x') = \mathcal{D}_x \mathcal{D}_{x'} f_{\alpha}(x;x') + 
\mathcal{D}_x f_{\beta}(x;x') + \mathcal{D}_{x'} f_{\gamma}(x;x') +
f_{\delta}(x;x') \; . \label{invM2again}
\end{equation}
The only contribution to Class $\delta$ is from the original, gauge-dependent
result $-i f_{\delta}(x;x') = -i M_{0}(x;x')$. In section 5.3 we derived the
$-i V_{\overline{\overline{4}}}$ contributions to Classes $\alpha$, $\beta$ 
and $\gamma$ in expressions (\ref{4alpha}), (\ref{4beta}) and (\ref{4gamma}),
respectively. Section 5.2 gives the Class $\beta$ and $\gamma$ contributions
from $-i V_{\overline{5}}$ in expressions (\ref{fbeta5}) and (\ref{fgamma5}),
respectively. (There are no Class $\alpha$ contributions from 
$-i V_{\overline{5}}$.) Finally, there are Class $\beta$ and $\gamma$ 
contributions from $-i V_{1b}(x;x')$ and $-i V_{1c}(x;x')$, respectively,
\begin{eqnarray}
-i f_{1\beta}(x;x') &\!\!\! = \!\!\!& i\kappa^2 \Bigl[ \tfrac{D}2 
(\tfrac{D-1}{2}) i\Delta_A + (\tfrac{D - 1}{2}) i\Delta_B + \tfrac12 
i\Delta_C \Bigr] \delta^D(x \!-\! x') \; , \qquad \label{1beta} \\
-i f_{1\gamma}(x;x') &\!\!\! = \!\!\!& i\kappa^2 \Bigl[ \tfrac{D}2 
(\tfrac{D-1}{2}) i\Delta_A + (\tfrac{D - 1}{2}) i\Delta_B + \tfrac12 
i\Delta_C \Bigr] \delta^D(x \!-\! x') \; . \qquad \label{1gamma}
\end{eqnarray}

To carry out renormalization we must first localize the primitive 
divergences using expressions (\ref{DAexp}-\ref{DCexp}) for the three 
propagators. The simplest case is the coincidence limits which appear 
in (\ref{1beta}-\ref{1gamma}),
\begin{eqnarray}
\lefteqn{\tfrac{D}{2} (\tfrac{D-1}{2}) i\Delta_A + (\tfrac{D - 1}{2}) 
i\Delta_B + \tfrac12 i\Delta_C = -\frac{H^{D-2}}{(4 \pi)^{\frac{D}2}} 
\frac{\Gamma(D\!-\!1)}{\Gamma(\frac{D}2)} \Bigl[ \pi {\rm cot}(
\tfrac{D \pi}{2}) \!-\! 2 \ln(a)\Bigr] } \nonumber \\
& & \hspace{1.5cm} \times  \tfrac{D}{2} (\tfrac{D-1}{2}) - 
\frac{H^{D-2}}{(4 \pi)^{\frac{D}2}} \frac{\Gamma(D\!-\!2)}{
\Gamma(\frac{D}2)} \!\times\! (\tfrac{D-1}{2}) + \frac{H^{D-2}}{
(4 \pi)^{\frac{D}2}} \frac{\Gamma(D\!-\!3)}{\Gamma(\frac{D}2)}
\!\times\! \tfrac12 \; . \qquad \label{coinc}
\end{eqnarray}
There are also divergences which appear in two products of 
noncoincident propagators,
\begin{equation}
i\Delta_A(x;x') i\Delta_C(x;x') \qquad , \qquad i\Delta_A(x;x') 
\Delta \eta_{x x'} \widetilde{D}_{x} i\Delta_C(x;x') \; , 
\label{products}
\end{equation}
where we recall,
\begin{equation}
\widetilde{D}_x \equiv \frac{\partial}{\partial x^0} + (\tfrac{D-2}{2})
a_x H \; . \label{tildeDdef}
\end{equation}
There are no divergences in the product,
\begin{equation}
a_x a_{x'} H^2 \Delta \eta_{x x'}^2 i\Delta_A(x;x') i\Delta_C(x;x') \; .
\end{equation}

The two products are reduced by first expanding, retaining only the 
potentially divergent terms in $D$ dimensions and setting $D=4$ for the
others,
\begin{equation}
i\Delta_A i\Delta_C = \frac{\Gamma^2(\frac{D}2 \!-\! 1)}{16 \pi^D} \Biggl\{
\frac1{[ a a' \Delta x^2]^{D-2}} - \frac{H^2 \ln(\frac14 H^2 \Delta x^2)}{2 
a a' \Delta x^2} + O(D\!-\!4)\Biggr\} . \label{DADCexp}
\end{equation}
The next step is to localize the ultraviolet divergence by extracting a 
derivative and then adding zero in the form of the flat space propagator 
equation \cite{Onemli:2002hr,Onemli:2004mb},
\begin{eqnarray}
\lefteqn{ \frac1{\Delta x^{2D-4}} = \frac{\partial^2}{2 (D\!-\!3) (D\!-\!4)} 
\Biggl[ \frac1{\Delta x^{2D - 6}} \Biggr] \; , } \\
& & \hspace{0cm} = \frac{\mu^{D-4}}{2 (D\!-\!3) (D\!-\!4)} 
\frac{4 \pi^{\frac{D}2} i \delta^D(x \!-\! x')}{\Gamma(\frac{D}2 \!-\! 1)} 
\nonumber \\
& & \hspace{5cm} + \frac{\partial^2}{2 (D\!-\!3) (D\!-\!4)} \Biggl[ 
\frac1{\Delta x^{2D -6}} - \frac{\mu^{D-4}}{\Delta x^{D-2}} \Biggr] , 
\qquad \\
& & \hspace{0cm} = \frac{\mu^{D-4}}{2 (D\!-\!3) (D\!-\!4)} 
\frac{4 \pi^{\frac{D}2} i \delta^D(x \!-\! x')}{\Gamma(\frac{D}2 \!-\! 1)} -
\frac{\partial^2}{4} \Biggl[ \frac{\ln( \mu^2 \Delta x^2)}{\Delta x^2}
\Biggr] + O(D \!-\! 4) \; . \qquad 
\end{eqnarray}
The final result for the first product is,
\begin{eqnarray}
\lefteqn{i\Delta_A i\Delta_C \longrightarrow \frac{\mu^{D-4} \Gamma(\frac{D}2 
\!-\! 1)}{8 (D\!-\!3) (D\!-\!4) \pi^{\frac{D}2}} \frac{i \delta^D(x\!-\!x')}{
(a a')^{D-2}} } \nonumber \\
& & \hspace{4cm} - \frac{\partial^2}{64 \pi^4 (a a')^2} \Bigl[ \frac{\ln(\mu^2 
\Delta x^2)}{\Delta x^2} \Bigr] - \frac{H^2 \ln(\frac14 H^2 \Delta x^2)}{32 
\pi^4 a a' \Delta x^2} . \qquad \label{product1}
\end{eqnarray}
Using expression (\ref{tildeDdef}) in the second product of (\ref{products})
gives,
\begin{eqnarray}
\lefteqn{ i\Delta_A \!\!\times\!\! \Delta \eta \widetilde{D}_x i\Delta_C =
\frac{\Gamma^2(\frac{D}2 \!-\! 1)}{16 \pi^D} \Biggl\{ \!\!\frac{(D \!-\! 2)
\Delta \eta^2}{(a a')^{D-2} \Delta x^{2D - 2}} \!-\! \frac{H^2 \Delta \eta^2
\ln(\frac14 H^2 \Delta x^2)}{a a' \Delta x^4} \!\! \Biggr\} , } \\
& & \hspace{-0.5cm} = \frac{\Gamma^2(\frac{D}2 \!-\! 1)}{16 \pi^D} \Biggl\{
-\frac1{2 (a a')^{D-2} \Delta x^{2D-4}} + \frac{H^2 [\ln(\frac14 H^2 
\Delta x^2) \!+\! 1]}{2 a a' \Delta x^2} \nonumber \\
& & \hspace{1.6cm} + \frac{\partial_0^2}{4 (a a')^2} \Bigl[\frac1{\Delta x^2} 
\Bigr] + \frac{H^2 \partial_0^2}{8 a a'} \Bigl[\ln^2(\tfrac14 H^2 \Delta x^2) 
\!+\! 2 \ln(\tfrac14 H^2 \Delta x^2)\Bigr] \Biggr\} , \qquad \\
& & \hspace{-0.5cm} \longrightarrow -\frac{\mu^{D-4} \Gamma(\frac{D}2 \!-\! 1)
}{16 (D\!-\!3) (D\!-\!4) \pi^{\frac{D}2}} \frac{i \delta^D(x\!-\!x')}{
(a a')^{D-2}} \nonumber \\
& & \hspace{1cm} + \frac{\partial^2}{128 \pi^4 (a a')^2} \Bigl[ 
\frac{\ln(\mu^2 \Delta x^2)}{\Delta x^2}\Bigr] + \frac{H^2 [\ln(\frac14 H^2
\Delta x^2) \!+\! 1]}{32 \pi^4 a a' \Delta x^2} \nonumber \\
& & \hspace{0.5cm} + \frac{\partial_0^2}{64 \pi^4 (a a')^2} \Bigl[
\frac1{\Delta x^2} \Bigr] + \frac{H^2 \partial_0^2}{128 \pi^4 a a'} \Bigl[
\ln^2(\tfrac14 H^2 \Delta x^2) \!+\! 2 \ln(\tfrac14 H^2 \Delta x^2)\Bigr] . 
\qquad \label{product2}
\end{eqnarray}

The 1-loop divergences of the invariant self-mass are canceled by three
counterterms,
\begin{equation}
\Delta \mathcal{L} = -\tfrac12 \alpha_1 \square \phi \square \phi \sqrt{-g}
- \tfrac12 \alpha_2 R \partial_{\mu} \phi \partial_{\nu} \phi g^{\mu\nu}
\sqrt{-g} - \tfrac12 \alpha_3 R \partial_0 \phi \partial_0 \phi g^{00}
\sqrt{-g} \; , \label{DL}
\end{equation}
where the Ricci scalar in de Sitter is $R = D (D-1) H^2$. The associated
contribution to the renormalized self-mass is,
\begin{eqnarray}
\lefteqn{-i M^2_{\rm ctm}(x;x') = -i \alpha_1 \mathcal{D}_x \mathcal{D}_{x'} 
\Bigl[\frac{\delta^D(x \!-\! x')}{a_x^D} \Bigr] } \nonumber \\
& & \hspace{3.3cm} + i \alpha_2 R \mathcal{D}_x \delta^D(x \!-\! x') - i 
\alpha_3 R \partial_0 \Bigl[a^{D-2} \partial_0 \delta^D(x \!-\! x')\Bigr] 
. \qquad \label{Mctm}
\end{eqnarray}
From relations (\ref{coinc}), (\ref{product1}) and (\ref{product2}) we see
that all divergences can be expressed in terms of two constants,
\begin{eqnarray}
K &\!\!\! \equiv \!\!\!& \frac{\mu^{D-4} \Gamma(\frac{D}2 \!-\! 1)}{8 
(D\!-\!3) (D \!-\! 4) \pi^{\frac{D}2}} = \frac{\mu^{D-4}}{8 \pi^2} \!\times\!
\frac1{D \!-\! 4} + O\Bigl( (D\!-\!4)^0\Bigr) \; , \qquad \label{Kdef} \\
\mathcal{K} &\!\!\! \equiv \!\!\!& \frac{\mu^{D-4}}{(4 \pi)^{\frac{D}2}}
\frac{\Gamma(D \!-\! 1)}{\Gamma(\frac{D}2)} \!\times\! \tfrac{\pi}{2} 
{\rm cot}(\tfrac{D \pi}{2}) = \frac{\mu^{D-4}}{8 \pi^2} \!\times\! 
\frac1{D \!-\!4} + O\Bigl( (D\!-\!4)^0\Bigr) \; . \qquad \label{scriptKdef}
\end{eqnarray}
The counterterms required to renormalize the original, gauge-dependent
self-mass are $(\alpha_1)_{0} = 0$ and \cite{Glavan:2021adm},
\begin{equation}
R (\alpha_2)_0 = -\kappa^2 H^2 \!\times\! (D \!-\! 2) K \quad , \quad 
R (\alpha_3)_0 = -\kappa^2 H^2 \Bigl[ 2 D \mathcal{K} - (D\!-\!2)
(D\!+\!4) K \Bigr] \; . \label{ctm0}
\end{equation}
From expressions (\ref{1beta}-\ref{1gamma}) we see that the $-i V_{1b,c}$
diagrams have $(\alpha_1)_1 = 0 = (\alpha_3)_1$ and,
\begin{equation}
R (\alpha_2)_1 = \kappa^2 H^2 \!\times\! D (D\!-\!1) \mathcal{K} \; . 
\label{ctm1}
\end{equation}
The $-i V_{\overline{\overline{4}}}$ diagrams (\ref{4alpha}), (\ref{4beta}) 
and (\ref{4gamma}) imply $(\alpha_3)_4 = 0$ and,
\begin{equation}
(\alpha_1)_4 = -\tfrac{\kappa^2}{3} (\tfrac{13 D - 46}{D-2}) K \qquad , 
\qquad R (\alpha_2)_4 = \kappa^2 H^2 \Bigl[-2 G_1(D) + (D\!-\!2)\Bigr] K 
\; , \label{ctm4}
\end{equation}
where $G_1(D)$ was defined in expression (\ref{G1def}). Finally, the $-i 
V_{\overline{5}}$ diagrams (\ref{fbeta5}-\ref{fgamma5}) imply 
$(\alpha_3)_5 = 0$ and,
\begin{equation}
(\alpha_1)_5 = -\kappa^2 K \qquad , \qquad R (\alpha_2)_5
= 2 \kappa^2 H^2 \mathcal{K} \; . \label{ctm5}
\end{equation}
Summing the contributions (\ref{ctm0}-\ref{ctm5}) gives,
\begin{eqnarray}
\alpha_1 &\!\!\! = \!\!\!& -\frac{2 \kappa^2}{8 \pi^2} \!\times\! 
\frac{\mu^{D-4}}{D \!-\! 4} + O(1) \; , \qquad \label{alpha1} \\
R \alpha_2 &\!\!\! = \!\!\!& -\frac{15 \kappa^2 H^2}{8 \pi^2} \!\times\! 
\frac{\mu^{D-4}}{D \!-\! 4} + O(1) \; , \qquad \label{alpha2} \\
R \alpha_3 &\!\!\! = \!\!\!& +\frac{8 \kappa^2 H^2}{8 \pi^2} \!\times\! 
\frac{\mu^{D-4}}{D \!-\! 4} + O(1) \; . \qquad \label{alpha3}
\end{eqnarray}

\section{Conclusions}

No one disputes that an epoch of primordial inflation would produce a vast
ensemble of cosmological scale gravitons \cite{Starobinsky:1979ty}. At some
level these quanta must interact with themselves and with other particles
to induce changes in kinematics and in long range forces. A fascinating 
aspect of these changes is that the continual production of infrared 
gravitons causes them to grow without bound in time and sometimes also in 
space. During a very long period of inflation, this growth can overwhelm the 
small coupling constants to produce significant effects which might persist
to the present day \cite{Woodard:2023cqi}. However, explicit computations 
\cite{Miao:2006gj,Glavan:2013jca,Wang:2014tza,Tan:2021lza,Glavan:2021adm,
Tan:2022xpn} have been criticized as potentially unphysical owing to gauge 
dependence \cite{Higuchi:2011vw,Miao:2011ng,Morrison:2013rqa,Miao:2013isa}.

Opinions differ on how to deal with gauge dependence. Some people favor 
replacing gauge-dependent Green's functions with expectation values of 
generally coordinate invariant operators \cite{Tsamis:1989yu,Modanese:1994wv,
Rovelli:2001my,Giddings:2005id,Green:2008kj,Donnelly:2015hta,Marolf:2015jha,
Miao:2017vly,Frob:2017apy,Frob:2017gyj,Wilson-Gerow:2020jmv}. We have 
instead pursued the approach of combining bits and pieces of gauge-dependent 
Green's functions so as to achieve a gauge-independent result based on the
flat space S-matrix.

In our view the usual effective field equations are gauge-dependent because 
they ignore quantum gravitational correlations with the source, which excites 
the effective field, and with the observer which detects it. We restore these 
correlations by first forming the position-space amplitudes that represent 
quantum gravitational corrections to the $t$-channel scattering between two 
massive fields by the exchange of the light field under study (in this case, 
a massless, minimally coupled scalar). These amplitudes involve 2-point, 
3-point and 4-point correlators. Our second step is to simplify the 3-point 
and 4-point correlators to 2-point form using a series of relations for 
extracting the $t$-channel contributions, first derived in flat space by 
Donoghue and collaborators \cite{Donoghue:1993eb,Donoghue:1994dn,
Bjerrum-Bohr:2002aqa,Donoghue:1996mt}. Our final step is to insert exchange 
propagators of the light field so that we can regard each of the 2-point 
contributions as a correction to the 1PI 2-point function of the light 
field. This has been explicitly carried out on flat space background, and 
verified to be independent of the gauge, for 1-graviton loop corrections to 
a massless, minimally coupled scalar \cite{Miao:2017feh} and for 1-graviton 
loop corrections to electromagnetism \cite{Katuwal:2021thy}. The current 
work generalizes this procedure to de Sitter background for 1-graviton loop 
corrections to a massless, minimally coupled scalar.

Our de Sitter analysis has been enormously simplified by correspondence 
with its flat space analog \cite{Miao:2017feh}. In particular, we used 
the very same Lagrangian (\ref{Lagrangian}), and the very same five
diagrams (Figures~\ref{Diagram1}-\ref{Diagram5}) which can potentially
enhance $t$-channel scattering between massive sources. The de Sitter
generalizations (\ref{Dono3dS}-\ref{der3ptDon}) and 
(\ref{Dono4dS}-\ref{Dono4bdS}) of the Donoghue Identities were motivated
by minimally preserving the transformation properties of the flat space
results. One new feature of our analysis is the recognition in section 4
of two classes of the five diagrams which can be consolidated to give
two residual amplitudes, $-i V_{\overline{\overline{4}}}(x;x';y;y')$
--- given in expressions (\ref{4abarbar}-\ref{4dbarbar}) --- and $-i 
V_{\overline{5}}(x;x';y;y')$ --- given in expressions 
(\ref{5abar}-\ref{5dbar}). This consolidation is exact, independent
of the Donoghue Identities, and valid for any gauge on any cosmological 
background. Of course the final reduction of these diagrams to 2-point 
form in sections 5.2 and 5.3 does require a specific gauge on de Sitter 
and the Donoghue Identities. Our final primitive results 
(\ref{fbeta5}-\ref{fgamma5}), (\ref{4alpha}), (\ref{4beta}), 
(\ref{4gamma}) and (\ref{1beta}-\ref{1gamma}) are expressed in terms of 
the four classes of 2-point diagrams defined in Figure~\ref{Reduced}. 
Renormalization is carried out in section 5.4 using the same three 
counterterms (\ref{DL}) of the original, gauge-dependent computation 
\cite{Glavan:2021adm}. Our final results for the gauge-independent
coefficients of these counterterms are given in equations 
(\ref{alpha1}-\ref{alpha3}).

Of course the reason for purging the scalar effective field equation 
of gauge dependence is to confirm or refute the reality of the large 
logarithmic correction that was found (using the simplest gauge) to the 
scalar exchange potential \cite{Glavan:2021adm}. Recall the relation that 
was found in the gauge-dependent calculation between large logarithms in 
the exchange potential and the curvature-dependent field strength
renormalization $-\frac12 \alpha_2 R \partial_{\mu} \phi \partial_{\nu} 
\phi g^{\mu\nu} \sqrt{-g}$ \cite{Glavan:2021adm}. Because the 
renormalized, gauge invariant self-mass has the same functional form as 
its gauge-dependent ancestor, this relation should continue to pertain. 
That means we can read off the gauge-independent logarithm from our result
(\ref{alpha2}) for the coefficient $\alpha_2 R$. The fact that it is 
nonzero contradicts the hypothesis that graviton-induced logarithms are 
a gauge artifact.

We stress the importance of making a direct check of gauge dependence
by re-doing the computation in the de Sitter generalization 
\cite{Glavan:2019msf} of the same 2-parameter family of gauges which 
was employed to check the flat space result \cite{Miao:2017feh}. One
reason is that there may be diagrams, such as external line corrections,
which do not enhance $t$-channel exchange in flat space but may do so in
de Sitter. Another reason is that we know of no way to derive the de 
Sitter Donoghue Identities (\ref{Dono3dS}-\ref{der3ptDon}) and 
(\ref{Dono4dS}-\ref{Dono4bdS}). Their flat space antecedents could be
(and were) derived \cite{Donoghue:1993eb,Donoghue:1994dn,
Bjerrum-Bohr:2002aqa,Donoghue:1996mt} by computing the scattering 
amplitudes in momentum space and then extracting the leading 
contributions for small $t$. What we did is to first express the same 
relations in flat position space (which is exact) and then generalize 
them to de Sitter using minimal coupling (which is a guess). We will 
not feel completely confident until the gauge independence of the 
final result has been confirmed.

\vspace{0cm}

\centerline{\bf Acknowledgements}

DG was supported by the European Union and the Czech Ministry of
Education, Youth and Sports (Project: MSCA Fellowship CZ FZU I ---
CZ.02.01.01/00/22 010/0002906). SPM was supported by Taiwan NSTC 
grants 111-2112-M-006-038 and 112-2112-M-006-017. TP was supported by 
the D-ITP consortium, a program of the Neth\-erlands Organization for 
Scientific Research (NWO) that is funded by the Dutch Ministry of 
Education, Culture and Science (OCW). RPW was supported by NSF grant 
PHY-2207514 and by the Institute for Fundamental Theory at the 
U. of Florida.

\end{document}